%% file: M82.tex
\documentclass[article]{emulateapj}

\usepackage{natbib}
\usepackage{graphics}
\usepackage{threeparttable}
\usepackage{amsmath}
\usepackage{bm}
\usepackage{subfigure}

\shorttitle{MOLECULAR GAS IN M82}
\shortauthors{KAMENETZKY ET AL.}

\newcommand{\unit}[1]{\ensuremath{\, \mathrm{#1}}}
\newcommand{\ms}{M$_{\odot}$}
\newcommand{\ls}{L$_{\odot}$}
\newcommand{\pyr}{yr$^{-1}$}
\newcommand{\kmspc}{$\,\rm km\,s^{-1}\,pc^{-1}$}
\newcommand{\kms}{$\,\rm km\,s^{-1}$}

\newcommand{\as}{$^{\prime\prime}$}

\newcommand{\eqq}{\!=\!}  
\newcommand{\too}{\!\rightarrow\!} 
\newcommand{\jone}{{$\rm J\eqq 1\too0$}}
\newcommand{\jtwo}{{$ \rm J\eqq2\too1$}}

\newcommand{\jfour}{{$\rm J\eqq4\too3$}}
\newcommand{\jfive}{{$\rm J\eqq5\too4$}}
\newcommand{\jsix}{{$\rm J\eqq6\too5$}}
\newcommand{\jseven}{{$\rm J\eqq7\too6$}}
\newcommand{\jeight}{{$\rm J\eqq8\too7$}}
\newcommand{\jnine}{{$\rm J\eqq9\too8$}}
\newcommand{\jten}{{$\rm J\eqq10\too9$}}
\newcommand{\jeleven}{{$\rm J\eqq11\too10$}}
\newcommand{\jtwelve}{{$\rm J\eqq12\too11$}}
\newcommand{\jthirteen}{{$\rm J\eqq13\too12$}}

\begin{document}
\bibliographystyle{apj_w_etal}
\title{{\it Herschel}-SPIRE Imaging Spectroscopy of Molecular Gas in M82}

\author{J.~Kamenetzky\altaffilmark{1},
 J.~Glenn\altaffilmark{1},
 N.~Rangwala\altaffilmark{1},
 P.~Maloney\altaffilmark{1},
 M.~Bradford\altaffilmark{2},
 C.D.~Wilson\altaffilmark{3},
 G.J.~Bendo\altaffilmark{4},
 M.~Baes\altaffilmark{5}
 A.~Boselli\altaffilmark{6},
 A.~Cooray\altaffilmark{7},
 K.G.~Isaak\altaffilmark{8},
 V.~Lebouteiller\altaffilmark{9},
 S.~Madden\altaffilmark{9},
 P.~Panuzzo\altaffilmark{9},
 M.R.P.~Schirm\altaffilmark{3},
 L. Spinoglio\altaffilmark{10},
 R.~Wu\altaffilmark{9}}

\altaffiltext{1}{Center for Astrophysics and Space Astronomy, 389-UCB, 
University of Colorado, Boulder, CO, 80303} 

\altaffiltext{2}{NASA Jet Propulsion Laboratory, Pasadena, CA, 91109} 

\altaffiltext{3}{Dept. of Physics \& Astronomy, McMaster University, Hamilton, 
Ontario, L8S 4M1, Canada}

\altaffiltext{4}{UK ALMA Regional Centre Node, Jordell Bank Center for Astrophysics, School of Physics and Astronomy, University of Manchester, Oxford Road, Manchester M13 9PL, U.K.}

\altaffiltext{5}{Sterrenkundig Observatorium, Universiteit Gent, Krijgslaan 281 S9, 9000 Gent, Belgium}

\altaffiltext{6}{Laboratoire d'Astrophysique de Marseille, UMR6110 CNRS, 38
rue F. Joliot-Curie, 13388 Marseille, France}

\altaffiltext{7}{Dept. of Physics \& Astronomy, University of California, Irvine, CA
92697, USA}


\altaffiltext{8}{ESA Astrophysics Missions Division, ESTEC, PO Box 299, 2200
AG Noordwijk, The Netherlands}

\altaffiltext{9}{CEA, Laboratoire AIM, Irfu/SAp, Orme des Merisiers, 91191 Gif-sur-Yvette, France}

\altaffiltext{10}{Istituto di Fisica dello Spazio Interplanetario, INAF, Via del Fosso
del Cavaliere 100, 00133 Roma, Italy}

\begin{abstract}

We present new {\it Herschel}-SPIRE imaging spectroscopy (194-671 $\mu$m) of the bright starburst galaxy M82.  Covering the CO ladder from \jfour\ to \jthirteen, spectra were obtained at multiple positions for a fully sampled $\sim$ 3 x 3 arcminute map, including a longer exposure at the central position.  We present measurements of $^{12}$CO, $^{13}$CO, [C {\sc{I}}], [N {\sc{ii}}], HCN, and HCO$^+$ in emission, along with OH$^+$, H$_2$O$^+$ and HF in absorption and H$_2$O in both emission and absorption, with discussion.  We use a radiative transfer code and Bayesian likelihood analysis to model the temperature, density, column density, and filling factor of multiple components of molecular gas traced by $^{12}$CO and $^{13}$CO, adding further evidence to the high-J lines tracing a much warmer ($\sim$ 500 K), less massive component than the low-J lines.  The addition of $^{13}$CO (and [C {\sc{I}}]) is new and indicates that [C {\sc{I}}] may be tracing different gas than $^{12}$CO.  No temperature/density gradients can be inferred from the map, indicating that the single-pointing spectrum is descriptive of the bulk properties of the galaxy.  At such a high temperature, cooling is dominated by molecular hydrogen.  Photon-dominated region (PDR) models require higher densities than those indicated by our Bayesian likelihood analysis in order to explain the high-J CO line ratios, though cosmic-ray enhanced PDR models can do a better job reproducing the emission at lower densities.  Shocks and turbulent heating are likely required to explain the bright high-J emission.
\end{abstract}

\subjectheadings{}


\section{Introduction}\label{sec:intro}

M82 is a nearly edge-on galaxy, notable for its spectacular bipolar outflow and high IR luminosity \citep[$5.6 \times 10^{10}$ \ls,][]{Sanders:2003}.  Though its high inclination angle of 77$^\circ$ makes it difficult to determine, M82 is likely a SBc barred spiral galaxy with two trailing arms \citep{Mayya:2005}.  Its redshift-independent distance is about 3.4 $\pm$ 0.2 Mpc \citep{Dalcanton:2009}, and after correcting the commonly cited redshift (0.000677, \citet{deVaucouleurs:1999}) with WMAP-7 parameters to the 3K CMB reference frame, we find a redshift of 0.000939.  Given this distance we assume a conversion of 17 pc/\as.

Due to its proximity, M82 is an exceptionally well-studied starburst galaxy.  High star formation rates \citep[9.8 \ms\ \pyr, likely enhanced by interactions with M81,][]{Yun:1993} and a large gas reservoir produce bright molecular and atomic emission lines.  Such lines can yield important information on the interaction between the interstellar medium (ISM) and star formation (SF) processes, such as the influence of SF on the ISM through photon-dominated region (PDR) or other excitation, as traced by intermediate to high-J CO rotational lines.


Ground-based studies of CO in M82 are numerous \citep{Wild:1992,Mao:2000,Weiss:2001,Ward:2003,Weiss:2005}, examining both morphology and physical conditions of the gas.  High-resolution CO maps of the 1 kpc disk indicate that the emission is largely concentrated in three areas: a northeast lobe, southwest lobe, and to a lesser extent, a central region \citep[see Figure 1 of][]{Weiss:2001}.  The two lobes are separated dynamically, as can be seen in position-velocity diagrams \citep[Figure 3 of][]{Weiss:2001}.  Outside of the disk, molecular gas emission is also detected in the halo/outflow \citep{Taylor:2001,Walter:2002}.

In addition to examining the morphology of molecular gas, CO emission lines can be used to determine the physical conditions of the molecular gas in galaxies.  Previously, due to terrestrial atmospheric opacity, only the first few lines in the CO emission ladder could be studied.  The first studies of higher-J lines (described below) have indicated that they can trace components of gas separate from those measurable with low-J lines.  Many of the most interesting questions about galaxy formation, evolution, and star formation concern the balance of different energy sources, i.e. what role might cosmic rays, ultraviolet light from stars, X-rays from powerful AGN, or turbulent motion play in the star formation history of various galaxies?  In what way does star formation influence the molecular gas, and vice versa?  Estimating the influence of these various energy sources, however, often depends on knowing the physical conditions of the gas.  We therefore model physical conditions of these high-J lines first in order to inform our later discussion on energy sources.  Other molecules are also useful in this study; in a ground-based survey of 18 different molecular species, \citet{Aladro:2011} also found that some molecules trace different temperature components than others and that the different chemical abundances in M82 and NGC253 may indicate different evolutionary stages of starbursts.

The {\it Herschel Space Observatory} \citep{Pilbratt:2010} is the unique facility that can measure the submillimeter properties of nearby galaxies in a frequency range that cannot be observed from the ground.  As one of the brightest extragalactic submillimeter sources, M82 has been studied extensively with {\it Herschel}.  For example, the imaging photometer of the Spectral and Photometric Imaging REceiver \citep[SPIRE,][]{Griffin:2010} has been used to study the cool dust of M82, revealing wind/halo temperatures that decrease with distance from the center with warmer starburst-like filaments between dust spurs \citep{Roussel:2010}.  The tidal interaction with M81 was likely very effective in removing cold interstellar dust from the disk; more than two thirds of the extraplanar dust follows the tidal streams.  \citet{Panuzzo:2010} used the SPIRE Fourier-Transform Spectrometer (FTS) to study a single spectrum of the $^{12}$CO emission from \jfour\ to \jthirteen\ to find that these higher-J CO lines likely trace a $\sim$ 500 K gas component not seen in the $\sim$ 30 K component that can be observed from ground-based studies.  Also on-board {\it Herschel} is the Heterodyne Instrument for the Far Infrared (HIFI, \citet{deGraauw:2010}), which consists of a set of 7 heterodyne receivers with resolution of 125 kHz to 1 MHz for electronically tuneable frequency coverage of 2 x 4 GHz; it covers 480 - 1910 GHz.  HIFI found ionized water absorption from diffuse gas \citep{Weiss:2010} and high-J transitions of the CO ladder.  These CO transitions indicated a combination of one low and two high density gas components via comparison to PDR models \citep{Loenen:2010}.  

We confirm the presence of multiple molecular hydrogen thermal components in M82 by performing a more in-depth modeling analysis on a deeper dataset as part of the {\it Herschel} Very Nearby Galaxies Survey.  We add to existing data by using the SPIRE FTS mapping mode, providing spectroscopic imaging of a region approximately 3$^{\prime}$ x 3$^{\prime}$, which helps us confirm our source-beam coupling corrections.  We also present a deep pointed spectrum \citep[64 scans vs. 10 scans in][]{Panuzzo:2010} in order to detect fainter lines, such as $^{13}$CO, H$_2$O, OH$^{+}$, HF, and more.  

We add depth to the analysis by modeling both the cool and warm components of molecular gas, and simultaneously accounting for $^{12}$CO, $^{13}$CO, and [C {\sc{I}}].  We also use [C {\sc{I}}] emission as a separate estimate of total hydrogen mass and other absorption lines for column density estimates.  We first analyze the CO excitation using likelihood analysis to determine the physical conditions, and then compare to possible energy sources.  Our likelihoods test the uniqueness and uncertainty in the conditions, as has also been done in \citet{Naylor:2010,Kamenetzky:2011,Scott:2011,Bradford:2009,Panuzzo:2010,Rangwala:2011}.

Our observations are described in Section \ref{sec:obs}.  We describe the Bayesian likelihood analysis used to find the best physical properties of the molecular gas in Section \ref{sec:like} and present the results in Sections \ref{sec:discdeep} and \ref{sec:discmap}.  In the remainder of Section \ref{sec:disc}, we discuss absorption results that are new to this study, possible excitation mechanisms of the warm gas, and comparisons to other galaxies.  Conclusions are presented in Section \ref{sec:concl}.

\section{Observations with SPIRE}\label{sec:obs}

\subsection{The SPIRE Spectrometer}\label{sec:spire}

The SPIRE instrument \citep{Griffin:2010} is on-board the {\it Herschel Space Observatory} \citep{Pilbratt:2010}.  It consists of a three-band imaging photometer (at 250, 350, and 500 $\mu$m) and an imaging Fourier-transform spectrometer (FTS).  We are presenting observations from the FTS, which operates in the range of 194-671 $\mu$m (447-1550 GHz).  The bandwidth is  split into two arrays of detectors: the Spectrometer Long Wave (SLW, 303-671 $\mu$m) and the Spectrometer Short Wave (SSW, 194-313 $\mu$m).  The SPIRE spectrometer array consists of 7 (17) operational unvignetted bolometers for the SLW (SSW) detector, arranged in a hexagonal pattern.  In the SLW, the beam FWHM is about 43\as\ at its largest, dropping to 30\as\ and then rising again to 35\as\ at higher frequency.  The SSW beam is consistently around 19\as.

Two SPIRE FTS observations from Operational Day 543 were utilized in this study: one long integration single pointing of 64 scans total (``deep", Observation ID 1342208389, 84 min [71 min integration time], AOR ``SSpec-m82  -deep") and one fully-sampled map (``map", Observation ID 1342208388, $\sim$ 5 hrs, AOR ``Sspec-m82").  The map observation was conducted in high-resolution (HR) mode and the deep observation was conducted in high+low-resolution (H+LR) mode.  Both were processed in high-resolution mode ($\Delta \nu \sim$ 1.19 GHz) with the {\it Herschel} Interactive Processing Environment (HIPE) 7.2.0 and the version 7.0 SPIRE calibration derived from Uranus \citep{Swinyard:2010,Fulton:2010}.

\subsection{Spectral Map Making Procedure}\label{sec:mapmaking}

\begin{figure}
\begin{center}
\includegraphics[width=\columnwidth]{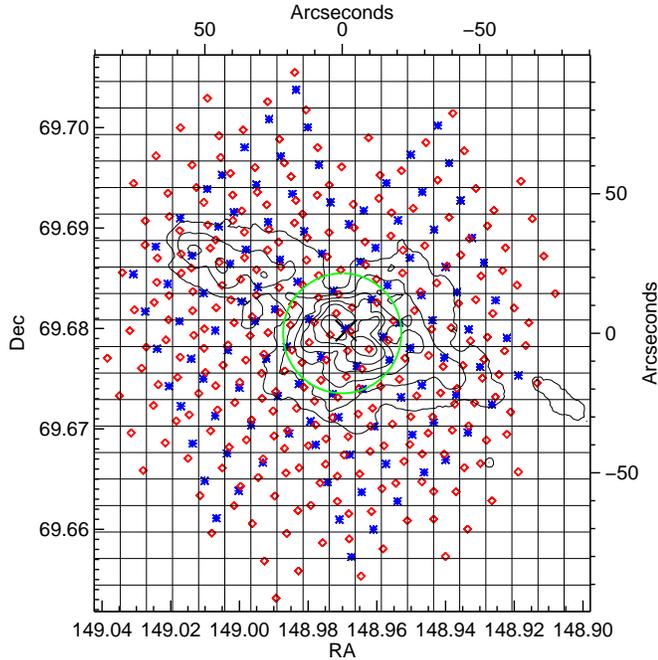}
\caption[Spectrometer mapping point locations.]{Spectrometer Mapping Point Locations.  SLW spectra locations are indicated by blue asterisks, SSW by red diamonds.  The pixel boundaries, spaced 9.5\as\ apart, are indicated by solid black lines.  H$\alpha$ contours are also plotted in black to indicate the orientation of the (nearly edge-on) galactic disk, from the Mount Laguna 40 inch telescope \citep{Cheng:1997}.  Contours are in decreasing intervals of 0.2 log(maximum), i.e. 10$^{0}$, 10$^{-0.2}$, 10$^{-0.4}$ ... 10$^{-1.2}$.  The green circle indicates the size of the $^{12}$CO \jfour\ 43\as\ beam FWHM.  The black X marks the position of the single pointing (``deep") spectrum.}
\label{fig:map}
\end{center}
\end{figure}

In mapping mode, the SPIRE detector arrays are moved around the sky to 16 different jiggle positions, creating 112 and 272 spectra of 16 scans each for SLW and SSW, respectively, covering an area of the sky approximately 3 x 3 arcminutes.  The positions of these scans on the sky are presented in Figure \ref{fig:map}, with blue asterisks for SLW and red diamonds for SSW.

The recommended map making method bins the spectra into pixels approximately one half the FWHM of the beam for each detector, which are about 35\as\ for SLW and 19\as\ for SSW, leading to pixel sizes of 17.5\as and 9.5\as.\footnotetext{SPIRE Data Reduction Guide, http://herschel.esac.esa.int/hcss-doc-7.0/print/spire\_dum/spire\_dum.pdf, December 2, 2011.}  We wrote a custom script to create the map, based largely on the {\sc NaiveProjection} method described in the SPIRE documentation for HIPE.  Each of the 256 scans per detector were processed individually.  All scans for a given detector and jiggle position falling within a given pixel were then averaged and an error bar for each wavenumber bin average is determined as the standard deviation of the scans divided by the square root of number of scans.  All detector averaged spectra that fall into a pixel are then averaged using a weighted mean (where the weight is the inverse of the square of the error bar).  We used the same 9.5\as\ grid for both bands.  

Using a 9.5\as\ grid introduced more blank pixels in the SLW map, because the SLW map is more sparsely sampled because of its larger beam, as can be seen in Figure \ref{fig:map}.  However, this enabled the comparison of the same spectral locations across both bands (where the data were available), without averaging together spatially discrepant spectra in the SSW, as would happen if pixels were made larger.  We emphasize that the blank pixels are somewhat artificial; the whole galaxy has been mapped, and the pixels locations are simply meant to indicate the central location of the detectors, though the beam size is larger than the pixel boundaries.  

\subsection{Line Fitting and Convolution Procedure}\label{sec:fitting}

The mirror scan length determines the spectral resolution of the spectrum.  Because the scan length is necessarily finite, the Fourier-Transformed spectrum contains ringing; therefore, the instrument's line profile is a sinc function, as can be clearly seen in Figures \ref{fig:spec} and \ref{fig:fitting}.  The spectrum also contains the underlying continuum which must be removed before fitting the lines, which we do sequentially rather than simultaneously (see exceptions below).  We isolate $\pm$ 25 GHz around the expected line center, and mask out the $\pm$ 6 GHz around the line center.  We then fit the remaining signal with a second order polynomial fit to determine the continuum shape.  After subtracting this continuum fit, we then use a Levenberg-Marquardt least-squares method to fit each line as a sinc function with the following free parameters: central frequency, line width, maximum amplitude, and residual (flat) baseline value.  The baseline value stays around zero because the continuum has already been subtracted.  The central frequency is limited such that the line center is no greater than $\pm$ 300 \kms\ from the expected frequency given the nominal redshift of M82.  For comparison, the resolution varies from 230 to 810 \kms, from the shorter to longer wavelength ends of the band.

For the deep spectrum, we detect weaker lines than in the map spectra.  However, ringing from the strongest lines can interfere with the signal; therefore we first fit the strong lines ($^{12}$CO, [C {\sc{I}}], and [N {\sc{ii}}]) using the procedure outlined above and subtract their fitted line profiles from the spectrum.  After all of the $^{12}$CO, [C {\sc{I}}], and [N {\sc{ii}}] lines are fitted and subtracted from the spectrum, we then do a second pass to fit the weaker lines.  An illustration of the difference this process can make for the $^{13}$CO \jfive\ line is in Figure \ref{fig:fitting}.

In general, all of the lines are fit independently, with a few exceptions: the $^{12}$CO \jseven\ and [C {\sc{I}}] \jtwo\ lines are a mere 2.7 GHz away in rest frequency, and ground state p-H$_2$O and o-H$_2$O$^+$ lines are separated by only 2 GHz.  These two pairs must be fit simultaneously.  Both are fit independently to supply initial guesses, which are then used to fit both lines as the sum of two sinc functions, each with their own central wavelength, width, and amplitude, but with one baseline value.

The integrated flux is simply the area under the fitted sinc function, which is proportional to the product of the amplitude and line width (converted to km/s).  The error in the integrated flux is based on propagating the errors from the fitted parameters themselves.  We note that the error estimation assumes all wavenumber bins are independent of one another, but that is in fact not entirely true in a FT spectrometer.  Though lines that are separated spectrally do not affect one another greatly (hence why we fit most lines independently), within each line fit the data points used in the 50 GHz range around the line center are not independent.

\begin{figure*}[th]
\centering
\subfigure{\includegraphics[width=0.8\textwidth]{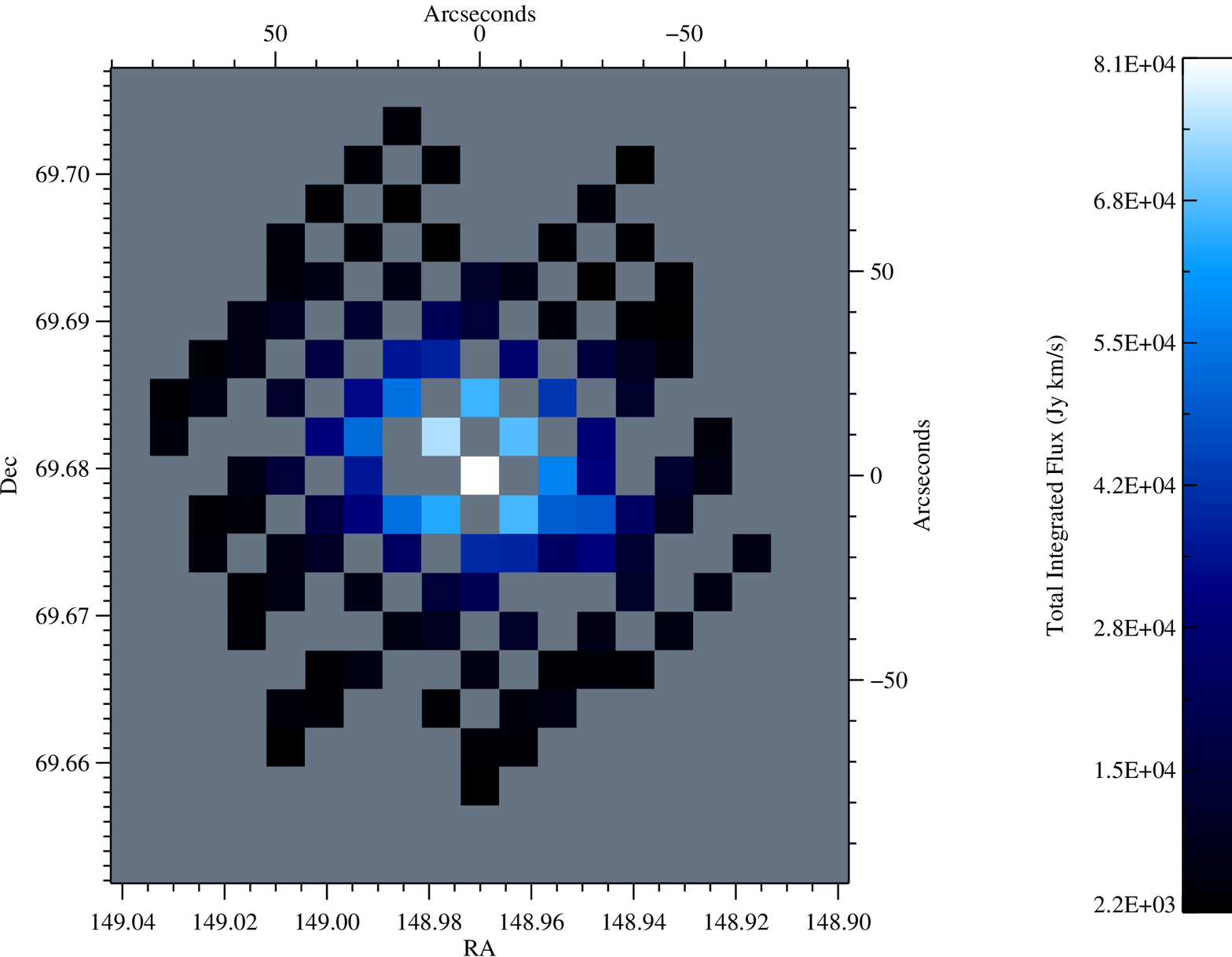}}
\subfigure{\includegraphics[width=0.8\textwidth]{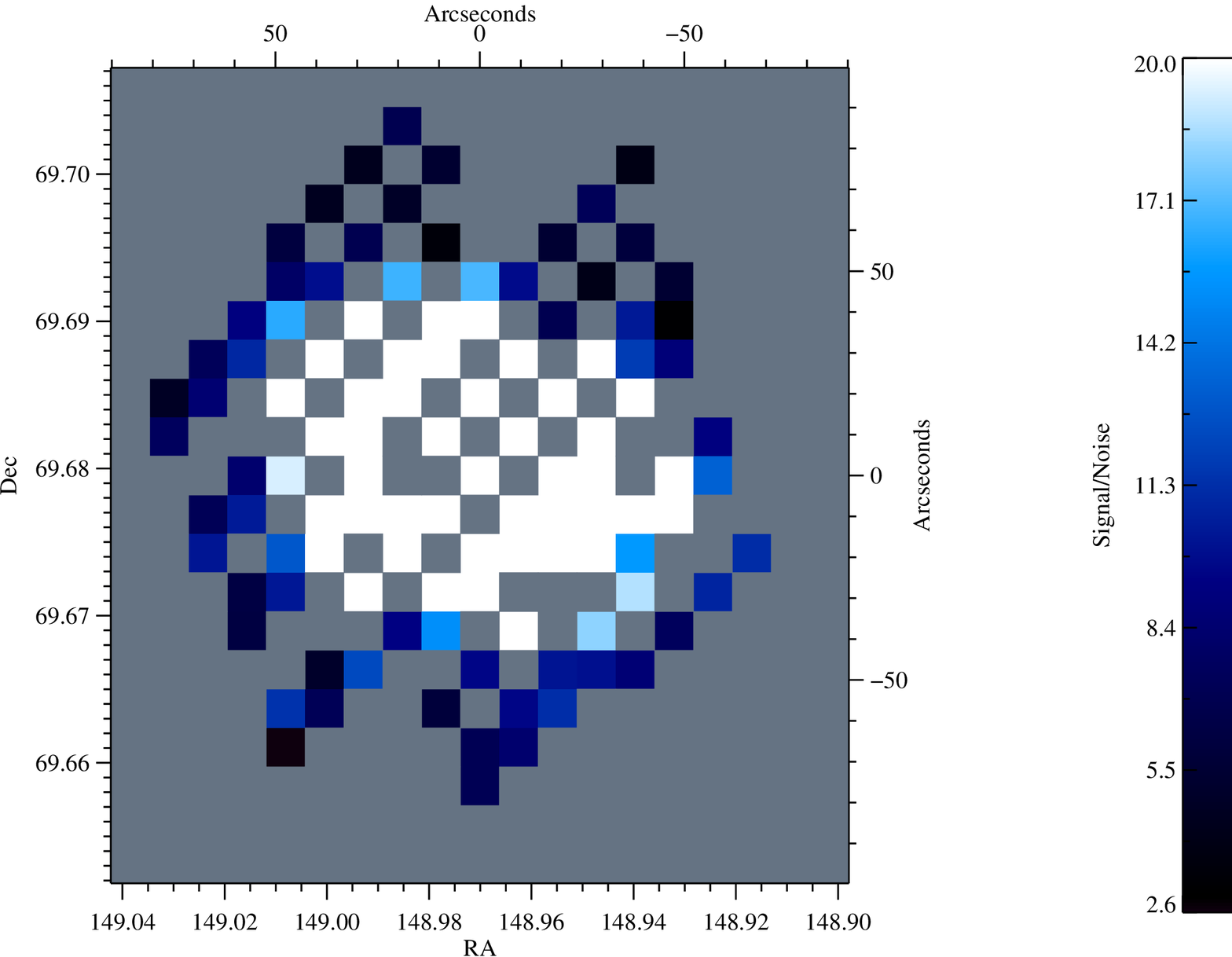}}
\caption{Integrated Flux (top) and Signal/Noise (bottom) maps for CO \jfour.  Figures 2.1-2.19 are available in the online version of the Journal.  The top half includes no beam correction or convolution.  Black corresponds to the lowest flux or zero if any fluxes are negative, at which point the colorbar becomes purple.  The bottom half is a map of signal/noise, though the color bar tops out at 20 in order to better illustrate which pixels are near the threshold of detectability.  On the color bar, black corresponds to the lowest signal/noise or three if any pixels have S/N less than three, at which point the colorbar becomes purple.}\label{fig:intflux1}
\end{figure*}

\begin{figure*}[t]
\centering
\subfigure{\includegraphics[width=\textwidth]{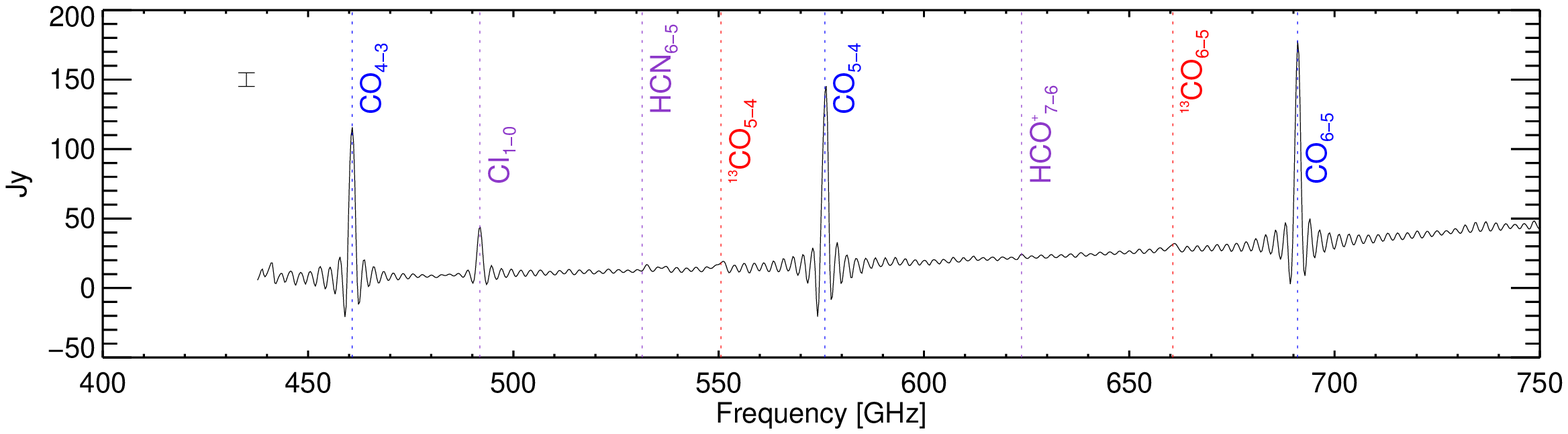}}
\subfigure{\includegraphics[width=\textwidth]{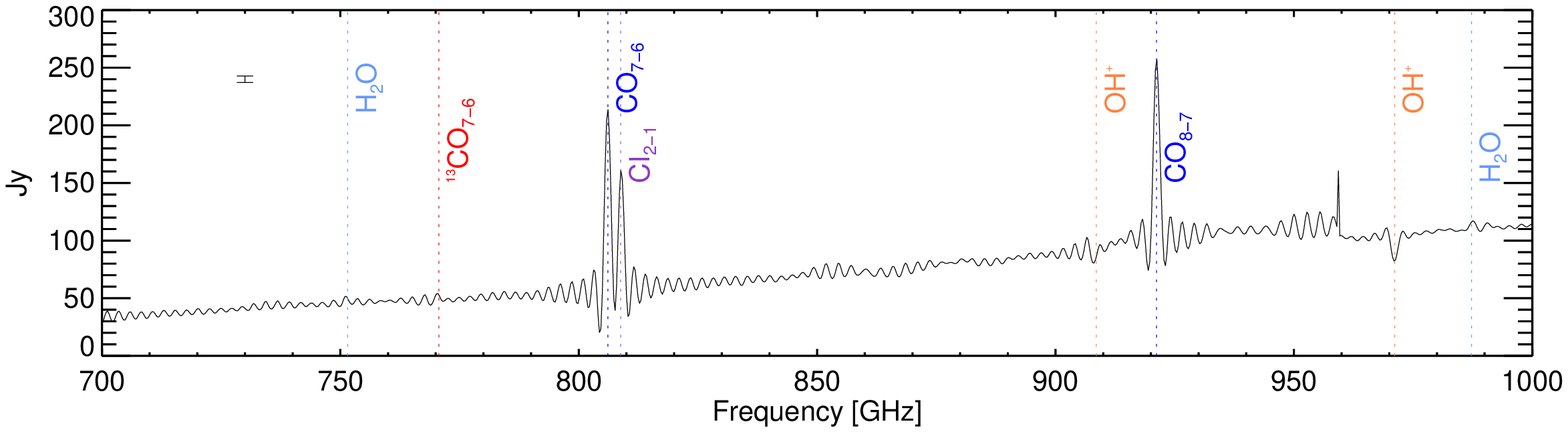}}
\subfigure{\includegraphics[width=\textwidth]{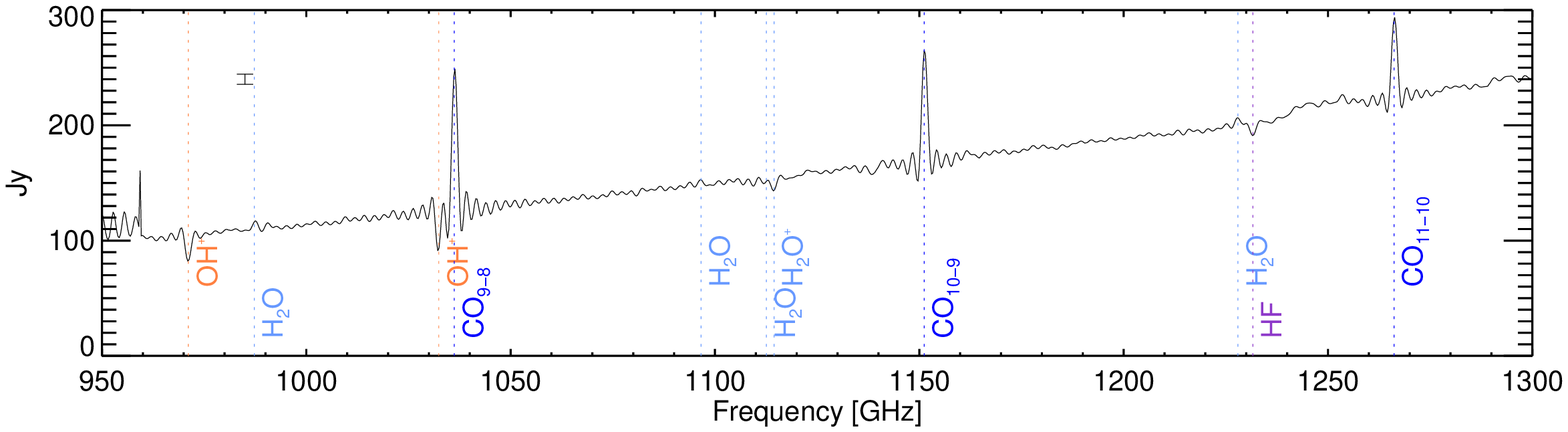}}
\subfigure{\includegraphics[width=\textwidth]{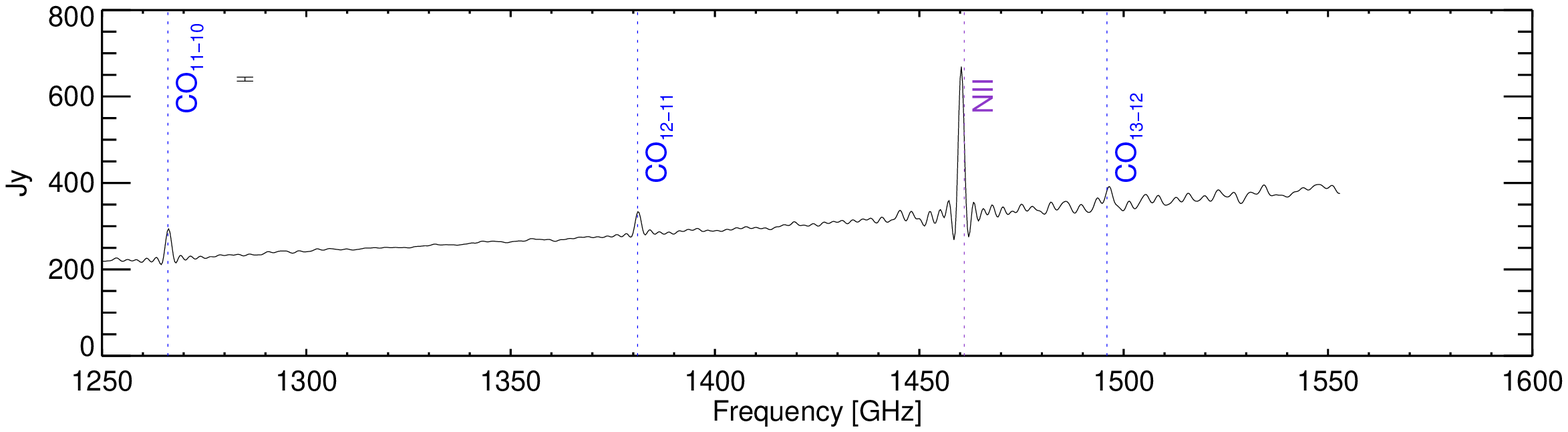}}
\caption{Deep Spectrum, split up by frequency.  The switch from SLW to SSW bands occurs at approximately 950 GHz.  Error bars are not shown on the spectrum for clarity, though the median error bar of each panel (times a factor of 20) is shown in the upper left corner.  The spectrum contains ringing because the line profile of the FTS is a sinc function, as discussed in Section \ref{sec:fitting}.  Emission/absorption line locations are color-coded by molecular species: blue for $^{12}$CO, red for $^{13}$CO, light blue for water and its ion, orange for OH$^+$, and violet for all others.}\label{fig:spec}
\end{figure*}

\begin{figure}[t]
\centering
\includegraphics[width=\columnwidth]{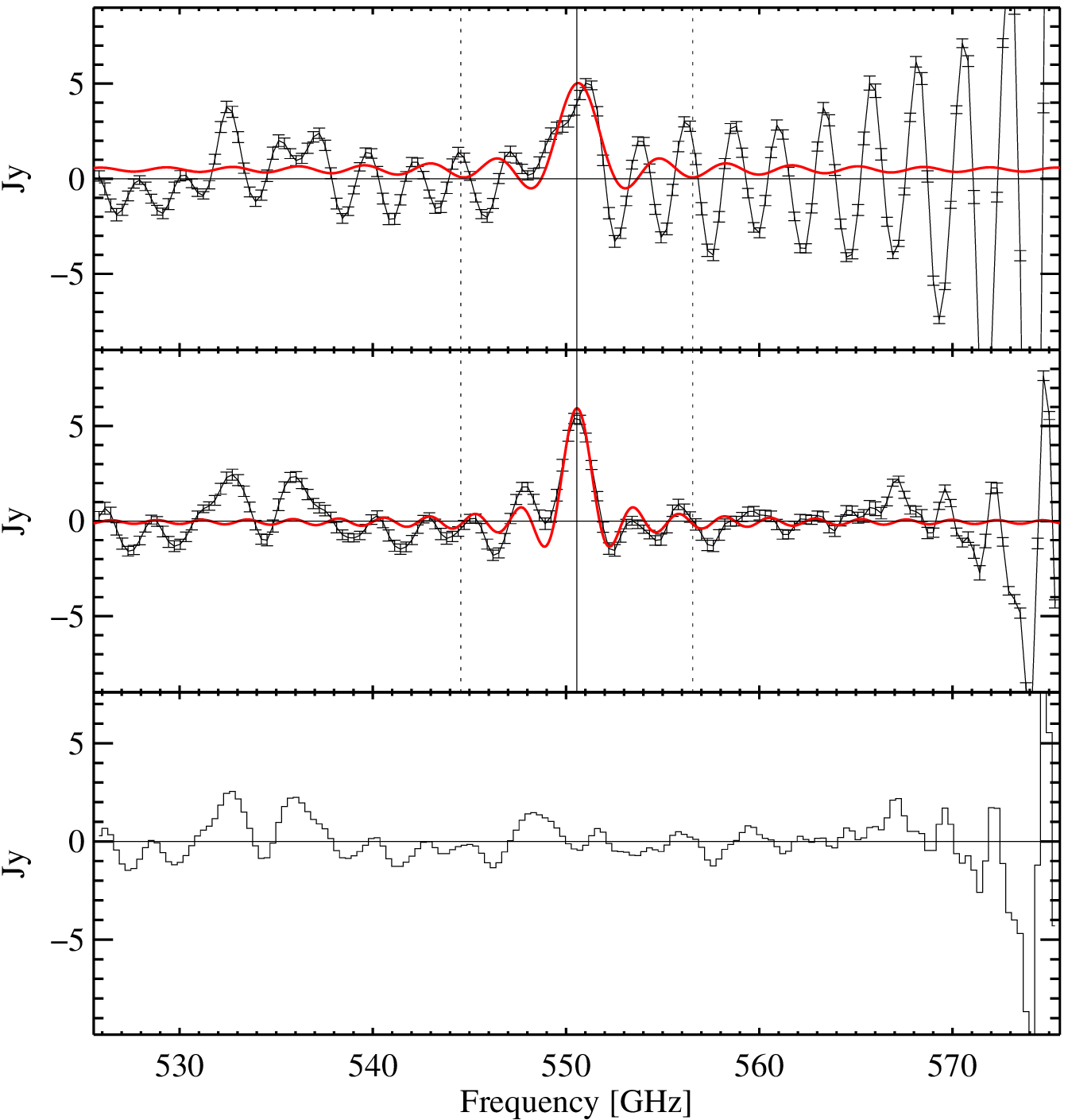}
\caption{Example of line fit, using the $^{13}$CO \jfive\ line.  The top panel shows the continuum-subtracted spectrum (black); the ringing of the nearby $^{12}$CO \jfive\ line is clearly visible on the right and interfering with the fit, overplotted in solid red.  The middle panel shows the spectrum after subtracting out the strong $^{12}$CO, [C {\sc{I}}], and [N {\sc{ii}}] lines (though only the subtraction of the $^{12}$CO \jfive\ line is visible in this figure).   In this example, the signal to noise ratio of the fit was improved from 1.1 to 6.6 (where the signal is the integrated flux and the noise is based on propagating the statistical errors of the fit parameters in the integrated flux calculation).  The bottom panel is the residual of the fit.  In all panels, a horizontal line indicates zero.  In the top two panels, the solid vertical line indicates the expected line center, and the two dashed vertical lines demarcate the area within that was not used for fitting the continuum.}\label{fig:fitting}
\end{figure}

The beam size of the SPIRE spectrometer varies between the two detector arrays.  In addition, it varies across the spectral range of the SLW, as described in Section \ref{sec:spire}, and is not strictly proportional to wavenumber due to the presence of multiple modes in the SLW detectors \citep{Chattopadhyay:2003}.  When examining the spectral line energy distribution (SLED), it is imperative to scale all fluxes to a single beam size.  For the map observation, we first fit the spectra as they were (with no correction factor).  An example of integrated flux map, prior to any convolution or beam correction, is presented in Figure \ref{fig:intflux1}, with all other integrated flux maps available in the online version of the Journal.  For the SLW detector, we present integrated flux maps using both 9.5\as\ and 17.5\as\ pixels.  

The integrated flux maps for each line were then convolved with a kernel that matched the beams to the beam of the $^{12}$CO \jfour\ map, which has a FWHM of 43\as.  The kernel was created using a modified version of the procedure decribed by \citet{Bendo:2011} \citep[see also][]{Gordon:2008}.  However, instead of directly applying Equation 3 from \citet{Bendo:2011} to the images of the beams, we applied the equation to one-dimensional slices of the beams to create the radial profile of convolution kernels, which we then used to create two-dimensional kernels.  The approach worked very effectively for creating kernels to match beams observed with SSW to the $^{12}$CO \jfour\ map.  However, in cases where we created convolution kernels for matching beams measured at two different wavelengths by the SLW array, we needed to manually edit individual values in the kernel radial profiles to produce effective two dimensional kernels.

After the mathematical convolution of the integrated flux map with the kernel (resampled to match our map sizes), the fluxes were all converted to the units of Jy km s$^{-1}$/beam, referring to the $^{12}$CO \jfour\ beam.  The ratio of beam areas was determined empirically by convolving the kernel with the smaller (observed) beam peak normalized to 1 and determining the ratio required to produce the larger $\Omega_{\rm CO \rm J\eqq4\too3}$ beam with the same peak (simulating the observation of a point source of 1 Jy km s$^{-1}$).  To account for blank pixels, only the portion of $\Omega_{\rm kernel}$ that encompasses data was used in the aforementioned conversion.


For the deep spectrum, we used the same source-beam coupling factor ($\eta_c(\nu)$) as in \citet{Panuzzo:2010}, which was derived by convolving the M82 SPIRE photometer 250 $\mu$m map \citep{Roussel:2010} with appropriate profiles to produce the continuum light distribution seen with the FTS.  The deep spectrum was multiplied by this factor before the line fitting procedure.  The deep spectrum is presented in four parts in Figure \ref{fig:spec}, and the measured lines fluxes ($\geq 3 \sigma$) are in Table \ref{table:fit}.


The central pixel of the convolved maps offers a direct comparison to the source-beam coupling corrected deep spectrum.  In the SLW, these two SLEDS are within $\pm$ 16\% of one another.  Later, we assume calibration error of 20\%, so these differences are within those bounds.  There is greater variation in the SSW band, though this is the region in which the signal to noise of the lines greatly drops.  When 20\% calibration error is included, all the line measurements have overlapping error bars with the exception of $^{12}$CO \jthirteen, where the deep spectrum measurement is more than twice that of the map.  The $^{12}$CO \jthirteen\ map, however, has a S/N of only 4 in the central pixel, with only 60/399 pixels having S/N greater than 3.  We primarily use the deep spectrum for our likelihood analysis because of the higher S/N and access to more faint lines, but compare with using the convolved map central pixel in Section \ref{sec:discmap}.  The similarity of the two SLEDs, within error bars, using two independent methods (the derived $\eta_c(\nu)$ from photometry comparisons vs. map convolution) to account for source-beam coupling, indicates that both methods are robust.

Though the maps do not provide adequately high spatial/spectral resolution for a detailed study of the morphology of the emission, some qualitative assessments can be made.  For example, the line centroids do trace the relative redshift/blueshift of the northeast and southwest components \citep[$v_{hel} \sim 300$ \kms\ and 160 \kms, respectively,][]{Loenen:2010}.  However, we do not resolve the two separate peaks in flux.  The capabilities of these maps to resolve gradients in the physical parameters modeled in this work will be discussed in Section \ref{sec:discmap}.

\begin{deluxetable}{ c c c }
   \tablecaption{Measured Fluxes of Detected Lines in Deep Spectrum\label{table:fit}}
   \tabletypesize{\footnotesize}
   \tablewidth{0pt}
   \tablehead{
   \colhead{Transition} & \colhead{$\nu_{rest}$} & \colhead{$\int F_\nu dv$\tablenotemark{a}} \\
   \colhead{}        & \colhead{[GHz]}        & \colhead{[10$^3$ Jy \kms]\tablenotemark{b}}       }
   \startdata
$^{12}$CO \jfour    &  461.041 & 85.66 $\pm$ 0.90 \\
$^{12}$CO \jfive    &  576.268 & 79.44 $\pm$ 0.94\\
$^{12}$CO \jsix     &  691.473 & 73.54 $\pm$ 0.44\\
$^{12}$CO \jseven   &  806.652 & 66.04 $\pm$ 0.65\\
$^{12}$CO \jeight   &  921.800 & 58.44 $\pm$ 0.87\\
$^{12}$CO \jnine    & 1036.912 & 42.90 $\pm$ 0.74\\
$^{12}$CO \jten     & 1151.985 & 29.60 $\pm$ 0.41\\
$^{12}$CO \jeleven  & 1267.014 & 19.35 $\pm$ 0.33\\
$^{12}$CO \jtwelve  & 1381.995 & 13.25 $\pm$ 0.35\\
$^{12}$CO \jthirteen& 1496.923 & 10.4 $\pm$ 1.1\\
$[\rm C {\sc{I}} ] \  ^3P_1 \too ^3P_0$ & 492.161 & 23.93 $\pm$ 0.55\\
$[\rm C {\sc{I}} ] \ ^3P_2 \too ^3P_1$ & 809.342 & 38.88 $\pm$ 0.58\\
$[\rm N {\sc{II}} ] \ ^3P_1 \too ^3P_0$ & 1462.000 & 82.4 $\pm$ 1.2\\
$^{13}$CO \jfive    &  550.926 & 3.83 $\pm$ 0.58\\
$^{13}$CO \jsix	    &  661.067 & 3.02 $\pm$ 0.15\\
$^{13}$CO \jseven   &  771.184 & 1.84 $\pm$ 0.31\\
$^{13}$CO \jeight \tablenotemark{c}  &  881.273 & 1.16 $\pm$ 0.45\\
HCN \jsix \tablenotemark{c}& 531.716 & 1.15 $\pm$ 0.42\\
HCO$^+$ \jseven	& 624.208 & 1.08 $\pm$ 0.17\\
OH$^+$ $\rm N\eqq1\too0$, $\rm J\eqq0\too1$ &  909.159 & -5.2 $\pm$ 1.1\\
OH$^+$ $\rm N\eqq1\too0$, $\rm J\eqq2\too1$ &  971.805 & -8.88 $\pm$ 0.41\\
OH$^+$ $\rm N\eqq1\too0$, $\rm J\eqq1\too1$ & 1033.118 & -9.94 $\pm$ 0.39\\
HF \jone & 1232.476 & -3.64 $\pm$ 0.34\\
o-H$_2$O  $3_{12} \too 3_{03}$  & 1097.365 & 1.39 $\pm$ 0.30\\
o-H$_2$O  $3_{12} \too 2_{21}$  & 1153.127 & 2.57 $\pm$ 0.37\\
p-H$_2$O $2_{11}\too 2_{02}$    &  752.033 & 1.86 $\pm$ 0.36\\
p-H$_2$O $2_{02} \too 1_{11}$ &  987.927 & 3.09 $\pm$ 0.56\\
p-H$_2$O $1_{11} \too 0_{00}$ & 1113.343 & -2.69 $\pm$ 0.44\\
p-H$_2$O  $2_{20} \too 2_{11}$  & 1228.789 & 2.01 $\pm$ 0.34\\
o-H$_2$O$^+$ $1_{11} \too 0_{00}$ &  1115.204 & -2.82 $\pm$ 0.42
   \enddata
	\tablenotetext{a}{All fits have been referenced to the 43\as\
 beam of $^{12}$CO \jfour, as described in Section \ref{sec:fitting}.  Uncertainties do not include calibration error.}
	\tablenotetext{b}{To convert to other units, we use these equations.  
L [\ls] = $\int F_\nu dv$ [Jy \kms] $\times$ 0.012 $\nu_{GHz}$.  
$\int T dv$ [K \kms] = $\int F_\nu dv$ [Jy \kms] 660.8 $\nu_{GHz}^{-2}$.  F [W/m$^2$] = $\int F_\nu dv$ [Jy \kms] $\times$ 3.3 $\times 10^{-23} \nu_{GHz}$.}
 	\tablenotetext{c}{The $^{13}$CO \jeight\ and HCN \jsix\ lines are detected at slightly less than S/N of 3; 2-$\sigma$ upper limits would be $9.0$ and $8.4 \times 10^2$ Jy \kms, respectively.}
\end{deluxetable}

\section{Bayesian Likelihood Analysis}\label{sec:like}

We follow the method described in \citet{Ward:2003} for the $^{12}$CO \jsix\ map of M82, used frequently in the analysis of ground based molecular observations of galaxies \citep[e.g. the Z-spec collaboration,][]{Naylor:2010,Kamenetzky:2011,Scott:2011,Bradford:2009} and also of the single pointing SPIRE spectrum of M82 \citep{Panuzzo:2010} and Arp220 \citep{Rangwala:2011}.  The goal of our Bayesian likelihood analysis is to map the relative probabilities of physical conditions over a large parameter space; this provides a more complete statistical analysis of the physical conditions as opposed to simply finding one best-fit solution.


For each molecular species, we first calculate a grid of expected line emission temperatures for various combinations of temperature ($T_{\rm kin}$), density ($n(H_2)$), and column density per unit linewidth ($N_{^{12}CO}/dv$) using RADEX \citep{vanderTak:2007}.  We use the uniform sphere approximation for calculating the escape probabilities; the actual morphology in M82 shows a more complex velocity structure, therefore this approximation is considered an average of the bulk properties of the gas (and the results are not sensitive to uniform sphere vs. LVG approximation).  RADEX performs statistical equilibrium calculations of the level populations, including the effects of radiative trapping, for a specified gas temperature, density, and column density per unit linewidth. The resulting solutions are output in the form of background-subtracted Rayleigh-Jeans equivalent line radiation temperatures.

We use a 2.73 K blackbody to represent the cosmic microwave background (CMB).  We also experimented with using the continuum flux measured in our deep spectrum as a background; we fit the continuum in Jy across both bands, masking out the lines, with a third-order polynomial.  The choice of this fit was to accurately represent the continuum background as a function of frequency; it was not meant to represent any physical conditions.  Because the relevant radiation background is the specific intensity (Jy/sr), we divide our continuum by the area corresponding to a 43\as\ beam (as the entire deep spectrum has been corrected to that).  Such a background is in fact orders of magnitude higher than the CMB at the highest frequencies that we are modeling.  However, a grid with this background vs. the CMB produces the same likelihood results.  This is because at a kinetic temperature of $\sim$ 500 K (the warm component we will model), collisional excitation greatly dominates over radiative excitation.  In other words, at high temperatures, the modeled line intensities do not depend on the background radiation field.  The cool component we will model is traced by low-J lines, whose background is not as affected, and so this component also finds the same results with either background.  Therefore we are presenting results using the CMB background.
 
In addition to $^{12}$CO, we also model $^{13}$CO and [C {\sc{I}}] as a function of the same temperatures, densities, and column density of $^{12}$CO.  For those molecular/atomic species, we also add the parameter of the relative abundance, e.g. [$^{13}$CO]/[$^{12}$CO] or ${X}_{^{13}CO} / {X}_{^{12}CO}$.  When modeling the intensity, the column density of $^{13}$CO is simply that of $^{12}$CO times the relative abundance.  [C {\sc{I}}] is modeled with the parameter of [C {\sc{I}}]/[$^{12}$CO].    Finally, we introduce one more parameter, the area fractional filling factor $\Phi_A$.  The modeled emission may not entirely fill the beam, so the flux may be reduced by this factor.  All model grid points are therefore multiplied by each value of $\Phi_A$.  The ranges of parameters, as well as the number of grid points, are presented in Table \ref{tab:radex}.  

\begin{deluxetable}{c c c}
   \tablewidth{0pt}
   \tablecaption{RADEX Model Parameters and Ranges\label{tab:radex}}
   \tablehead{
   \colhead{Parameter} & \colhead{Range} & \colhead{\# of Points}
	}
   \startdata
$T_{kin}$ [K] & $10^{0.7} - 10^{3.7}$ & 61\\
$n(H_2) $ [cm$^{-3}$] & $10^{2} - 10^{6}$ &  41\\
$N_{^{12}CO}$/$\Delta V$ [cm$^{-2}$] & $10^{14} - 10^{18}$ &  41\\
$\Phi_A$ & $10^{-3} - 10^{0}$ &  41\\
$X_{^{13}CO}/X_{^{12}CO}$ & $10^{-2} - 10^{-1}$ &   11\\
$X_{[C {\sc{I}}]}/X_{^{12}CO}$ & $10^{-2} - 10^{2}$ &   11\\
$\Delta V$ [km s$^{-1}$] & 1.0 &  fixed\\
$T_{background}$ [K] & 2.73 &  fixed
   \enddata 
\tablecomments{All parameters are sampled evenly in log space.  Velocity is fixed because all modeling is column per unit linewidth.}
\end{deluxetable}

The RADEX grid gives us a set of line intensities as a function of model parameters $\bm{p}  = (N_{^{12}CO}/dv,n(H_2),T_{\rm kin},\Phi_A, \bm{X}_{mol} / \bm{X}_{^{12}CO})$, which we then compare to our measured intensities $\bm{x}$.  To compare to column density per unit linewidth, we divide the measured intensities by 180 \kms, so that they are also per unit linewidth.  The optical depth, and in turn the RADEX results, depend only on the ratio of column density to line width.  \citet{Ward:2003} found linewidths of 180 \kms\ for the NE component and 160 \kms\ for the SW by resolving the structure in position-velocity diagrams for their study from CO \jone\ to \jseven, but we do not resolve the difference between the two and so we use the larger value.  The Bayesian likelihood of the model parameters given the measurements is

\begin{equation}P(\bm{p} | \bm{x}) = \frac{P(\bm{p}) P(\bm{x} | \bm{p})}{P(\bm{x})},\end{equation}

\noindent where $P(\bm{p})$ is the prior probability of the model parameters (see Section \ref{sec:prior}), $P(\bm{x})$  is for normalization, and $P(\bm{x} | \bm{p})$ is the probability of obtaining the observed data set given that the source follows the model described by $\bm{p}$.  $P(\bm{x} | \bm{p})$ is the product of Gaussian distributions in each observation,

\begin{equation}P(\bm{x} | \bm{p}) = \prod_i \frac{1}{\sqrt{2 \pi \sigma_i^2}} \rm exp \it \bigg[- \frac{(x_i - I_i (\bm{p}))^2}{2 \sigma_i^2} \bigg] \end{equation}

\noindent where $\sigma_i$ is the standard deviation of the observational measurement for transition $i$ and $I_i({\bf p})$ is the RADEX-predicted line intensity for that transition and model.  For the total uncertainty, we take the statistical uncertainty in the total integrated intensity from the line fitting procedure and add 20\%/10\% calibration error for SSW/SLW in quadrature.  To find the likelihood distribution of one parameter out of all of $\bm{p}$, we integrate over all other parameters to find, for example, $P(T_{kin})$.

\subsection{Separate Components}\label{sec:components}

We divide the lines fluxes into two components, one warmer and one cooler.  \citet{Panuzzo:2010} already showed that the high-J lines of $^{12}$CO trace a warmer component than those transitions available from the ground.  However, some of the mid-J lines (especially CO \jfour) may have significant contributions from both components.  To separate the $^{12}$CO fluxes from each component, we follow an iterative procedure.  We first model the lowest three $^{12}$CO lines from \citet{Ward:2003}; we take the sum of their measurements for the two observed lobes and scale the result by the ratio of their beam area to our 43\as\ beam.  The best-fit SLED is then subtracted from our SPIRE measurements, producing the black triangles in Figures \ref{fig:deep_seds_co} and \ref{fig:deep_seds_multi}.  These triangles comprise the ``warm component."  The best fit warm SLED is then subtracted from the low-J lines, producing the asterisks in the aforementioned figure.  These fluxes are refit to produce our results for the ``cool component."

We present a two-component model using just $^{12}$CO, as well as one including our high-J lines of $^{13}$CO along with the warm component.  We did not model the low-J $^{13}$CO lines reported in \citet{Ward:2003} due to the uncertainties presented in their Table 2 footnotes.  We instead predicted the $^{13}$CO spectrum of the cool component, given its best fit results and a $^{12}$CO/$^{13}$CO ratio of of 35 \citep[inbetween previously found 30 to 40][]{Ward:2003}, and found a very small contribution to the higher-J lines we are modeling.  These small contributions are subtracted from the warm component, as with $^{12}$CO in the previous paragraph.

We also sought to include [C {\sc{I}}], which was assumed to be associated with the cool component, due to the low excitation temperature ($\sim$ 30 K) derived from the line ratios ($n_u/n_l = g_u/g_l \exp(-\Delta T/T_{ex})$, $n_i \propto I_i \unit{[W/cm^2]} / (A_i \nu_i)$).  Here, we use $\Delta T = 38.84$, $g_u = 5$, $g_l = 3$, $A_u=2.65 \times 10^6$ s$^{-1}$, $A_l=7.88 \times 10^8$ s$^{-1}$.  See Table \ref{table:fit} for the frequencies and unit conversion. However, this model produced some unphysical situations.  We discuss our findings and implications of them in Section \ref{sec:discdeep}.

This analysis necessarily assumes that all of the line emission for a given component is coming from one portion of gas described in bulk by the model parameters.  In reality, there is likely a variety of physical conditions, existing in a continuum of parameters.  However, the high-J SPIRE data does not provide justification for modeling more than one warm component because the SLEDs are well-described by one component.  We did attempt a procedure to model multiple warm components of CO by first fitting the highest-J lines and subtracting the predicted line fluxes for the mid-J lines.  Such a procedure has been used in \citet{Rangwala:2011}, for example.  However, the predictions for the mid-J lines either matched or were an overestimate of the observed fluxes, leaving no second component to be modeled.  A range of conditions is definitely present in the molecular gas, yet these two (warm and cool) components are dominating the emission, within the observational and modeling uncertainties.  We note that, with regards to the different molecules/atoms being modeled here, all three species have similar profiles, as shown from the HIFI (higher spectral resolution) spectra in \citet{Loenen:2010}.

\subsection{Prior Probabilities}\label{sec:prior}

\begin{deluxetable}{c c c}
\tablewidth{0pt}
\tablecaption{Likelihood Parameters Used\label{table:likeparams}}
\tablehead{\colhead{Parameter} & \colhead{Value\tablenotemark{a}} & \colhead{Units}}
\startdata
Line width\tablenotemark{b} &  180 & [km s$^{-1}$]\\
Abundance ($X_{^{12}CO}/X_{H_2}$) & 3.0 $\times 10^{-4}$ & \\
Angular Size Scale\tablenotemark{c} &  17 & [pc/\as]\\
Emission Size &  43.0 & [\as]\\
Length Limit & 900 & [pc]\\
Dynamical Mass Limit & 2 $\times 10^{9}$ & [\ms]
\enddata
	\tablenotetext{a}{Citations for parameters are in Section \ref{sec:prior}.}
	\tablenotetext{b}{Used for scaling the line intensities.  All other parameters used for prior probabilities.}
	\tablenotetext{c}{A$_{region}$ = $\pi$ (Angular Size Scale [pc/\as] $\times$ Source Size [\as] / 2)$^2$ }
\end{deluxetable}

We use a binary prior probability, $P(\bm{p})$, to indicate either a physically allowed scenario ($P(p)$=1) or an unphysical and thus not allowed scenario ($P(p)$=0).  In other words, all combinations of parameters that are deemed physical based on the following three conditions were given equal prior probability, and all others are given zero prior probability.  The conditions are as follows:

\begin{enumerate}

\item The total length of the column ($L_{col}$) cannot exceed the length of the entire region.  This assumes the length in the plane of the sky is the same as that orthogonal to the plane of the sky; we chose an upper-limit to the length of 900 pc because of the observed size \citep{Ward:2003}.  This is the most significant of all the priors, placing an upper limit on the column density and a lower limit on the density.  This prior can be stated as 

\begin{equation}
\frac{N_{^{12}CO}}{n(H_2) \sqrt{\Phi_A} X_{^{12}CO}} \le 900 \unit{pc}.
\end{equation}

For the relative abundance $X_{^{12}CO}$ to molecular hydrogen, we assume 3.0 $\times 10^{-4}$ \citep{Ward:2003}.

\item The total mass in the emission region ($M_{region}$) cannot exceed the dynamical mass of the galaxy.  We use the expression 

\begin{equation}\label{eqn:mass}
M_{region} = \frac{A_{region} N_{^{12}CO} \Phi_A 1.5 m_{H_2}}{X_{^{12}CO}}
\end{equation}

\noindent to calculate the mass in the emission region, where the 1.5 accounts for helium and other heavy elements, and $\Phi_A$ is the filling factor.  We estimate the dynamical mass to be $2 \times 10^9$ \ms\ \citep{Ward:2003, Naylor:2010}, calculated using rotational velocity and radius.  The other assumed parameters in the above expression are listed in Table \ref{table:likeparams}.

\item The optical depth of a line must be less than 100, as recommended by the RADEX documentation.  The cloud excitation temperature can become too dependent on optical depth at high column densities, and so very high optical depths can lead to unreliable temperatures.  We found that in the presence of the other priors, this limit does not affect the likelihood results.

\end{enumerate}

\subsection{Likelihood Analysis of the Map}\label{sec:maplike}

We run each map pixel through the aforementioned likelihood analysis independently.  (Note that due to the beam size being larger than the pixels themselves, each pixel's data are not independent of its neighbors).  We only model $^{12}$CO in the spectral maps because they are of lower integration time and $^{13}$CO cannot be reliably measured.  We also cannot account for cool emission at different locations in our map.

To be run through the likelihood analysis, a pixel was required to have both an SLW and SSW spectrum and at least 5 $^{12}$CO lines with S/N $\geq$ 10 (the convolution tends to increase the signal/noise of each pixel).  90 pixels met this requirement (98 pixels would meet the requirement of 5 $^{12}$CO lines with S/N $\geq$ 3, so little would be gained by going to lower S/N).  Additionally, we do not find statistically significantly different results requiring only 3 lines, the minimum with which we could reasonably model the emission; to some extent, this requires at minimum the \jsix\ transition, which means we will be tracing higher temperatures.


\section{Modeling Results and Discussion}\label{sec:disc}

\begin{figure}
\begin{center}
\includegraphics[width=\columnwidth]{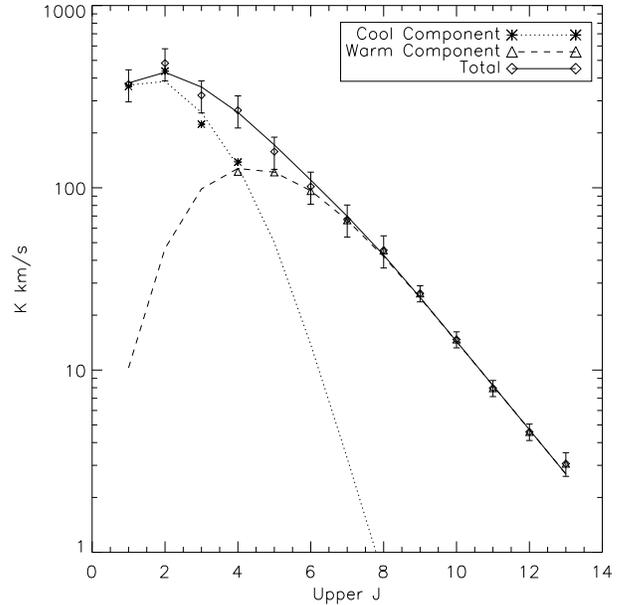}
\caption{Bayesian Likelihood Analysis, Spectral Line Energy Distributions, $^{12}$CO Only.  Asterisks represent the cool component, with its best fit SLED (``4D Max" column in Table \ref{table:like1}) shown by a dotted line.  Triangles represent the warm component, with its best fit SLED shown by a dashed line.  The total measurements are shown with diamonds with their associated error bars.  The total of both components is the solid line.}
\label{fig:deep_seds_co}
\end{center}
\end{figure}

\begin{figure}
\begin{center}
\includegraphics[width=\columnwidth]{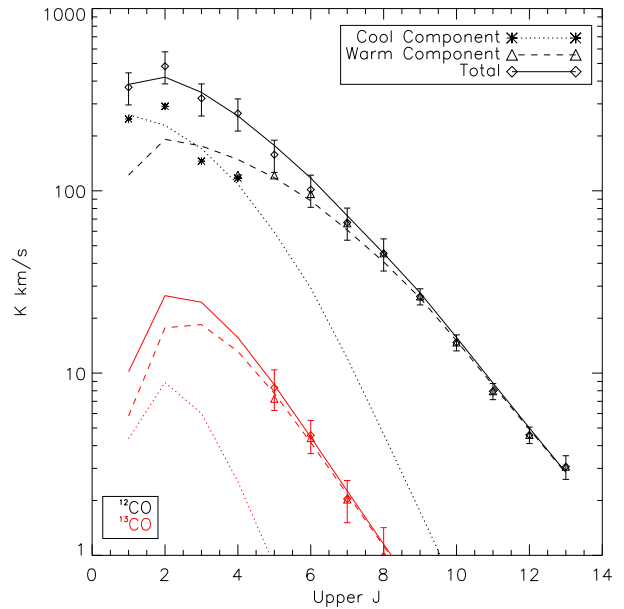}
\caption{Bayesian Likelihood Analysis, Spectral Line Energy Distributions, including $^{13}$CO.  Each color is a separate species: black for $^{12}CO$, red for $^{13}$CO (only warm component).  The total fluxes are shown by diamonds, but their separate components are in asterisks/triangles for the cool/warm components.  Best fit SLEDs (``4D Max" column in Table \ref{table:like2}) are shown with dotted/dashed lines for cool/warm components.  The total SLED, shown with a solid line, is the sum of the two components.}
\label{fig:deep_seds_multi}
\end{center}
\end{figure}

\begin{figure}
\begin{center}
\includegraphics[width=\columnwidth]{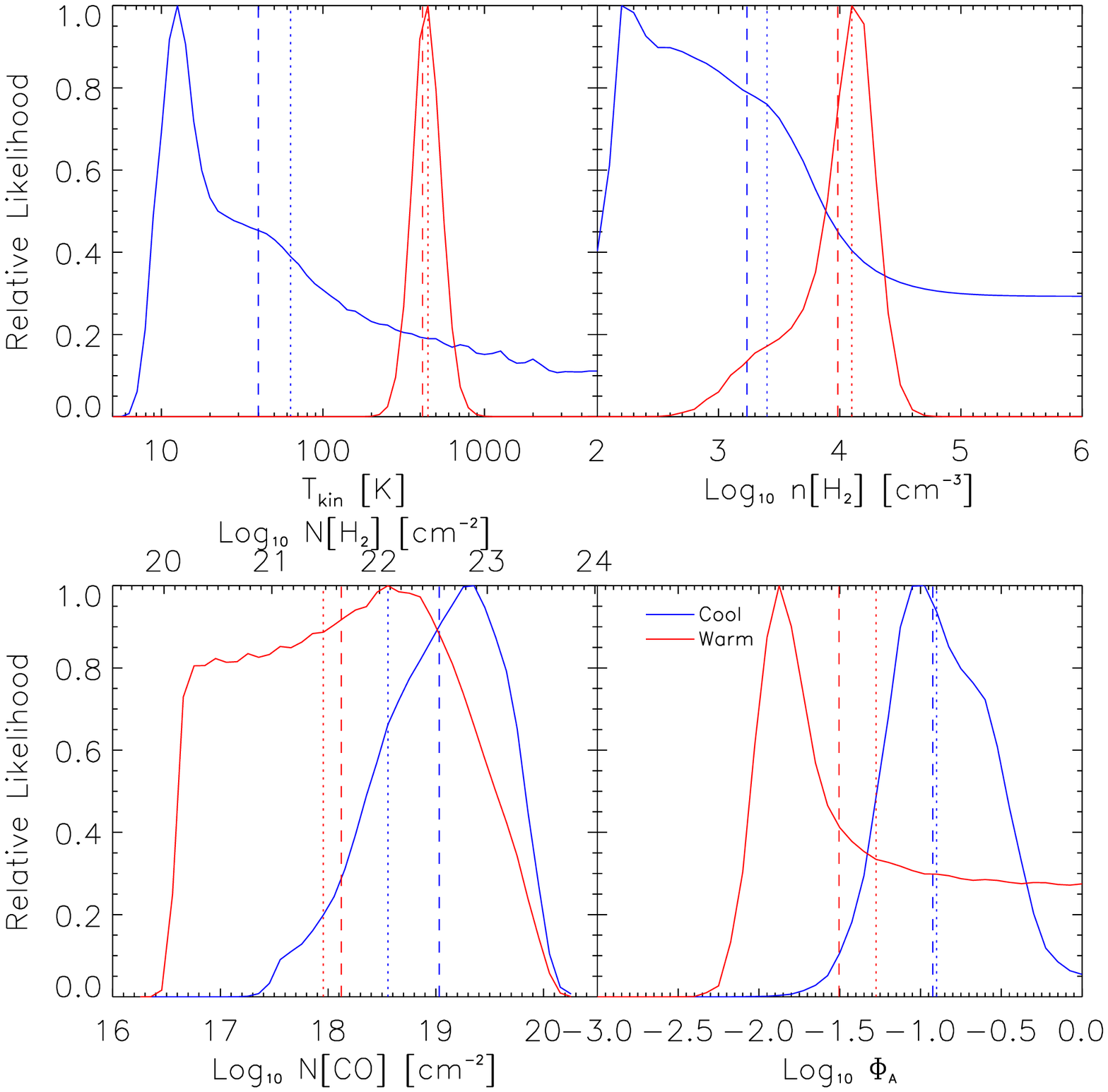}
\caption{Bayesian Likelihood Analysis, Primary Parameter Results, $^{12}$CO only.  Each color represents a separate component; blue for cool, red for warm (see Section \ref{sec:like}).  Dashed/dotted vertical lines indicate the median/4D maximum of the distribution (see Table \ref{table:like1}).}
\label{fig:deep_results1_co}
\end{center}
\end{figure}

\begin{figure}
\begin{center}
\includegraphics[width=\columnwidth]{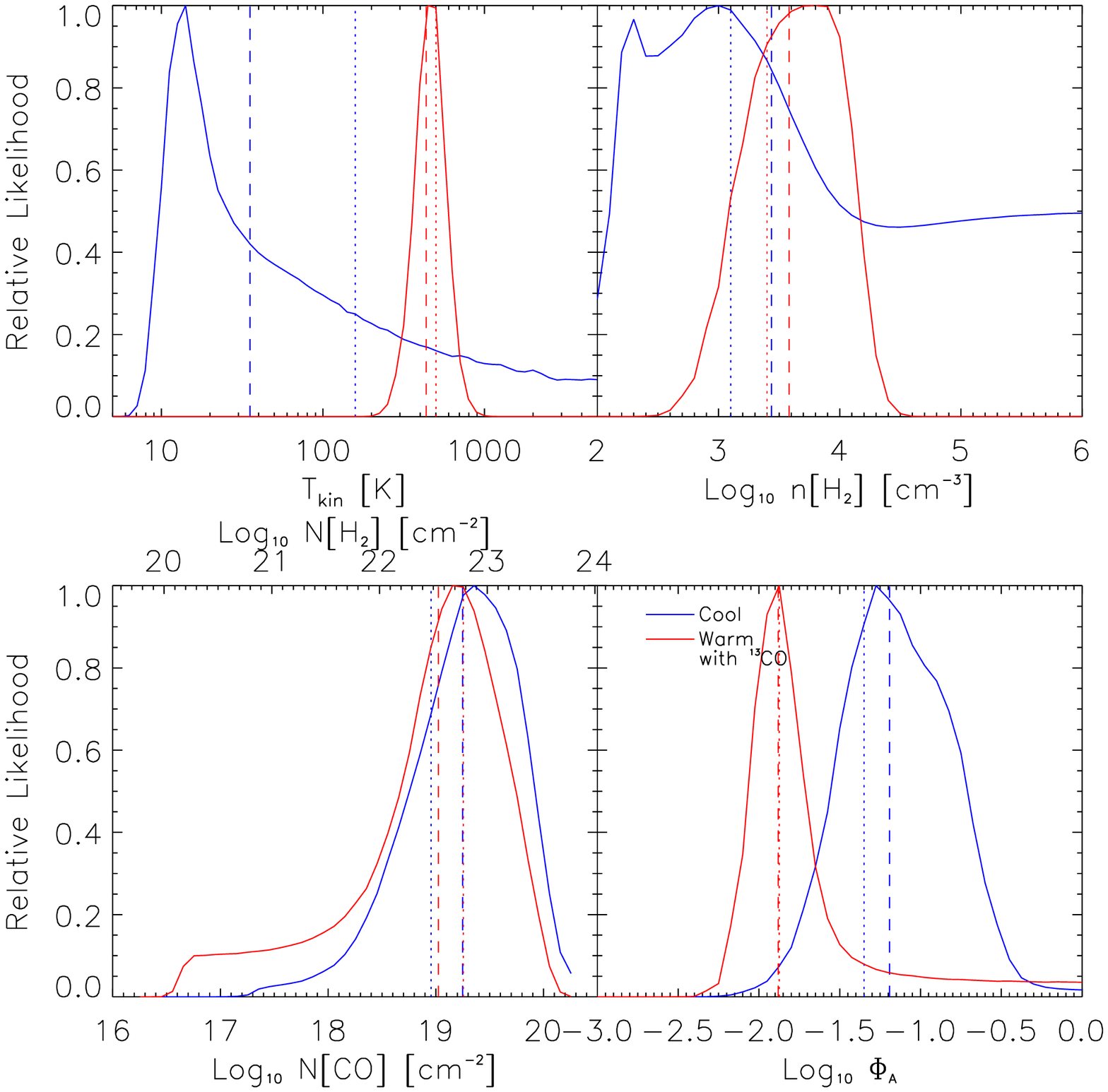}
\caption{Bayesian Likelihood Analysis, Primary Parameter Results, including $^{13}$CO.  Each color represents a separate component; blue for cool, red for warm (see Section \ref{sec:like}).  Dashed/dotted vertical lines indicate the median/4D maximum of the distribution (see Table \ref{table:like1}).}
\label{fig:deep_results1_multi}
\end{center}
\end{figure}

\begin{figure}
\begin{center}
\includegraphics[width=\columnwidth]{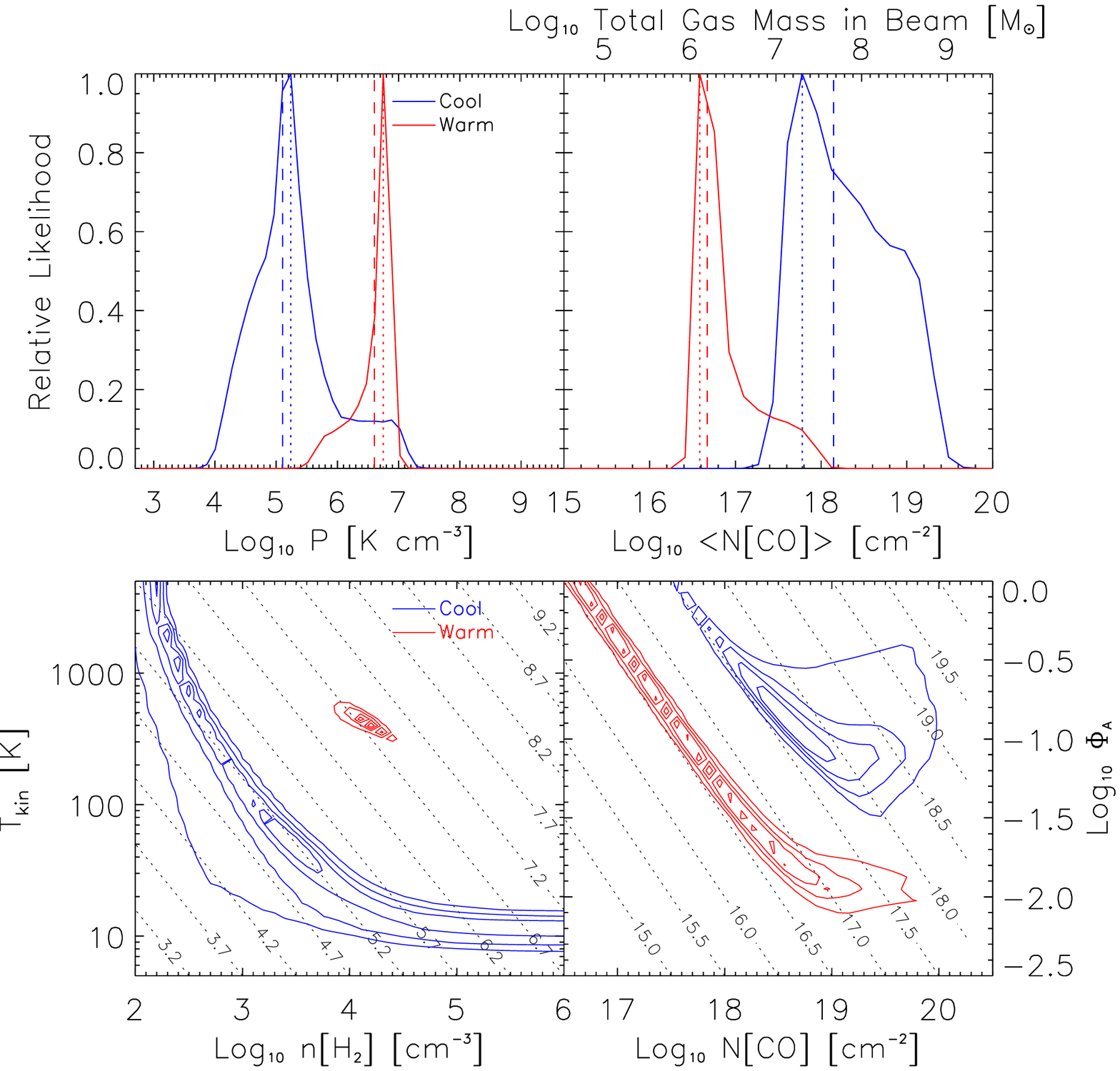}
\caption{Bayesian Likelihood Analysis, Secondary Parameter Results, $^{12}$CO only.  Top: Each color represents a separate component; blue for cool, red for warm (see Section \ref{sec:like})  Dashed/dotted vertical lines indicate the median/4D maximum of the distribution (see Table \ref{table:like1}).  Bottom: 2D distributions for the pairs of primary parameters from which the secondary parameters (top) were derived; colors correspond to component.  Diagonal lines indicate values of the top parameters (pressure on left and beam-averaged column density on right).  Contour levels are 20, 40, 60, and 80\% of the maximum likelihood.}
\label{fig:deep_results2_co}
\end{center}
\end{figure}

\begin{figure}
\begin{center}
\includegraphics[width=\columnwidth]{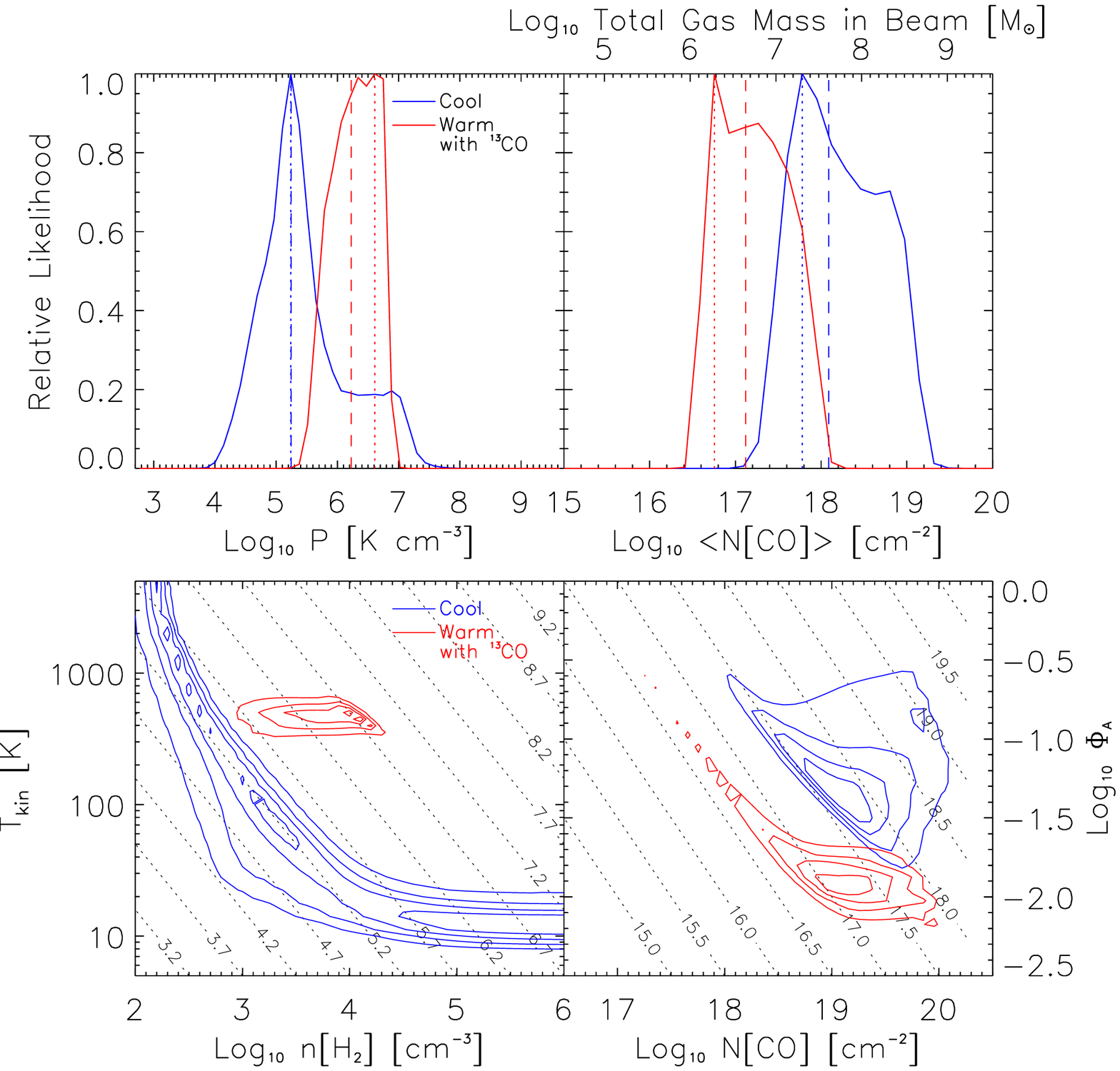}
\caption{Bayesian Likelihood Analysis, Secondary Parameter Results, including $^{13}$CO.  Top: Each color represents a separate component; blue for cool, red for warm (see Section \ref{sec:like})  Dashed/dotted vertical lines indicate the median/4D maximum of the distribution (see Table \ref{table:like1}).  Bottom: 2D distributions for the pairs of primary parameters from which the secondary parameters (top) were derived; colors correspond to component.  Diagonal lines indicate values of the top parameters (pressure on left and beam-averaged column density on right).  Contour levels are 20, 40, 60, and 80\% of the maximum likelihood.}
\label{fig:deep_results2_multi}
\end{center}
\end{figure}



\begin{deluxetable*}{c | c c c | c | c c c | c c}
\tablecaption{Likelihood Results: $^{12}$CO Only\label{table:like1}}
\startdata
\hline
 & \multicolumn{3}{|c|}{Integrated Likelihood: Cool} & Cool &\multicolumn{3}{|c|}{Integrated Likelihood: Warm} & Warm & \\
& Median & 1 Sigma Range & 1D Max & 4D Max & Median & 1 Sigma Range & 1D Max & 4D Max & \\
\hline
$T_{\rm kin}$  &   40 &   12 $-$  472 &   13 &   63                      & 414 &  335 $-$  518 &  447 &  447 & [K] \\
$n({\rm H_2})$ & $10^{3.23}$ & $10^{2.36} - 10^{4.81}$ & $10^{2.20}$ & $10^{3.40}$      & $10^{3.98}$ & $10^{3.53} - 10^{4.21}$ & $10^{4.10}$ & $10^{4.10}$ &[cm$^{-3}$] \\
$N_{co}$       & $10^{19.03}$ & $10^{18.38} - 10^{19.56}$ & $10^{19.36}$ & $10^{18.56}$ & $10^{18.12}$ & $10^{17.06} - 10^{19.09}$ & $10^{18.56}$ & $10^{17.96}$ &[cm$^{-2}$] \\
$\Phi_{\rm A}$ & $10^{-0.92}$ & $10^{-1.22} - 10^{-0.57}$ & $10^{-0.97}$ & $10^{-0.90}$ & $10^{-1.50}$ & $10^{-1.96} - 10^{-0.52}$ & $10^{-1.88}$ & $10^{-1.28}$ &\\
$P$            & $10^{ 5.11}$ & $10^{ 4.55} - 10^{ 5.76}$ & $10^{ 5.24}$ & $10^{ 5.24}$ & $10^{ 6.61}$ & $10^{ 6.16} - 10^{ 6.80}$ & $10^{ 6.75}$ & $10^{ 6.75}$ &[K cm$^{-2}$]\\
$<N_{\rm co}>$ & $10^{18.15}$ & $10^{17.63} - 10^{18.84}$ & $10^{17.78}$ & $10^{17.78}$ & $10^{16.67}$ & $10^{16.44} - 10^{17.20}$ & $10^{16.58}$ & $10^{16.58}$ &[cm$^{-2}$]\\
Mass           & $10^{ 7.67}$ & $10^{ 7.16} - 10^{ 8.36}$ & $10^{ 7.31}$ & $10^{ 7.31}$ & $10^{ 6.20}$ & $10^{ 5.97} - 10^{ 6.72}$ & $10^{ 6.11}$ & $10^{ 6.11}$ &[M$_\odot$]
\enddata
\tablecomments{1D Max refers to the maximum likelihood of that parameter after integrated over all the other parameters (the mode of the likelihood distributions).  4D Max refers to the single most probable grid point (mode of the entire multi-dimensional distribution).  Median, 1 sigma lower and upper values refer to the integrated distribution, when the cumulative distribution function equals 0.5, $\sim$ 0.16 and $\sim$ 0.84, respectively.  Note that the 1D Max may be outside the 1 sigma range because of asymmetry in the integrated likelihood.  $P$ and $\langle N_{\rm ^{12}CO} \rangle$ are calculated from the 2D distribution of $T_{\rm kin}$ and $n({\rm H_2})$ (or $\Phi_{\rm A}$ and $N_{\rm ^{12}CO}$) , which is why we give the same value in both the 1D Max and 4D Max columns.}
\end{deluxetable*}

\begin{table*}
\centering
\caption{Likelihood Results: $^{12}$CO and $^{13}$CO\label{table:like2}}
\begin{tabular}{c | c c c | c | c c c | c c}
\hline
 & \multicolumn{3}{|c|}{Integrated Likelihood: Cool} & Cool &\multicolumn{3}{|c|}{Integrated Likelihood: Warm} & Warm & \\
& Median & 1 Sigma Range & 1D Max & 4D Max & Median & 1 Sigma Range & 1D Max & 4D Max & \\
\hline
$T_{\rm kin}$   &   35         &   12  $-$  385       &   14      &  158       &436 &  344 $-$  548 &  447 & 501 &[K] \\
$n({\rm H_2})$  & $10^{3.44}$  & $10^{2.48} - 10^{5.15}$   & $10^{3.00}$  & $10^{3.10}$  &$10^{3.58}$ & $10^{3.17} - 10^{3.96}$ & $10^{3.80}$ & $10^{3.40}$ & [cm$^{-3}$] \\
$N_{co}$        & $10^{19.25}$ & $10^{18.69} - 10^{19.70}$ & $10^{19.36}$ & $10^{18.96}$ &$10^{19.02}$ & $10^{18.19} - 10^{19.51}$ & $10^{19.16}$ & $10^{19.26}$ & [cm$^{-2}$] \\
$\Phi_{\rm A}$  & $10^{-1.19}$ & $10^{-1.52} - 10^{-0.83}$ & $10^{-1.28}$ & $10^{-1.35}$ &$10^{-1.88}$ & $10^{-2.06} - 10^{-1.50}$ & $10^{-1.88}$ & $10^{-1.88}$ &  \\
$P$             & $10^{ 5.24}$ & $10^{ 4.73} - 10^{ 6.18}$ & $10^{ 5.24}$ & $10^{ 5.24}$ &$10^{ 6.23}$ & $10^{ 5.80} - 10^{ 6.60}$ & $10^{ 6.61}$ & $10^{ 6.61}$ & [K cm$^{-2}$]\\
$<N_{\rm co}>$  & $10^{18.09}$ & $10^{17.59} - 10^{18.71}$ & $10^{17.78}$ & $10^{17.78}$ &$10^{17.12}$ & $10^{16.69} - 10^{17.59}$ & $10^{16.76}$ & $10^{16.76}$ & [cm$^{-2}$]\\
Mass            & $10^{ 7.62}$ & $10^{ 7.11} - 10^{ 8.23}$ & $10^{ 7.31}$ & $10^{ 7.31}$ &$10^{ 6.65}$ & $10^{ 6.22} - 10^{ 7.12}$ & $10^{ 6.28}$ & $10^{ 6.28}$ & [M$_\odot$]\\
\hline
$X_{\rm 13co}/X_{\rm co}$ & - & - & - & - & $10^{-1.58}$ & $10^{-1.70} - 10^{-1.46}$ & $10^{-1.50}$ & $10^{-1.50}$ & \\
$N_{\rm 13co}$ & - & - & - & - & $10^{17.44}$ & $10^{16.62} - 10^{17.92}$ & $10^{17.62}$ & $10^{17.62}$ & [cm$^{-2}$] \\
\hline
\end{tabular}
\end{table*}

\subsection{Physical Conditions: Deep Spectrum}\label{sec:discdeep}

We present two different versions of our likelihood analysis for the deep spectrum: one using only $^{12}$CO, and one using $^{12}$CO and $^{13}$CO (``multiple molecule").  The motivation behind this is to investigate two questions: does the addition of different species change the modeled parameters of the gas, and/or does it better constrain the parameters?  Both versions contain a warm and cool component.  The modeling assumes all of the emission in a given component is coming from the same homogeneous gas, and by comparing these models we will investigate the validity of this assumption in this subsection.

The results for each of these versions are presented in Tables \ref{table:like1} and \ref{table:like2} respectively.  Figures \ref{fig:deep_seds_co} and \ref{fig:deep_seds_multi} show the input SLED as well as the best-fit model results.  The primary results (temperature, density, column density, and filling factor) are displayed graphically in Figures \ref{fig:deep_results1_co} and \ref{fig:deep_results1_multi}.  Secondary parameters, which are calculated from the aforementioned primary results, are displayed in Figures \ref{fig:deep_results2_co} and \ref{fig:deep_results2_multi}.  These include the pressure (the product of temperature and density) and the beam-averaged column density ($\langle N_{\rm ^{12}CO} \rangle$, the product of column density and filling factor).  We note that in the results, the parameters of the most likely grid point (``4D Max") is not necessarily the same as the median or the 	mode (``1D Max") of the integrated likelihood distributions.  The ``4D Max" is describing one specific point, but the median, 1D Max, and associated error range are representative of the larger likelihood across all other parameters in the grid.  

We first compare our two models, $^{12}$CO only vs. multiple molecule.  These are Tables \ref{table:like1} vs. \ref{table:like2}, Figures \ref{fig:deep_results1_co} vs. \ref{fig:deep_results1_multi}, and Figures \ref{fig:deep_results2_co} vs. \ref{fig:deep_results2_multi}.  

The addition of $^{13}$CO to the warm component does not significantly change the temperature, but it does increase the likelihood of lower densities.  It also decreases the likelihood of lower column densities.  The consequences of these changes can be seen in the pressure and mass distributions in Figures \ref{fig:deep_results2_co} and \ref{fig:deep_results2_multi}.  Adding $^{13}$CO increased the likelihoods of the ``shoulders" of these distributions; the lower half of pressure, and the upper half of mass.  An examination of the contour plots in the bottom half of these figures illustrates why, statistically.  In the $^{12}$CO only model, though column density and filling factor are not well constrained independently, they are highly correlated; their contours run along an almost constant line of beam-averaged column density.  Adding $^{13}$CO introduced the $X_{^{13}CO}/X_{^{12}CO}$ parameter, which also impacts the absolute flux level of the models, like column density and filling factor.  The result are likelihoods that are more constrained but not as highly correlated with one another.  Therefore the mass distribution is wider.  In the $^{12}$CO-only model, the mass of the warm component is about 3.4/6.3\% (median/4D Max) the mass of the cool component.  \citet{Rigopoulou:2002} noted that ``warm" gas is generally around 1 to 10\% of total gas mass for starburst galaxies, so this is about as expected.  The addition of $^{13}$CO creates somewhat overlapping likelihood distributions for mass (the 1-sigma ranges are just touching), but the median and 4D Max now warm/cool ratios of 11\% and 9\%, respectively.  One factor that may contribute to wider distributions is overestimated error; the error bars are dominated by our 20\% calibration error, not statistical error.

After this point, we compare frequently to \citet{Panuzzo:2010}.  The major factor responsible for the differences, just modeling $^{12}$CO alone, is the shape of the CO SLED at those lines with upper rotational number greater than J=8.  We also explicitely subtracted the cool component's contribution from $^{12}$CO \jfour, whereas \citet{Panuzzo:2010} simply underpredicted the total flux.  We will also compare with \citet{Loenen:2010}, another high-J CO study of M82, in Section \ref{sec:energy}.

Our results are similar to \citet{Panuzzo:2010}, who found that these high-J CO lines trace a very warm gas component that is separate from the cold molecular gas traced by those lines below \jfour.  Our best-fit temperature of the $^{12}$CO only model (at 447 K, Table \ref{table:like1}) is close to their value at 545 K, but the overall likelihood for temperature, integrated over all parameters, yields a slightly lower 414 K.  Given the size of the uncertainty (335-518 K) in the parameter, the two distributions are very similar, and therefore the result is not significantly different.  Such warm gas has also been traced in the S(1) and S(2) transitions of H$_2$, at 450 K with the Infrared Space Observatory \citep{Rigopoulou:2002} and 536 K with the Spitzer Infrared Spectrograph \citep{Beirao:2008}.

We do find a slightly higher density than \citet{Panuzzo:2010}, with our best-fit value of $10^{4.1}$ compared to $10^{3.7}$ cm$^{-3}$, though the integrated likelihood distributions do overlap (see Figure \ref{fig:deep_results1_co}).  However, the temperature and density are degenerate; higher temperatures and lower densities may produce the same fluxes as lower temperatures and higher densities.  Their product, the pressure, is better constrained.  We seem to have collapsed/constrained the pressure distribution to the upper half of that presented in \citet{Panuzzo:2010}.

The column density is not as well constrained as presented in \citet{Panuzzo:2010}; we found that they had an error in calculating the expected fluxes of the higher-J lines for lower column density values.  We have recalculated the fluxes for those column densities, and we find that in fact when properly calculated these column densities have a non-zero likelihood.  In the $^{12}$CO only model, the column density itself is not constrained.  However, the column density and filling factor are degenerate, so it is their product (beam-averaged column density, $\langle N_{\rm co} \rangle$) that is better constrained.  Our best-fit value is 10$^{16.6}$ cm$^{-2}$.

The total mass in the beam can be calculated using Equation \ref{eqn:mass} (and is presented as the top y-axis in Figures \ref{fig:deep_results2_co} and \ref{fig:deep_results2_multi}, upper right).  As previously discussed, the $^{12}$CO only model produces the expected result of less gas mass in the warm component.  In the cool component we find a best-fit mass of 2.0 $\times 10^7$ \ms\ (median 4.7 $\times 10^7$ \ms).  This is smaller than the 2.0 $\times 10^8$ \ms\ traced by the LVG analysis of \citet{Ward:2003} with lower-J CO lines.  This difference is due to the fact that we subtract the contribution to the low-J flux from the warm component; in our initial modeling of the cold component, before this subtraction, our best-fit mass is 9.8 $\times 10^7$ \ms\ (with a range from 0.2 to 2.2 $\times 10^8$ \ms).  The warm component is a smaller fraction of the gas, with a best-fit of 1.3 $\times 10^6$ \ms, about 6.3\% the mass of the cool component.  

The $^{12}$CO/$^{13}$CO relative abundance is also a free parameter in our multi-species model; we find a best-fit value of about 32, similar to the 40 and 30 found previously for the NE and SW lobes, respectively \citep{Ward:2003}. 

As mentioned in Section \ref{sec:components}, we also attempted to include our two [C {\sc{I}}] lines with the cool component.  We do not present the tables for this model because the mass distributions of the warm and cool components became overlapping, indicating the same amount of mass in both components, an unphysical situation.  Additionally, the derived relative abundance of [C {\sc{I}}] to $^{12}$CO was unusually high.  We found a ratio of 0.48 to 3.3, which is higher than \citet[average value $\sim 0.5$]{White:1994}, \citet[0.1-0.3]{Schilke:1993}, and \citet[0.5]{Stutzki:1997} using other methods.  Before subtracting the warm component's contribution to the $^{12}$CO flux (when we just fit all of the low-J $^{12}$CO flux and [C {\sc{I}}]), we find ratios more consistent with these values (best-fit 0.4, range 0.09 to 1.23).  These two problems could be indicating that the assumption of CO and [C {\sc{I}}] coming from the same component is flawed.  The column density, temperature, and mass developed somewhat of a double-peaked structure; specifically, the addition of [C {\sc{I}}] increases the likelihood of lower column densities and masses, but does not eliminiate the previous likelihood peak.  

It is unclear how much of the molecular CO and atomic C are truly cospatial and therefore how physical our results for modeling them all together as one bulk gas component may be.  \citet{Papadopoulos:2004a} presented results which argue that [C {\sc{I}}] and CO are cospatial and trace the same hydrogen gas mass ([C {\sc{I}}] doing so better than CO in many conditions).  This conflicts with the theoretical picture of gas clouds (especially in PDRs) as a structured transition between molecular, atomic, and ionized gas, but new observational and theoretical evidence indicates the types of gas are not so distinct \citep[see references within][]{Papadopoulos:2004a}.  For example, \citet{Howe:2000} and \citet{Li:2004} have found [C {\sc{I}}] to trace $^{13}$CO well.  If the ISM is clumpy, well mixed, and dynamic, the [C {\sc{I}}] and CO may be cospatial averaged over large scales.  Strong stellar winds (and possibly the interaction with M81) could be contributing to the dynamic nature of the gas, so it is not unreasonable to believe that the gas has not achieved the simple layered pattern.  Though the SPIRE FTS cannot resolve the two separate velocity components of M82, HIFI can, and observations of these two [C {\sc{I}}] lines indicate a generally similar shape to the CO lines, namely two Gaussians with the SW component demonstrating higher flux \citep{Loenen:2010}.  

\citet{Wolfire:2010} presented a model PDR which shows a cloud layer traced by atomic (not-yet-ionized) [C {\sc{I}}] where the hydrogen is still molecular due to self-shielding effects.  This ``dark molecular gas" (called so because it is not traced by CO) would be less-shielded and at a warmer temperature than the inner-most cloud layer of CO.  It is possible that [C {\sc{I}}] and CO are somewhat cospatial yet somewhat segregated as in the ``dark molecular gas model."  Our analysis is consistent with a picture in which the [C {\sc{I}}] and $^{12}$CO do not completely overlap spatially.  

The [C {\sc{I}}] \jone\ emission can also be used to estimate the total hydrogen mass using Equation 12 of \citet{Papadopoulos:2004a}.  Using the median $X_{[C {\sc{I}}]}/X_{H_2}$ of 1.5 $\times 10^{-4}$ from $X_{[C {\sc{I}}]}/X_{^{12}CO}$ = 0.5 \citep{White:1994,Stutzki:1997} and our assumed $X_{^{12}CO}/X_{H_2}$, we find a total gas mass of $4.4 \times 10^7 Q_{10}^{-1}$ \ms.  $Q_{10}$ is the ratio of the column of the \jone\ emission to the total [C {\sc{I}}] column \citep[see Appendix A of][]{Papadopoulos:2004a}, which depends on the excitation conditions of the gas; for $Q_{10} \sim 0.5$ \citep{Papadopoulos:2004b}, the gas mass is $8.8 \times 10^7$ \ms\ but is uncertain by modeled uncertainty in $X_{[C {\sc{I}}]}$ alone.  This method of estimating the mass is higher than total mass estimate of the cool component described earlier in this section.  [C {\sc{I}}] may be coming from a range of temperatures, but with only two lines, we cannot sort that out.

\subsection{Physical Conditions: Map}\label{sec:discmap}

\begin{figure*}
\begin{center}
\includegraphics[width=\textwidth]{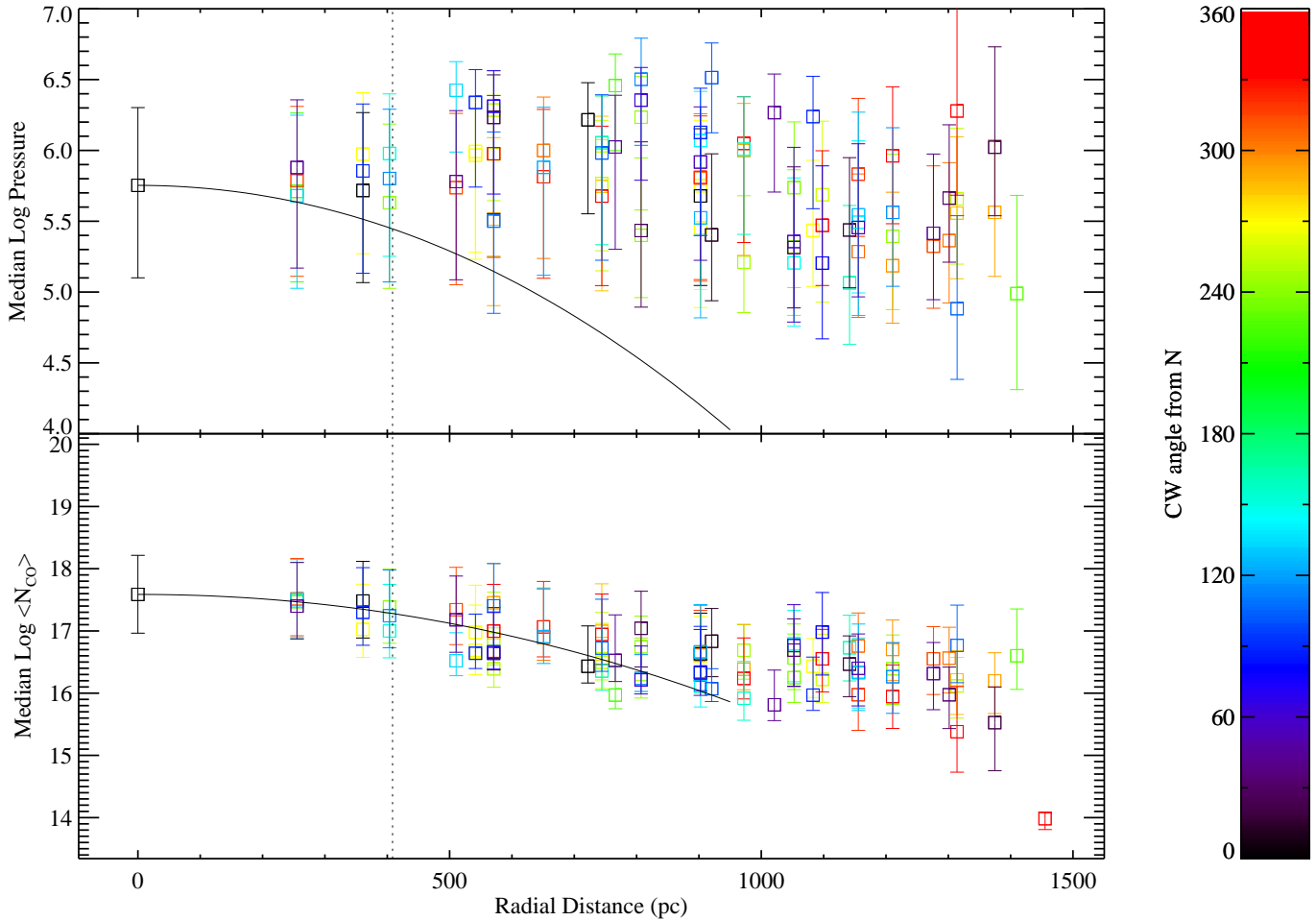}
\caption{Bayesian Likelihood Analysis for Mapping Observation.  The x-axis is the map pixel's radial distance (in pc) from the central detector's position, with its clockwise angle (meaning degrees from north through west) represented by the colorbar on the right.  The y-axis is the median value (after integrating over all other parameters) and associated 68\% error bars.  The solid line represents the logarithm of the $^{12}$CO \jfour\ beam profile such that the peak corresponds with the result at the central pixel; it is only plotted to 50\as\ from the center because of uncertainties in the profile beyond this region.  The dashed line represents one half the beam FWHM.}
\label{fig:map_results}
\end{center}
\end{figure*}

Results for all of the same parameters for the deep spectrum were produced for each of the pixels in the map (that met the criteria in Section \ref{sec:maplike}).  We find that the beam size of SPIRE cannot resolve the structure in M82 as has been done with interferometric maps or high spectral resolution observations (which can resolve the velocity components of the NE and SW lobes).  Because of the degeneracy between temperature/density and column density/filling factor, we present results of their products, pressure and beam-averaged column density in Figure \ref{fig:map_results}.  $\langle N_{\rm ^{12}CO} \rangle$ shows a radially decreasing trend, roughly corresponding to the decrease in the beam profile (plotted with a solid black line), implying an observational effect due to source-beam coupling.  We would expect an off-centered beam to be able to probe the pressure of the central region (because the relative ratios of the SLED would be preserved), so the lack of a radial trend also implies that we are not measuring different areas of M82 in each map pixel.  This indicates that the SPIRE FTS map cannot resolve M82's structure, and therefore the single ``deep" spectrum is an adequate representation of the galaxy as a whole.  Our results in Section \ref{sec:discdeep} are descriptive of the bulk properties of the galaxy and we do not see trends on the scale of our map.  This analysis is separate from the dust temperature gradients (which we also see in the continuum gradients of our map spectra) found in M82 by \citet{Roussel:2010} and indicates that the dust and molecular gas are not coupled.

One difficulty with the map is it has lower signal/noise; it must also be convolved up the largest beam size (43\as, like the deep spectrum).  However, as mentioned in Section \ref{sec:discdeep}, it is the highest-J fluxes in the SSW detector that largely constrain the results of the likelihood for the deep spectrum.  Therefore, we also attempted to model the map with just those lines with upper-J level of 9 or higher, without convolving.  These lines all were measured with a beam FWHM of $\sim$ 19\as, offering higher resolution.  However, these lines are also weaker, and with fewer, lower signal/noise lines this method does not constrain any parameters as well.

Though the off-axis pixels may not provide new information about the physical trends of the galaxy, we can compare the deep spectrum to the center pixel of our map as a test of the source-beam coupling corrections (see Section \ref{sec:mapmaking}).  The results of the central pixel are very similar to those presented in Section \ref{sec:discdeep}, though the integrated likelihoods are generally wider (the parameters are less constrained).  This is partially, but not entirely, due to larger error bars on the SSW lines.  

\subsection{Molecular Line Absorption}\label{sec:abs}

In addition to the Bayesian likelihood modeling, we can also briefly discuss the absorption lines presented in Table \ref{table:fit}.  

\subsubsection{Hydrogen Fluoride (HF)}

Hydrogen fluoride (HF) is potentially a sensitive probe of total molecular gas column, because the HF/H$_2$ ratio is more reliably constant than $^{12}$CO/H$_2$ because the formation of HF is dominated by a reaction of F with H$_2$ \citep{Monje:2011}.  Furthermore, HF \jone\ is generally seen in absorption because of its high A-coefficient, 2.42 $\times 10^{-2}$ s$^{-1}$ \citep{Monje:2011}.  Assuming all HF molecules are in the ground state \citep[generally true in the diffuse and dense ISM,][]{Monje:2011}, the HF \jone\ line yields the optical depth simply as $\tau = -ln(F_l/F_c)$, where $F_l/F_c$ is the line-to-continuum ratio.  In the case of HF, we mask out a nearby water emission line (1226 to 1229 GHz), though because the signal in each wavenumber bin is not independent due to the ringing, this can introduce added uncertainty.  Therefore the following discussion is meant to be approximate.  We estimate the HF column density using

\begin{equation}\label{eqn:column}
\int \tau dv = \frac{A_{ul} g_u \lambda^3}{8 \pi g_l} N(HF), 
\end{equation}

\noindent where $g_u$ = 3 and $g_l$ = 1.  This implies $\int \tau dv = 4.16 \times 10^{-13} N(HF) \unit{cm^2 \ km \ s^{-1}}$.  The HF line occurs in a part of our spectrum with some noticeable structure in the continuum (see Figure \ref{fig:spec}, third panel, around 1250 GHz).  If we only integrate the 6 GHz surrounding the line, we find a column density of HF of 6.61 $\times 10^{13}$ cm$^{-2}$.  Expanding the range over which we integrate increases the derived column density, but this may be due to other features in the spectrum, and so we consider our derived value a lower limit.  Assuming a predicted abundance of HF of $3.6 \times 10^{-8}$ \citep{Monje:2011}, this corresponds to a molecular hydrogen column density of $1.84 \times 10^{21}$ cm$^{-2}$.  The column density derived from this line is similar to that of $\langle N \rangle$ of the cool component of $^{12}$CO.  However, there are still some uncertainties to this calculation.  We are only able to see the HF in front of the continuum emission, and therefore we are not probing the total column density.  Higher spectral resolution could reveal the extent of spatial colocation of HF with CO.  There are also uncertainties associated with either molecular abundance assumed and whether or not all HF molecules are truly in the ground state.

\subsubsection{Water and Water Ion (H$_2$O$^{(+)}$)}

Water is fundamental to the energy balance of collapsing clouds and the subsequent formation of stars, planets, and life.  Many {\it Herschel} key programs are currently studying water and chemically related molecular species in a variety of conditions.  An excellent summary of water chemistry in star forming regions is available in \citet{vanDishoeck:2011}.  

\citet{Weiss:2010} studied the low-level water transitions in M82, and they detect the ground-state o-H$_2$O emission in two clearly resolved components, which we do not.  With the 41\as\ beam, the two components add to 370 $\pm$ 44 Jy km/s beam$^{-1}$, well below our threshold of detection, as can be seen by examining Table \ref{table:fit}.  Though we do not detect that line, we do detect four new water lines in addition to those presented in \citet{Weiss:2010}: two p-H$_2$O (752 and 1229 GHz) and two o-H$_2$O (1097 and 1153 GHz).  Combined with their ground-state transition of o-H$_2$O, we add to the picture of the water excitation in M82.

Our ground-state lines indicate similar column densities as \citet{Weiss:2010}, within a factor of 2.  Using Equation \ref{eqn:column} (low-excitation approximation), we find column densities of p-H$_2$O and o-H$_2$O$^+$ of $\sim 4 \times 10^{13}$ cm$^{-2}$, whereas \citet{Weiss:2010} finds 9.0 and 2.2 $\times 10^{13}$ cm$^{-2}$, respectively.  They found that the water absorption comes from a region northeast of the central CO peak; shocks related to the bar structure of M82 could be releasing water into the gas phase at such a location.  The fact that the water comes from a lower column density region seems to contradict the existence of a PDR, which would require high column densities to shield water from UV dissociation; however, the relative strength and similarity of absorption profiles of o-H$_2$O$^+$ compared to p-H$_2$O indicates some ionizing photons (see their work for complete interpretation).  Though these transitions are tracing a different region than CO, they add to the picture of a complicated mix of energy sources present in the gas, as addressed in Section \ref{sec:energy}.  Models of water emission from shocks have been investigated by others \citep[i.e.][]{Flower:2010}, but detailed modeling of the water spectrum of M82 is outside the scope of this work.

\subsection{Gas Excitation}\label{sec:energy}

At the high temperature of the warm component, the cooling will be dominated by hydrogen.  \citet{LeBourlot:1999} modeled the cooling rates for H$_2$, and made their tabulated rates available with an interpolation routine for desired values of density, temperature, ortho- to para-H$_2$ ratio, and H to H$_2$ density ratio\footnote{http://ccp7.dur.ac.uk/cooling\_by\_h2/}.  For our best-fit temperature and density, assuming n(H)/n(H$_2)$ = 1 \citep[recommended by][for PDRs]{LeBourlot:1999}, this corresponds to a cooling rate of $10^{-22.54}$ erg s$^{-1}$ per molecule, or 3 \ls/\ms (using o/p H$_2$ = 1, though the number is only $\sim$ 3\% lower for o/p H$_2$ = 3).  Given the warm mass of 1.3 $\times$ 10$^{6}$ \ms\ (using the $^{12}$CO model for the rest of this section), that implies a hydrogen luminosity of 3.9 $\times 10^{6}$ \ls.  The total observed hydrogen luminosity thus far has been higher than this number; adding the luminosities presented in \citet{Rigopoulou:2002} and correcting for extinction as in \citet{Draine:1989}, we find a total of 1.2 $\times 10^7$ \ls in the (0-0)S(0)-S(3), S(5), S(7), and (1-0)Q(3) lines.  However, some of these hydrogen lines are tracing lower or higher temperature gas.  We note that the mass range within one standard deviation of our likelihood results for the warm component is 0.93-5.2 $\times 10^6$ \ms, which corresponds to a predicted luminosity of 0.28-1.6 $\times 10^7$ \ls, encompassing the measured hydrogen luminosity.

There are a few possibilities for the source of the excitation of the gas: X-ray photons, cosmic rays, UV excitation of PDRs and shocks/collisional excitation.  Hard X-rays from an AGN have already been ruled out by others in the literature due to the lack of evidence for a strong AGN and low X-ray luminosity \citep[1.1 $\times 10^6$ \ls,][]{Strickland:2007}.

The CO emission from M82 has previously been interpreted using PDR models.  \citet{Beirao:2008} noted with the Spitzer Infrared Spectrograph that the H$_2$ emission is correlated with PAH emission, indicating that it is mainly excited by UV radiation in PDRs.  \citet{Loenen:2010} combined HIFI data with ground based detections in order to model $^{12}$CO \jone\ to \jthirteen\ and $^{13}$CO \jone\ to \jeight.  They reproduced the measured SLEDs with one low-density (log($n(H_2)$)=3) and two high-density (log($n(H_2)$)=5,6) components with relative proportions of 70\%, 29\%, and 1\%, respectively.  The low-density component is largely responsible for the low-J emission, while the highest-density component is responsible for the highest-J emission.  These high densities are not consistent with the results of our likelihood analysis detailed in Section \ref{sec:discdeep}; the likelihood of solutions for the warm component at log($n(H_2)$)$>$5 is essentially zero.  We note that \citet{Loenen:2010}'s Figure 3 shows the consistency between HIFI and SPIRE fluxes from \citet{Panuzzo:2010}.  In other words, the difference is not due to discrepant line fluxes, but different models (PDR vs. CO likelihood analysis).

There are two major differences in the approach of this work and \citet{Loenen:2010}.  First, the order in which we approach the problem is different.  We first analyze the CO excitation using likelihood analysis to determine the physical conditions.  Once we have these conditions, we then look to the possible energy sources \emph{based on} the conditions we have already derived, instead of first seeing under which conditions a certain energy source fits.  Second, we look beyond the one best fit solution: in addition to presenting the best-fit solution, our likelihoods analyze the relative probabilities for a larger parameter space.

\begin{figure}
\begin{center}
\includegraphics[width=\columnwidth]{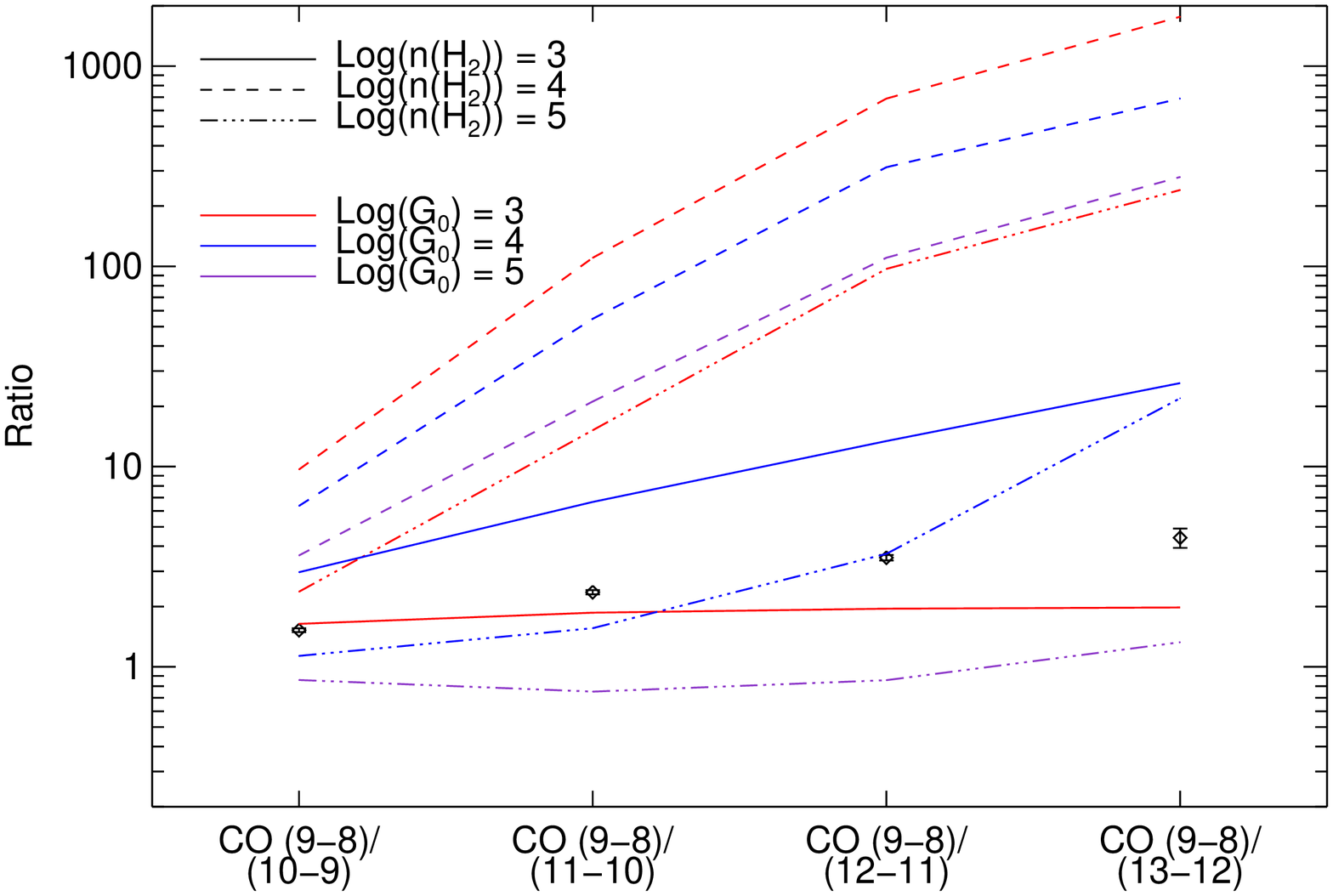}
\caption{Enhanced Cosmic Ray PDR Models.  The observed ratios are black diamonds.  For the models, the line style indicates gas density and the line color indicates incident flux.  Log(G$_0$)=2 is not plotted because the highest-J line fluxes are not reported for that model.}
\label{fig:PDR}
\end{center}
\end{figure}

We also attempted to reproduce our deep SLED with various PDR models.  \citet{Meijerink:2006} have added to their PDR and XDR models to include enhanced cosmic rays (at a rate of 5 $\times 10^{-15}$ s$^{-1}$), near our assumed rate discussed later in this section.  Such models are currently available for incident flux of log(G$_0$) of 2-4 (G$_0$ = 1.6 $\times 10^{-3}$ erg cm$^{-2}$ s$^{-1}$) for log($n(H_2)$)=3 and log(G$_0$) of 3-5 for log($n(H_2)$) of 4 and 5.  By examining the ratios of $^{12}$CO \jnine\ with all higher-J lines (those largely driving the likelihood results, and also measured from similar beam sizes), none of the available 9 PDR models are an ideal match, but the PDR scenario for log($n(H_2)$)=3, log(G$_0$)=3 is the best match.  Figure \ref{fig:PDR} compares the predicted and observed ratios.  The ratios used are without source-beam coupling correction because the SSW already has similar beam sizes, but the ratios with beam correction are within 4-8\% of those presented in Figure \ref{fig:PDR}.

Additionally, we used a higher-resolution (in density and incident flux) grid of $^{12}$CO PDR models \citep{Wolfire:2010}.  The same line ratios previously discussed (those shown in Figure \ref{fig:PDR}) can only be reproduced by higher densities.  For (\jnine)/(\jten), the observed ratio is only found for log($n(H_2)$) $>$ 4.5, log(G$_0$) $>$ 2, and by (\jnine)/(\jthirteen), only for log($n(H_2)$) $>$ 5, log(G$_0$) $>$ 2.5.

To summarize, current PDR models can only explain the observed high-J $^{12}$CO emission with densities higher than those indicated by the likelihood analysis (even when all priors are excluded).  The cosmic-ray enhanced PDR models, though sparse, can come closer to reproducing the high-J line ratios at a lower density, though these are also below our likelihoods.  There is also evidence that shocks enhance high-J lines far more than PDRs \citep{Pon:2012}.  In their models, almost all of the emission from the lowest-J lines came from unshocked gas, and most of the emission above \jseven\ was from shocked gas.  The combination of shocks and PDRs could be responsible for the high-J CO lines observed, while PDRs alone are adequate to explain the lower-J CO lines and PAH emission.  In summary, these high-J lines are not consistent with current PDR models, but improved modeling of shocks and the effects of cosmic rays on PDRs may help explain their emission.

Cosmic rays (CRs) are another possibility for the excitation.  The Very Energetic Radiation Imaging Telescope Array System (VERITAS) Collaboration recently reported that the cosmic-ray density in M82 is about 500 times the average Galactic (Milky Way)  density \citep{Veritas:2009}.  By using the cosmic ray ionization rate of the Galaxy \citep[2-7 $\times 10^{-17}$ s$^{-1}$,][]{Goldsmith:1978, vanDishoeck:1986}, multiplied by 500, and then multiplied by the average energy per ionization (20 eV), one finds an energy deposition rate of 2-7 $\times 10^{-13}$ eV/s per H$_2$ molecule in M82.  This implies a heating rate of 0.03 to 0.12 \ls/\ms, less than 5\% of the required molecular hydrogen cooling rate.  Thus, cosmic rays alone cannot excite the molecular gas.

Turbulent heating mechanisms may also play a role in M82.  From \citet{Bradford:2005}, the turbulent heating per unit mass can be expressed in \ls/\ms\ as

\begin{equation}
1.10 \bigg( \frac{v_{rms}}{25 \unit{km \ s}^{-1}} \bigg)^3 \bigg( \frac{1 \unit{pc}}{\Lambda_d} \bigg)
\end{equation}

\noindent where $v_{rms}$ is the turbulent velocity and $\Lambda_d$ is the typical size scale of turbulent structures.  We use the Jeans length for this size scale, calculated from the parameters of the likelihood results (Section \ref{sec:discdeep}), which is 0.9 pc.  Given this size scale, the observed cooling rate could be replicated with a turbulent velocity of 33.7 \kms.  This would imply a velocity gradient of approximately 37.5 \kmspc.  When we calculate the  velocity gradient using our model results ($dv/dr = \Delta v \ n(H_2) / N(H_2)$), we find a 68\% confidence lower limit of 16 \kmspc] (4 \kmspc\ if we use the model with $^{13}CO$) , but the upper limit is unphysically high.  The velocity required for turbulent heating seems reasonable in context of our likelihood results.  \citet{Panuzzo:2010} used their calculated velocity gradient of 35 \kmspc\ to determine that they could match the heating required with a sizescale of 0.3 to 1.6 pc.  These velocity gradients seem large compared to Galactic star-forming sites \citep[e.g.][]{Imara:2011}, but M82 is known to have powerful stellar winds.  According to \citet{Beirao:2008}, the starburst activity has decreased in the past few Myr, and this appears to be evidence of negative feedback (by stellar winds and supernovae), because M82 still has a large reservoir of gas available for star formation.  Therefore, there is evidence for turbulent heating mechanisms being in place.  Additionally, \citet{Downes:1998} found high turbulent velocities (30-140 \kms) in models of extreme star-forming galaxies.

None of the possibilities described seem to provide enough heating by themselves, with the exception of turbulent heating, which is based on a few approximations and assumptions.  Likely, there is a combination of factors, namely PDRs and shocks/turbulent mechanisms.  Such a situation has also been seen in other submillimeter-bright galaxies, discussed next.  Interestingly, even a more quiescent galaxy like NGC 891 requires a combination of PDRs and shocks to explain mid-J CO transitions \citep[\jsix, \jseven,][]{Nikola:2011}.

\subsection{Comparison to Other Starburst and Submillimeter Galaxies}

Because only the first few lines in the CO ladder are easily visible from the ground for nearby galaxies, the high-J lines detected by {\it Herschel} represent new territory.  Therefore, while adequate diagnostics of high-J CO lines are still being developed, it is useful to compare to other submillimeter-bright galaxies.

Mrk 231 contains a luminous (Seyfert 1) AGN.  It also shows a strong high-J CO ladder, such that only the emission up to \jeight\ is explained by UV irradiation from star formation.  Their high-J CO luminosity SLED however, is flat (though ours for M82 are stronger than predicted, they are still decreasing with higher-J).  \citet{vanderWerf:2010} can explain this trend with either an XDR or a dense PDR.  An additional difference between M82 and Mrk 231 is that OH$^+$ and H$_2$O$^+$ are both seen in strong emission in Mrk 231 (instead of absorption), indicative of X-ray driven chemistry.  Mrk 231 is also more face-on than M82.

The FTS spectrum of Arp 220 has many features not present in M82, such as strong HCN absorption, P-Cygni profiles of OH$^+$, H$_2$O$^+$ and H$_2$O, and evidence for an AGN \citep{Rangwala:2011}.  CO modeling, similar to the procedure done in this work, also indicates that the high-J lines trace a warmer component than low-J lines ($\sim$ 1350 K).  Mechanical energy likely plays a large role in the heating of this merger galaxy as well.  Though M82 has an outflow, it is not detected in P-Cygni profiles of the aforementioned lines.

The redshift of HLSW-01 \citep{Conley:2011} allows the CO \jseven\ to \jten\ lines to be observed from the ground, as has been done with Z-Spec \citep{Scott:2011}.  Unlike M82 (and others), the known CO SLED from \jone\ and up can be described by a single component at 227 K ($1.2 \times 10^3$ cm$^{-3}$ density).  If the velocity gradient is not constrained to be greater than or equal to that corresponding to virialized motion, the best fit solution becomes 566 K ($0.3 \times 10^3$ cm$^{-3}$ density), closer to our temperature.  We chose to exclude this prior due to uncertainties in the calculation of velocity gradient related to M82's turbulent morphology.  HLSW-01 appears to be unique in that a cold gas component is not required to fit the lower-J lines of the SLED, though two-component models can find a best-fit solution with a cold component.  

In summary, M82 (like Arp 220 and HLSW-01) does not have the high CO excitation dominated by an AGN as seen in Mrk 231.  Therefore, in addition to distinguishing between PDRs and shocks, high-J CO lines may also be used to indicate XDRs.

\section{Conclusions}\label{sec:concl}

We have presented a multitude of molecular and atomic lines from M82 in the wavelength range (194-671 $\mu$m) accessible by the {\it Herschel}-SPIRE FTS (Table \ref{table:fit}).  After modeling $^{12}$CO, $^{13}$CO and [C {\sc{I}}], we find support for the high-temperature molecular gas component presented in the results of \citet{Panuzzo:2010}.  The temperature traced by the warm component of $^{12}$CO is quite high (335-518 K), and the addition of $^{13}$CO slightly expands the likelihood ranges.  The addition of [C {\sc{I}}] produced results that indicate that these atom is not entirely tracing the same region as $^{12}$CO.  Some of the emission from these molecules (especially [C {\sc{I}}]) are likely tracing more diffuse gas less shielded from UV radiation.

The mapping observations did not resolve any significant gradients in physical parameters (except evidence for a slight drop-off in beam-averaged column density, consistent with the beam profile) indicating that the single ``deep" spectrum is an adequate representation of the galaxy when limited by our beam size.  However, the mapping observations were important in confirming the source-beam coupling factor utilized in \citet{Panuzzo:2010} and here, because through convolution of the maps we were able to confirm the central pixel's results matched with the deep spectrum.

Molecular absorption traces lower column regions of the disk than those traced by CO emission, but contribute to the interpretation of the molecular gas of M82 being excited by a combination of sources.  Despite the enhanced cosmic ray density in M82, we do not find evidence that cosmic rays alone are sufficient to heat the gas enough to match the modeled hydrogen cooling rate.  PDR models can only replicate the high-J CO line emission at high densities incompatible with those indicated by the likelihood analysis, though cosmic-ray enhanced PDRs may be a closer match at lower densities.   Turbulent heating from stellar winds and supernovae likely play a large role in the heating.  More specifically, shocks are required to explain bright high-J line emission \citep{Pon:2012}.

Like other submillimeter bright galaxies, {\it Herschel} has opened up new opportunities and questions about molecular and atomic lines that have never been observed before.  Because of this, the diagnostic power of high-J CO lines is still in development, and newer models currently being developed may be able to explain the emission seen from M82 and other extreme environments.

\paragraph{Acknowledgments}
SPIRE has been developed by a consortium of institutes led
by Cardiff Univ. (UK) and including: Univ. Lethbridge (Canada);
NAOC (China); CEA, LAM (France); IFSI, Univ. Padua (Italy);
IAC (Spain); Stockholm Observatory (Sweden); Imperial College
London, RAL, UCL-MSSL, UKATC, Univ. Sussex (UK); and Caltech,
JPL, NHSC, Univ. Colorado (USA). This development has been
supported by national funding agencies: CSA (Canada); NAOC
(China); CEA, CNES, CNRS (France); ASI (Italy); MCINN (Spain);
SNSB (Sweden); STFC, UKSA (UK); and NASA (USA).  J.K. also acknowledges the funding sources from the NSF GRFP.  The research of C.D.W. is supported by grants from the Natural Sciences and Engineering Research Council of Canada.  Thank you to the anonymous referee for comments which significantly improved this work.


\bibliography{M82}

\include{M82_appendix}

\end{document}

%% file: M82_appendix.tex
\appendix
\section{Integrated Line Flux Maps}\label{sec:appendix}




\setcounter{figure}{0}
\renewcommand{\thefigure}{2.\arabic{figure}}

\begin{figure*}[th]
\centering
\subfigure{\includegraphics[width=0.8\textwidth]{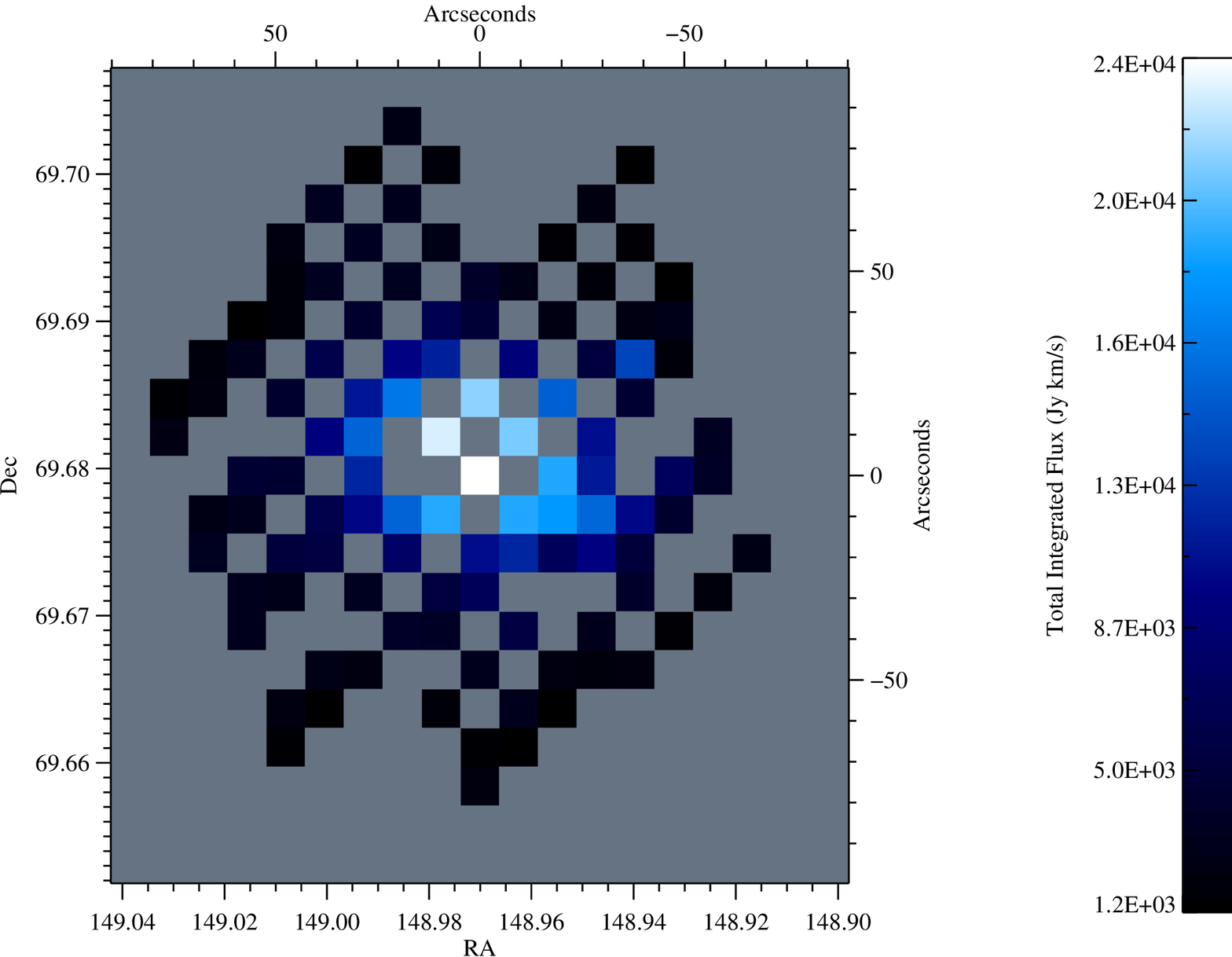}}
\subfigure{\includegraphics[width=0.8\textwidth]{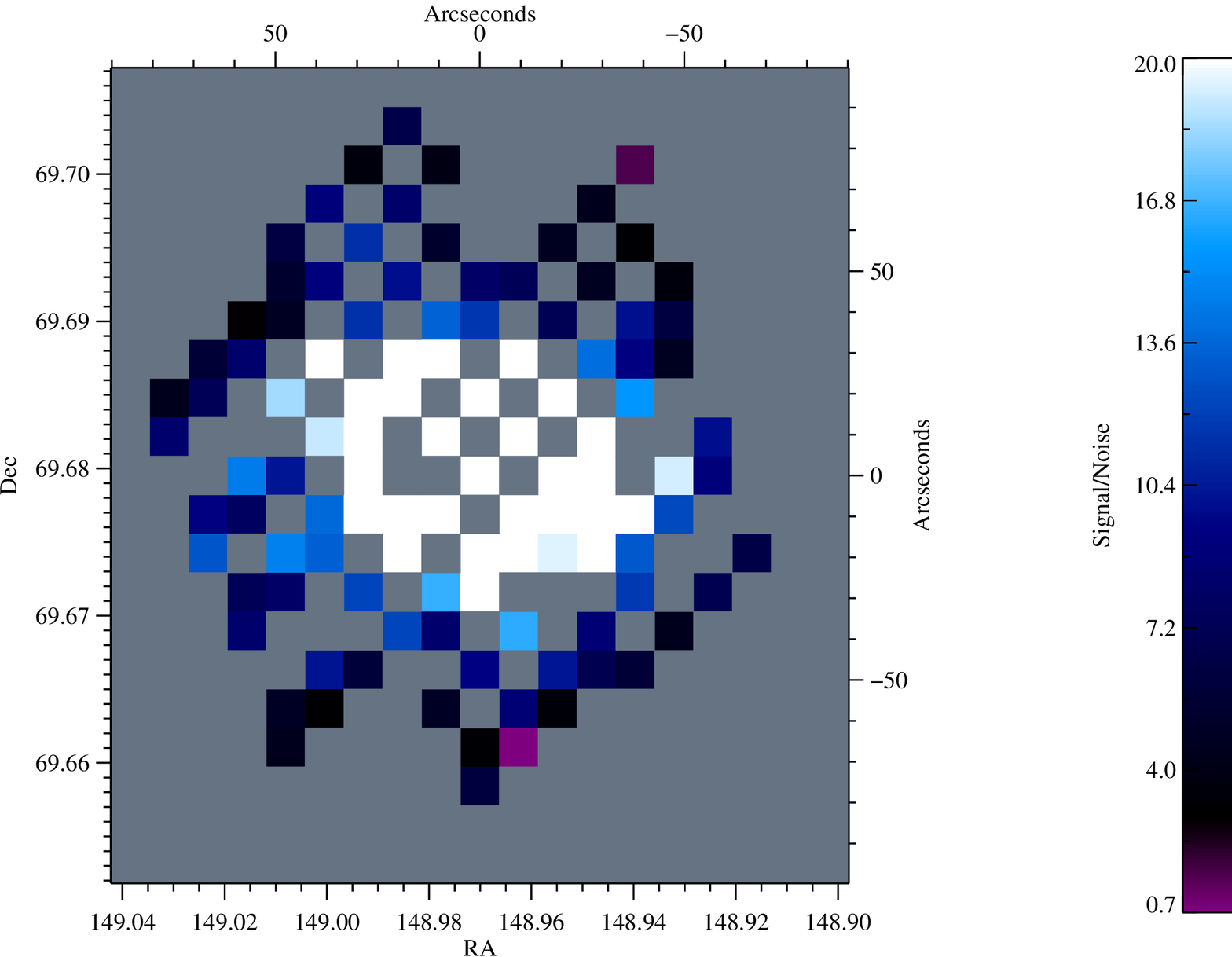}}
\caption{Integrated Flux (top) and Signal/Noise (bottom) maps for CI \jone.}\label{fig:intflux2}
\end{figure*}

\clearpage

\begin{figure*}[th]
\centering
\subfigure{\includegraphics[width=0.8\textwidth]{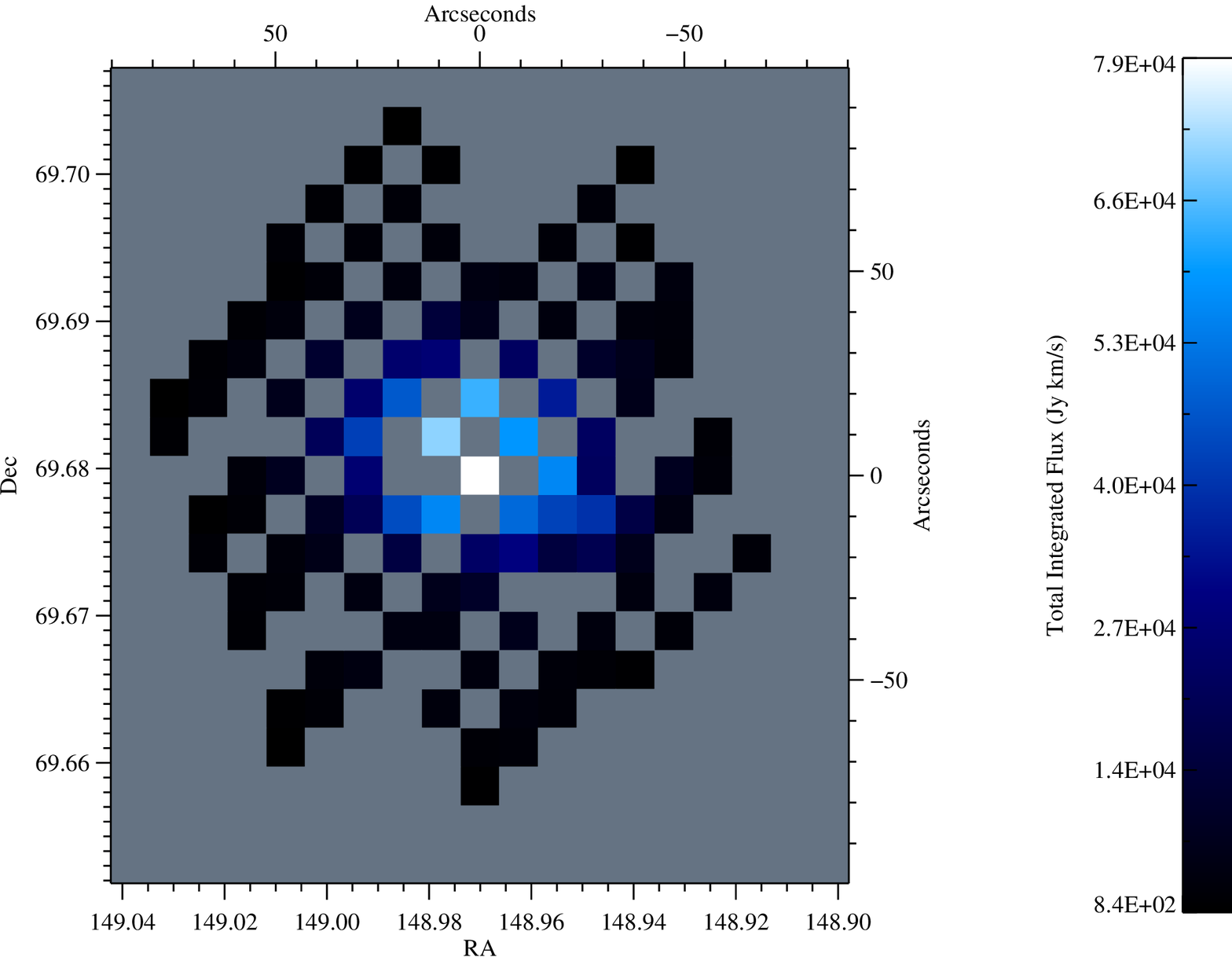}}
\subfigure{\includegraphics[width=0.8\textwidth]{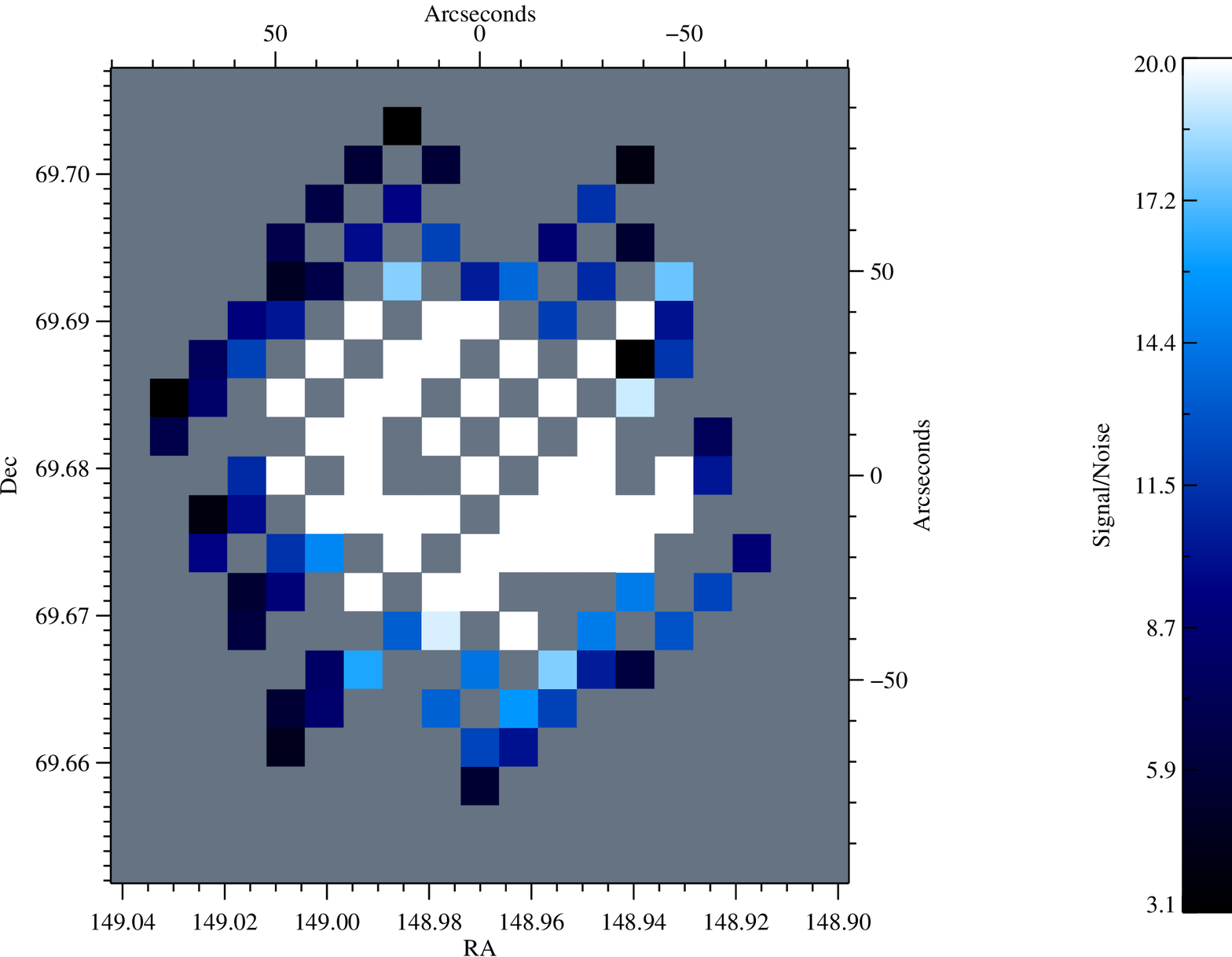}}
\caption{Integrated Flux (top) and Signal/Noise (bottom) maps for CO \jfive.}\label{fig:intflux3}
\end{figure*}

\clearpage

\begin{figure*}[th]
\centering
\subfigure{\includegraphics[width=0.8\textwidth]{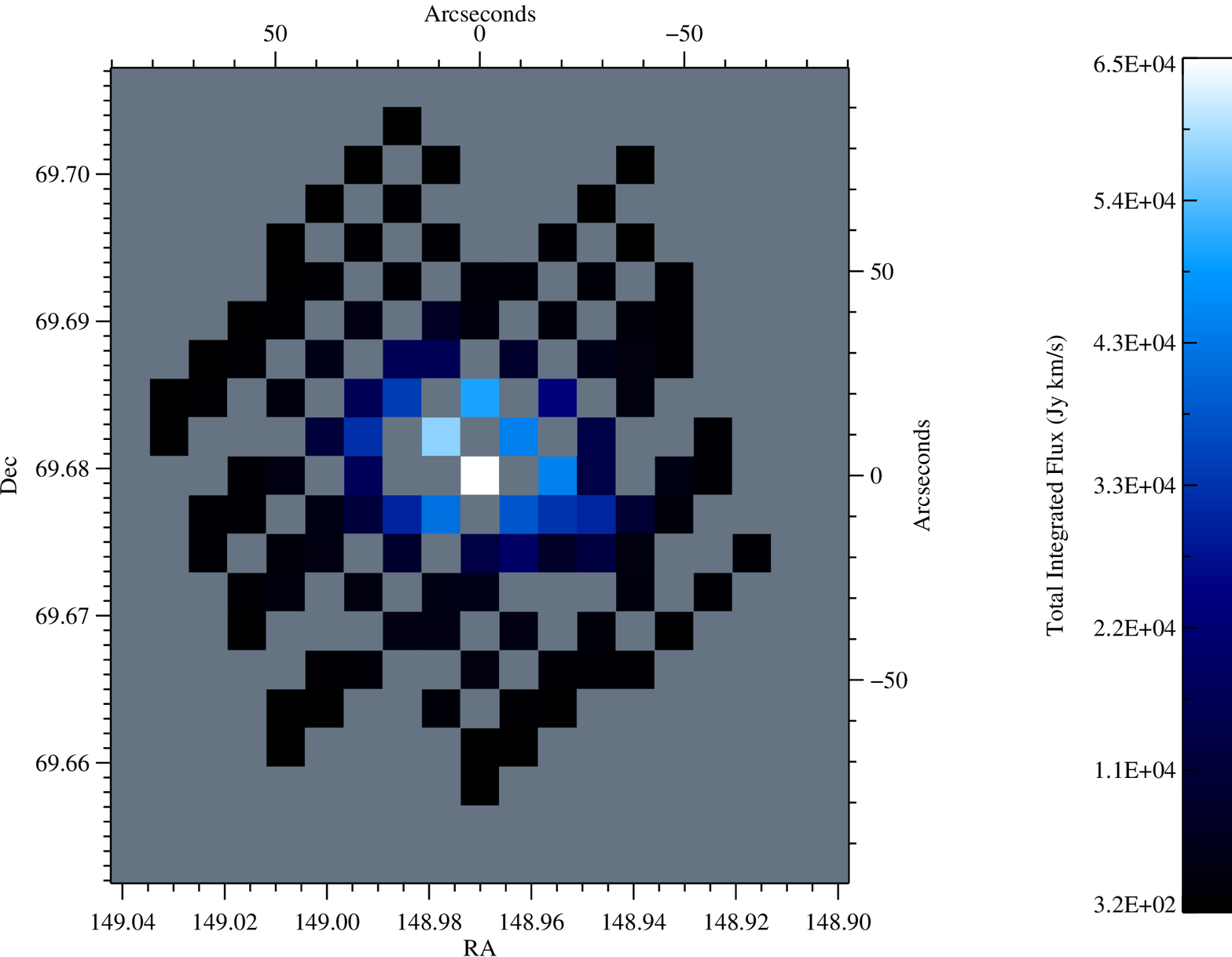}}
\subfigure{\includegraphics[width=0.8\textwidth]{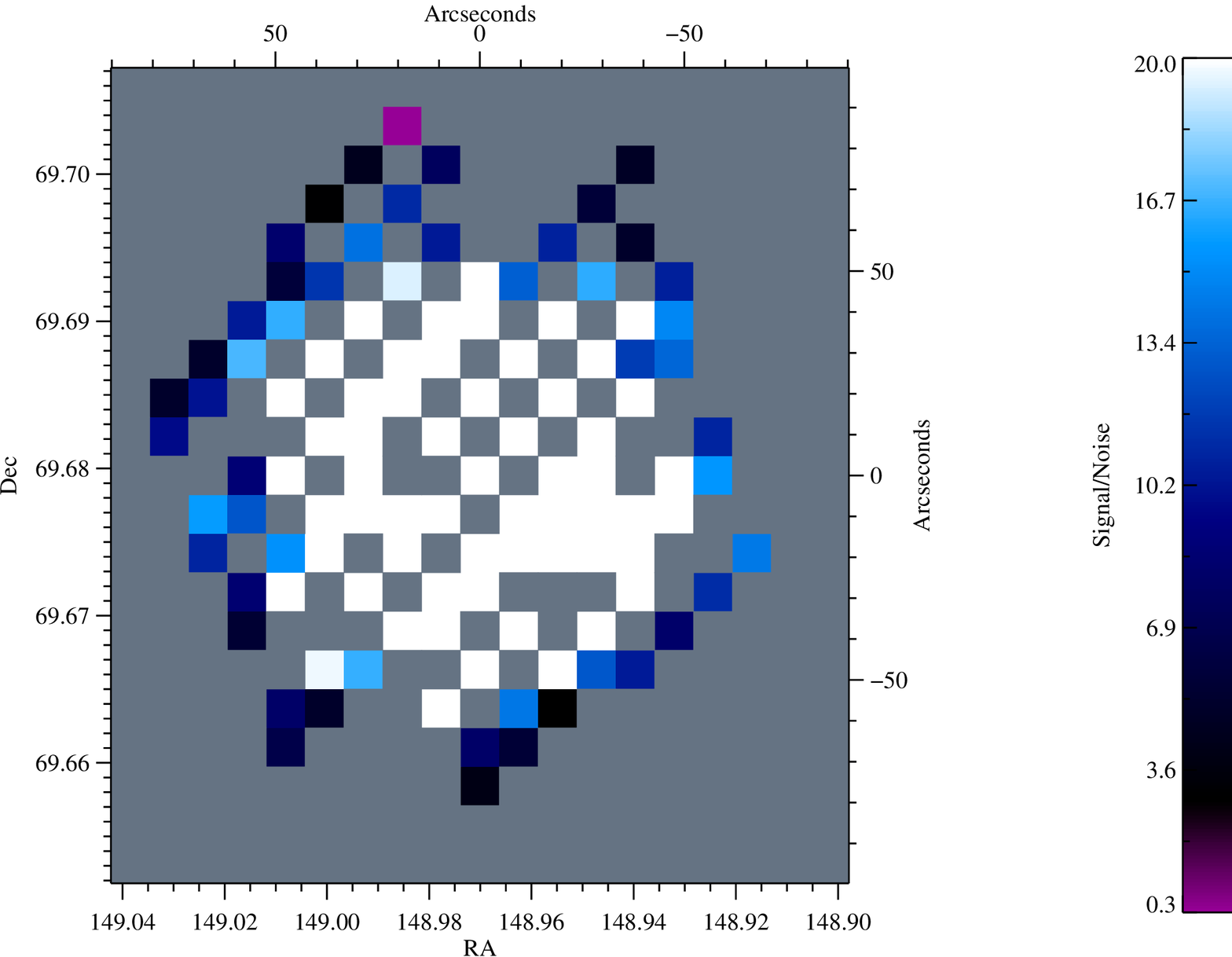}}
\caption{Integrated Flux (top) and Signal/Noise (bottom) maps for CO \jsix.}\label{fig:intflux4}
\end{figure*}

\clearpage

\begin{figure*}[th]
\centering
\subfigure{\includegraphics[width=0.8\textwidth]{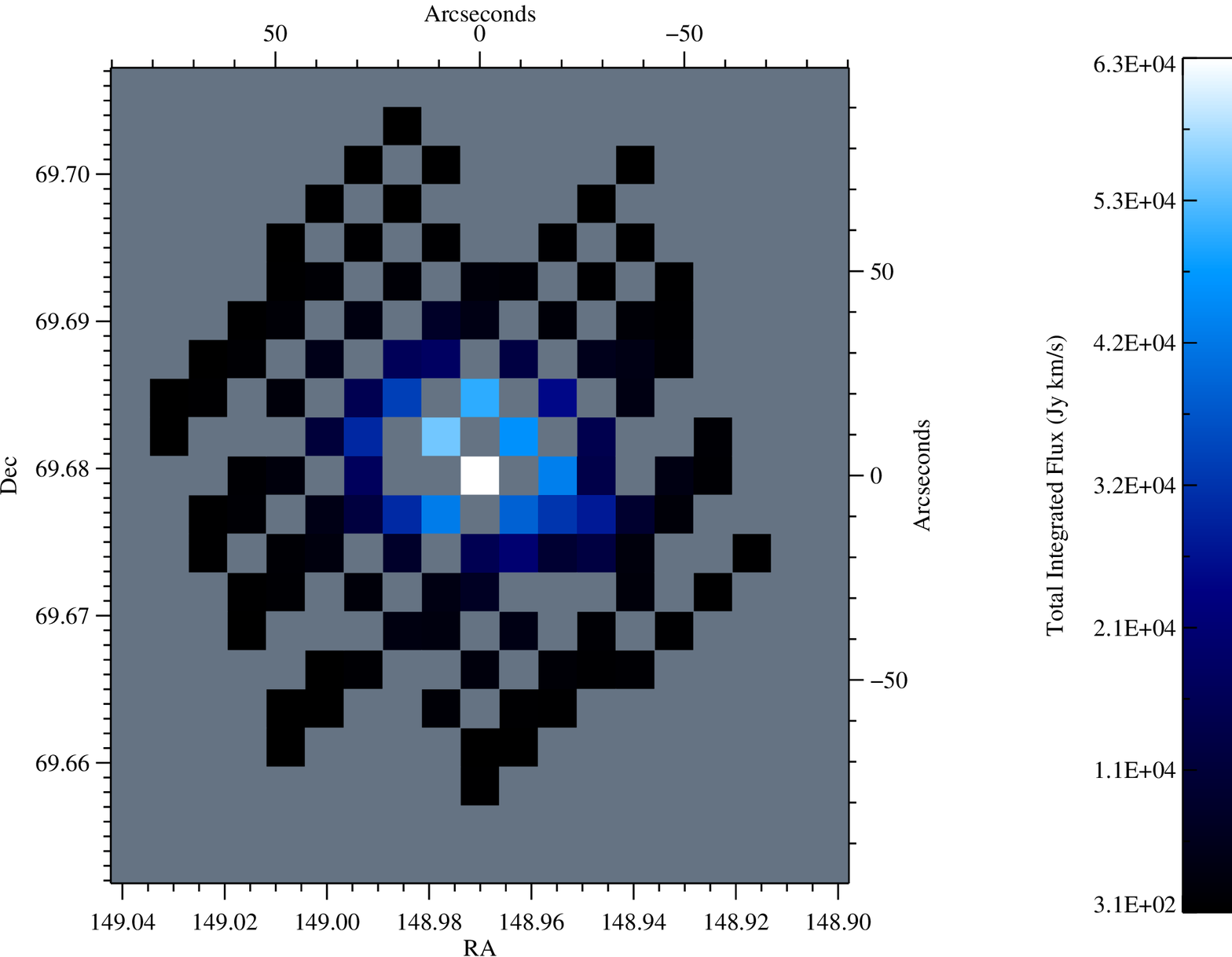}}
\subfigure{\includegraphics[width=0.8\textwidth]{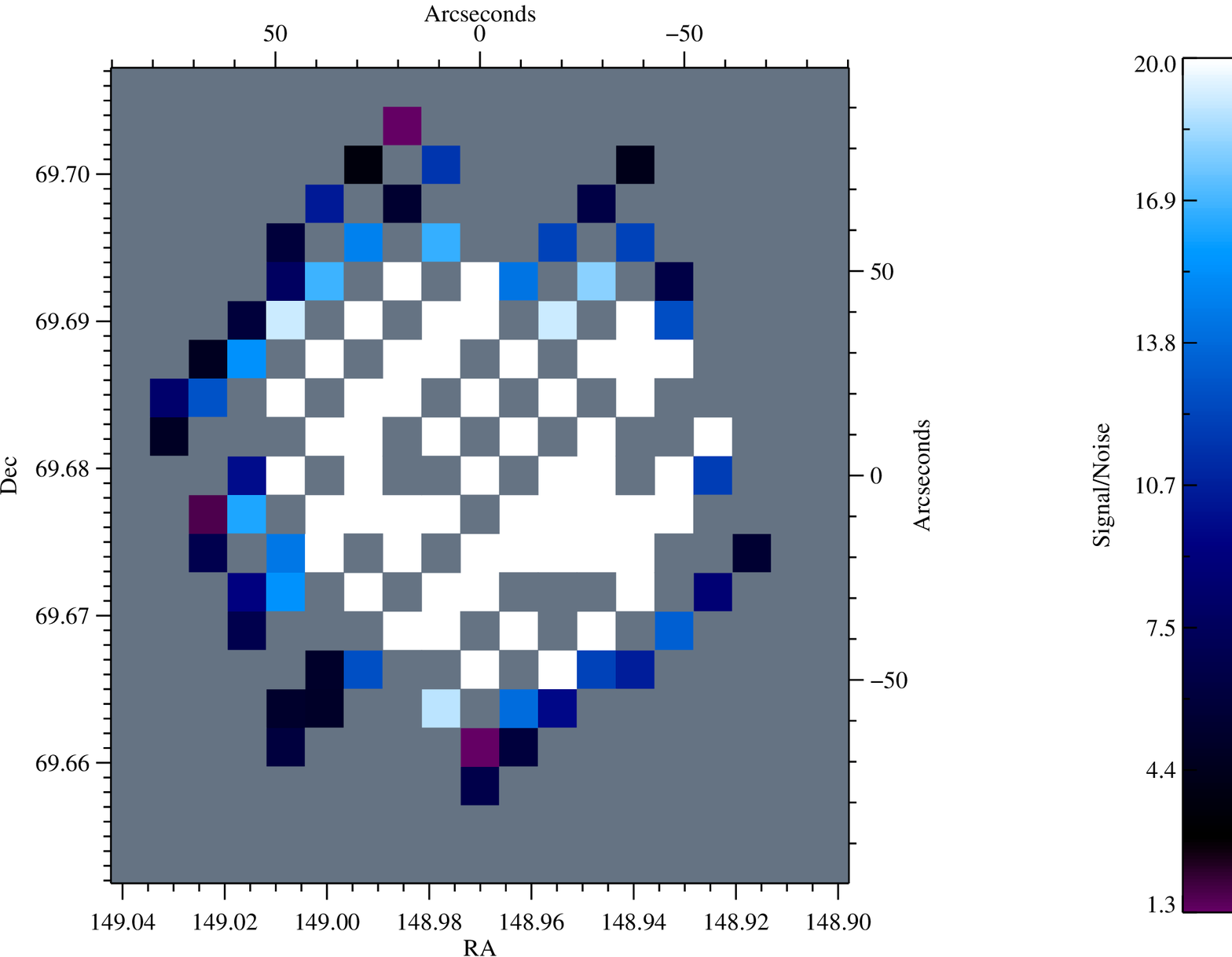}}
\caption{Integrated Flux (top) and Signal/Noise (bottom) maps for CO \jseven.}\label{fig:intflux5}
\end{figure*}

\clearpage

\begin{figure*}[th]
\centering
\subfigure{\includegraphics[width=0.8\textwidth]{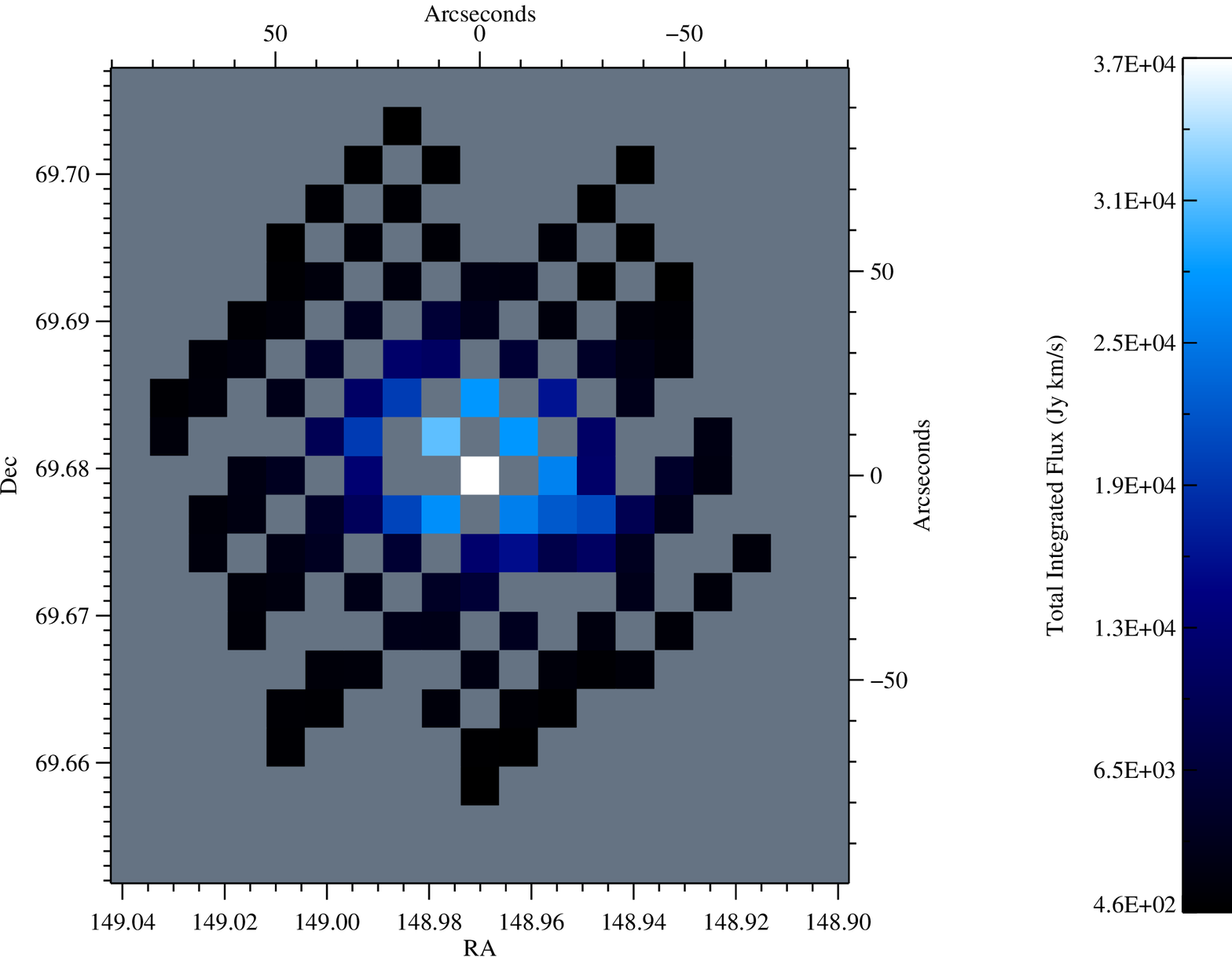}}
\subfigure{\includegraphics[width=0.8\textwidth]{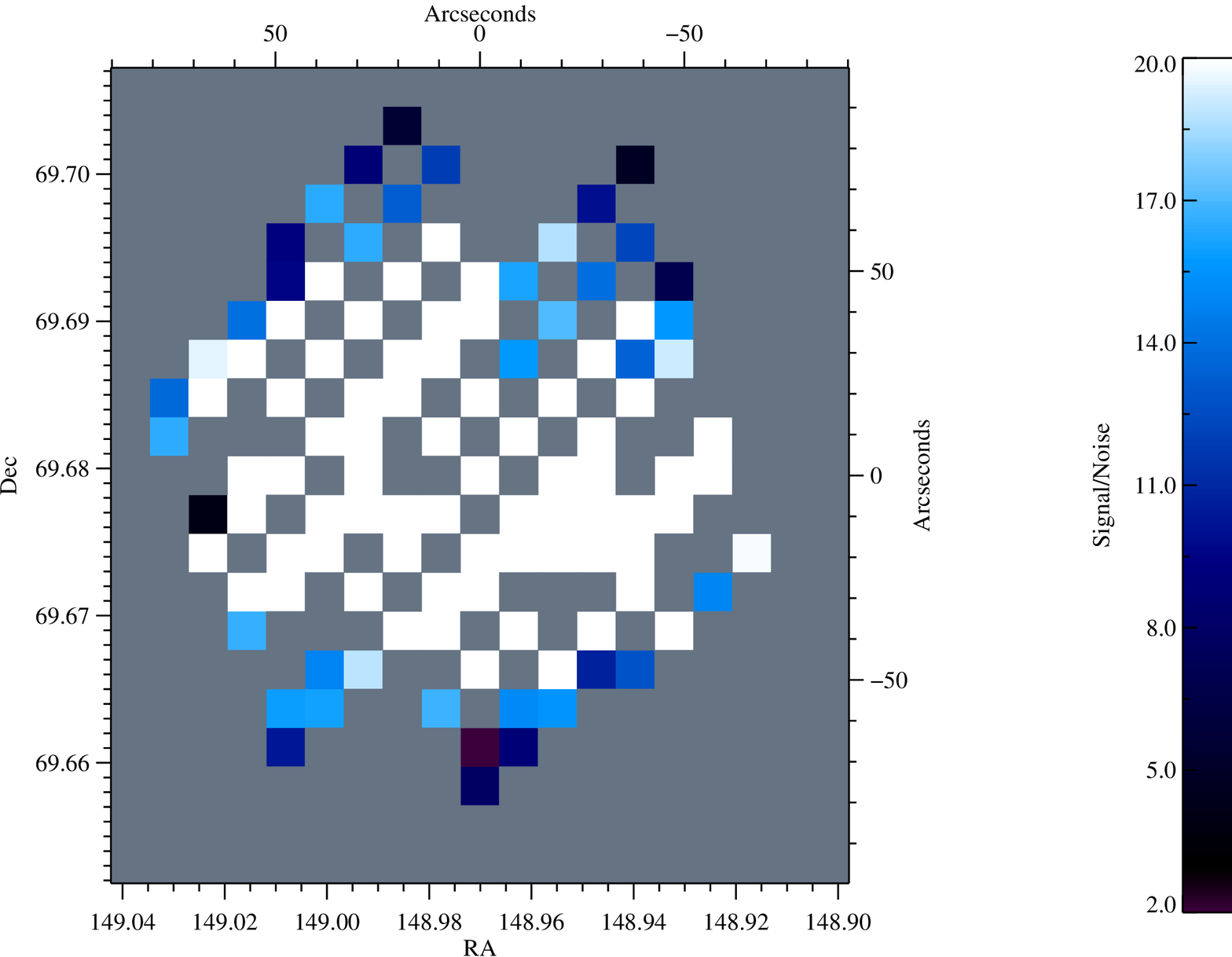}}
\caption{Integrated Flux (top) and Signal/Noise (bottom) maps for CI \jtwo.}\label{fig:intflux6}
\end{figure*}

\clearpage

\begin{figure*}[th]
\centering
\subfigure{\includegraphics[width=0.8\textwidth]{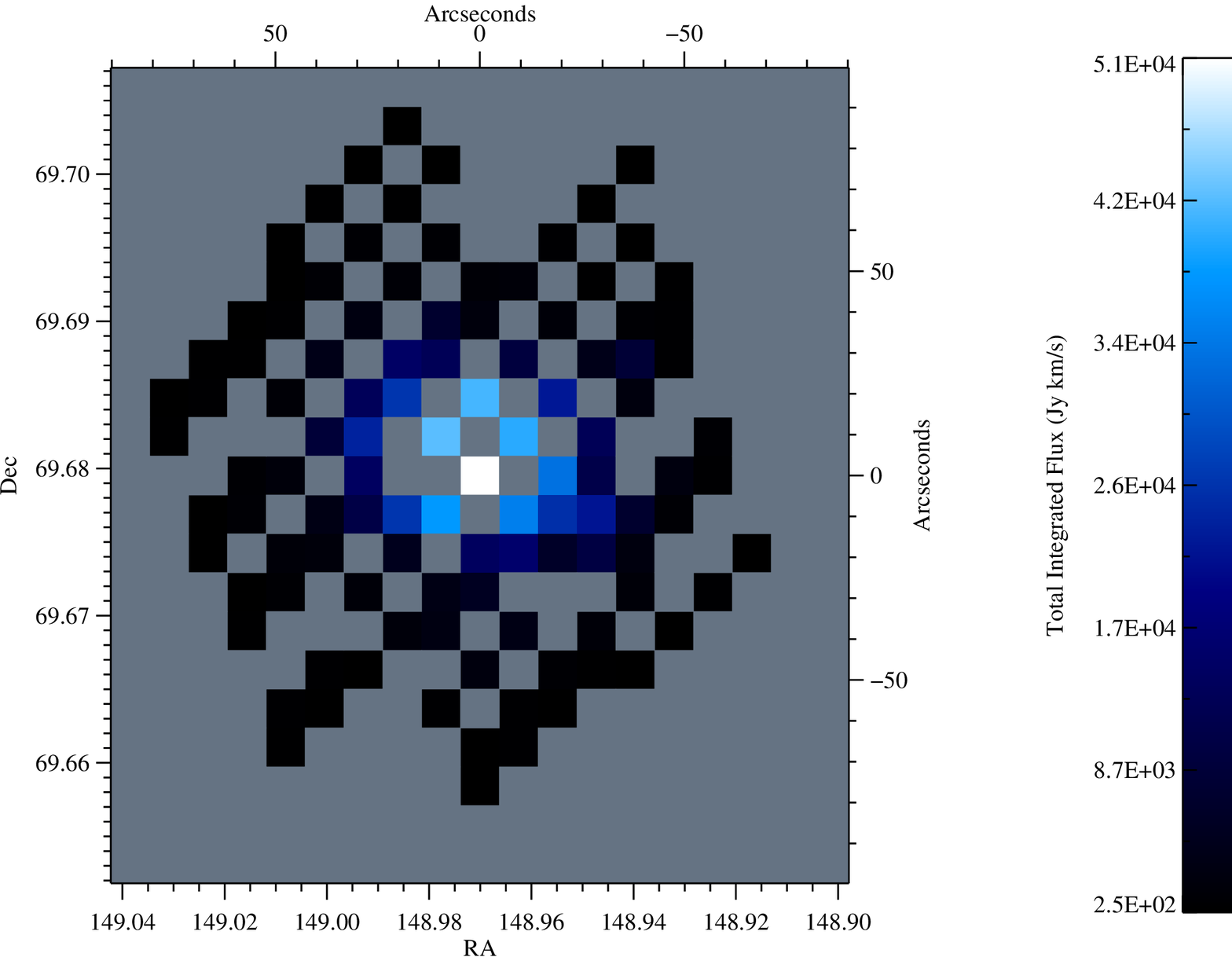}}
\subfigure{\includegraphics[width=0.8\textwidth]{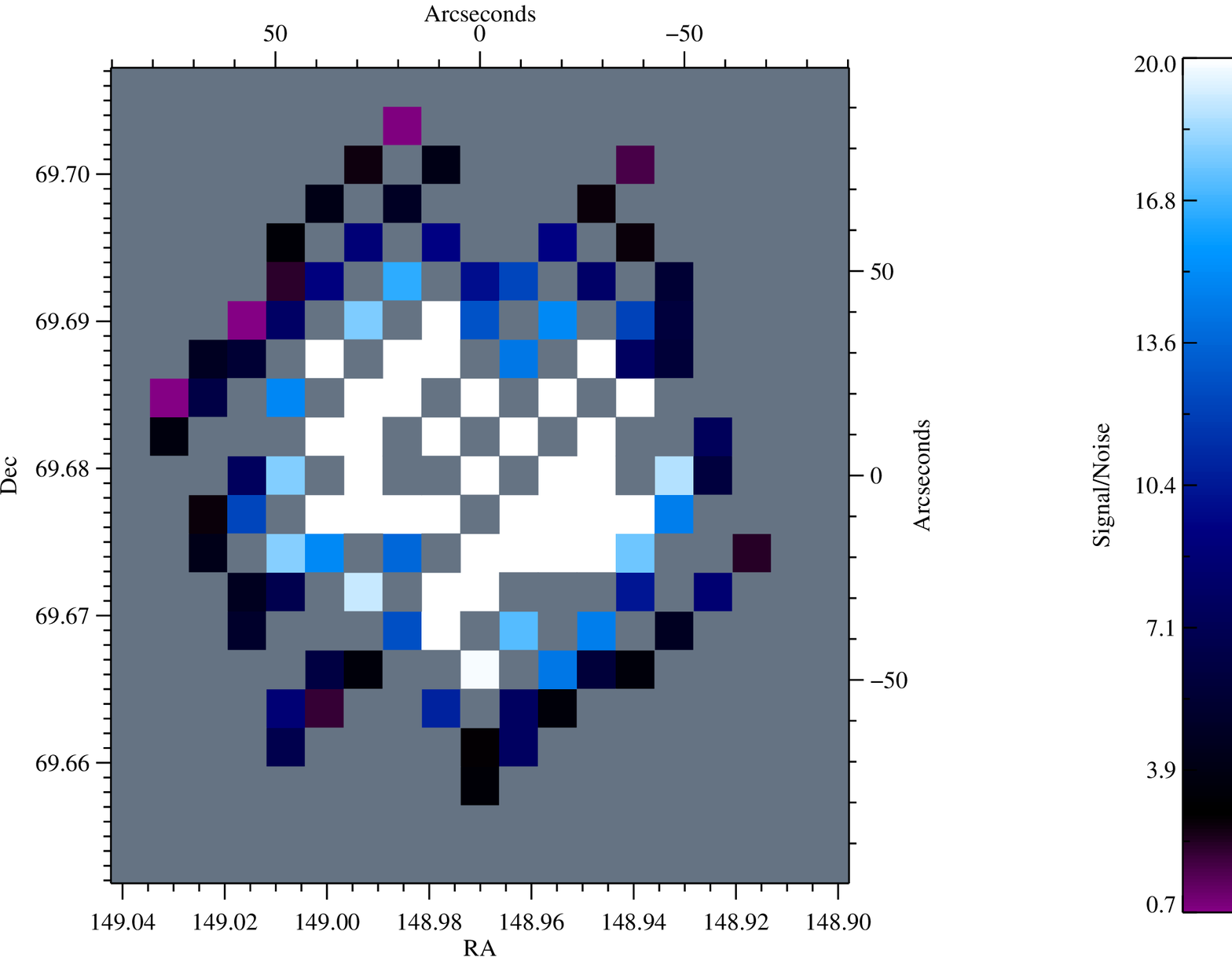}}
\caption{Integrated Flux (top) and Signal/Noise (bottom) maps for CO \jeight.}\label{fig:intflux7}
\end{figure*}

\clearpage

\begin{figure*}[th]
\centering
\subfigure{\includegraphics[width=0.8\textwidth]{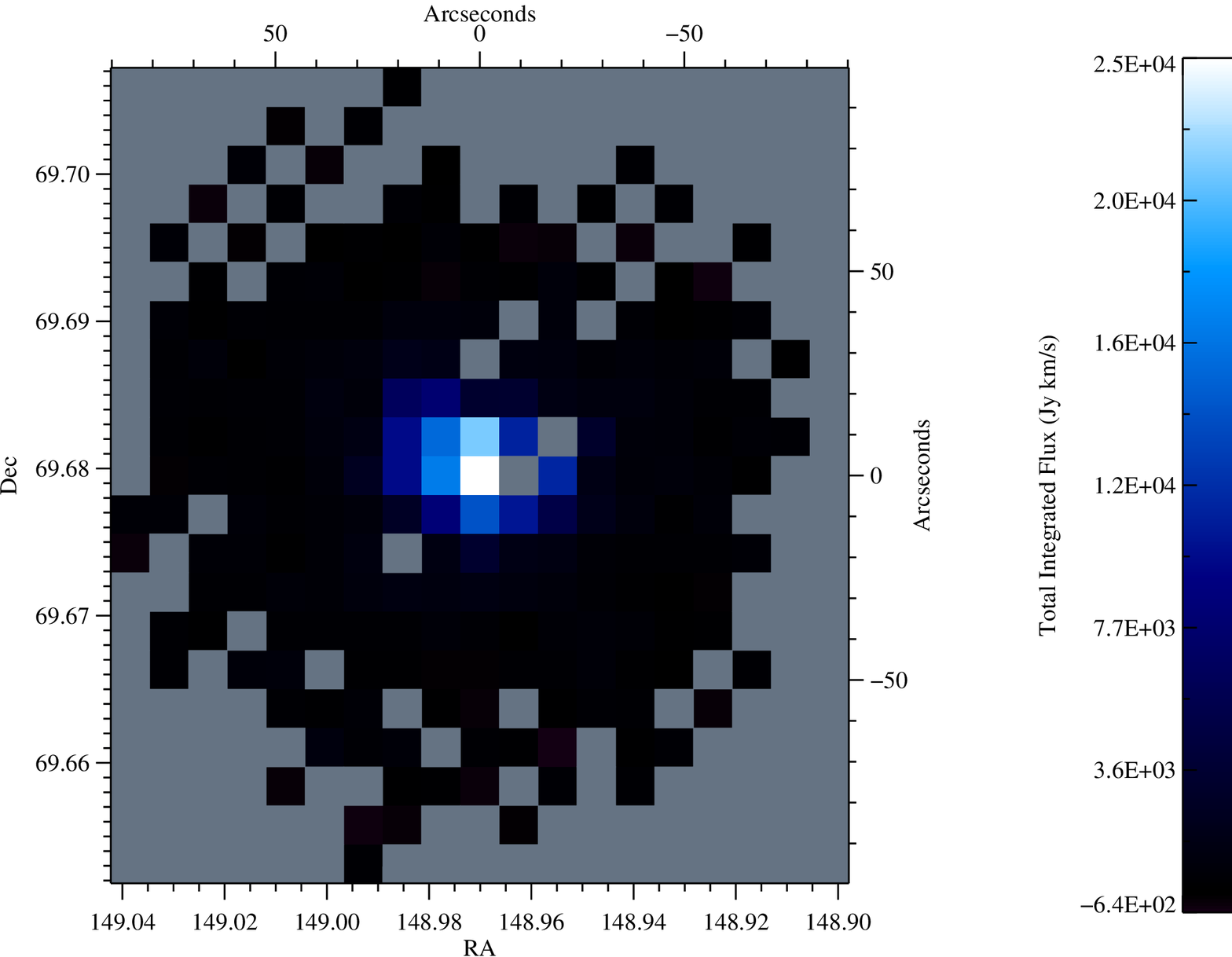}}
\subfigure{\includegraphics[width=0.8\textwidth]{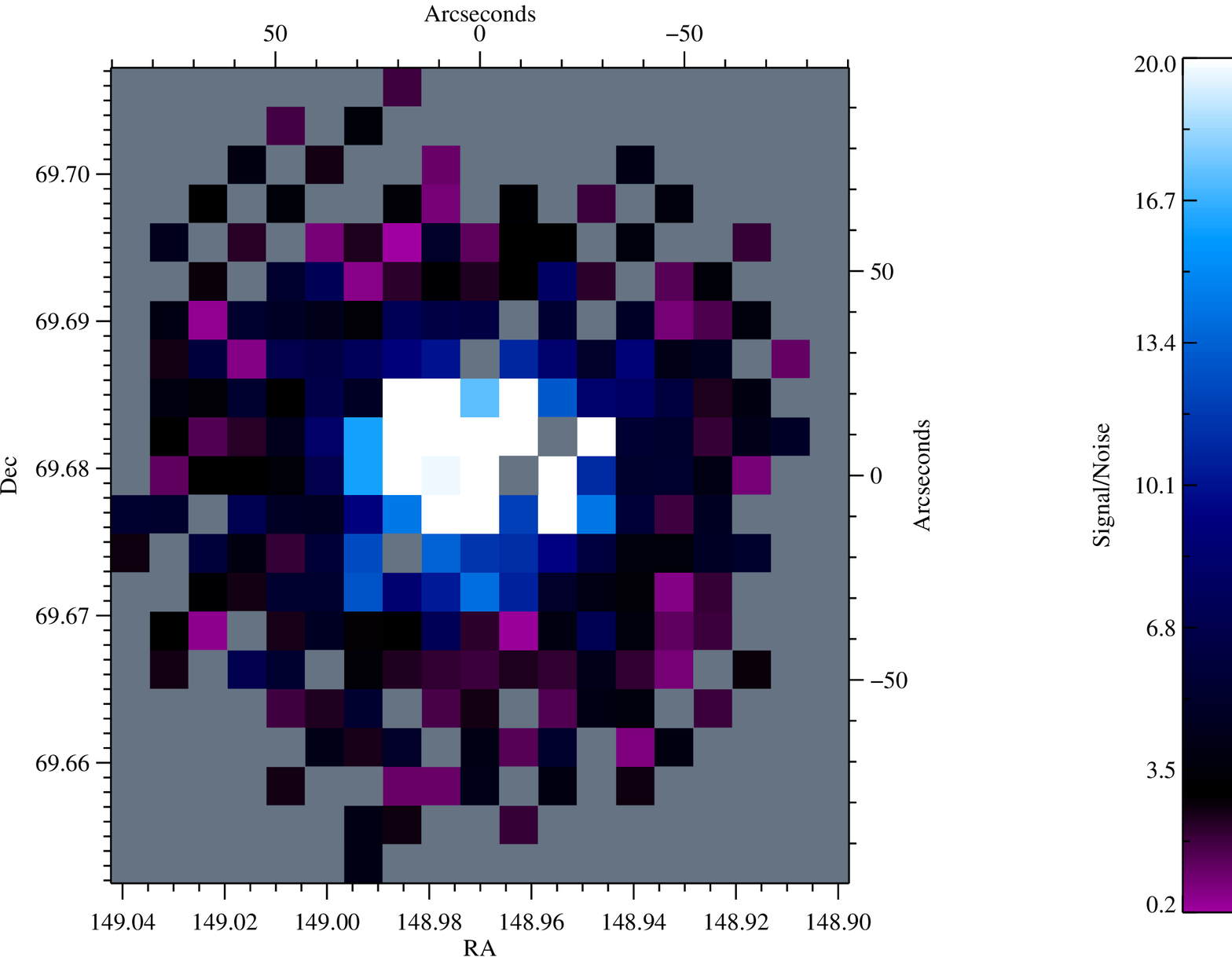}}
\caption{Integrated Flux (top) and Signal/Noise (bottom) maps for CO \jnine.}\label{fig:intflux8}
\end{figure*}

\clearpage

\begin{figure*}[th]
\centering
\subfigure{\includegraphics[width=0.8\textwidth]{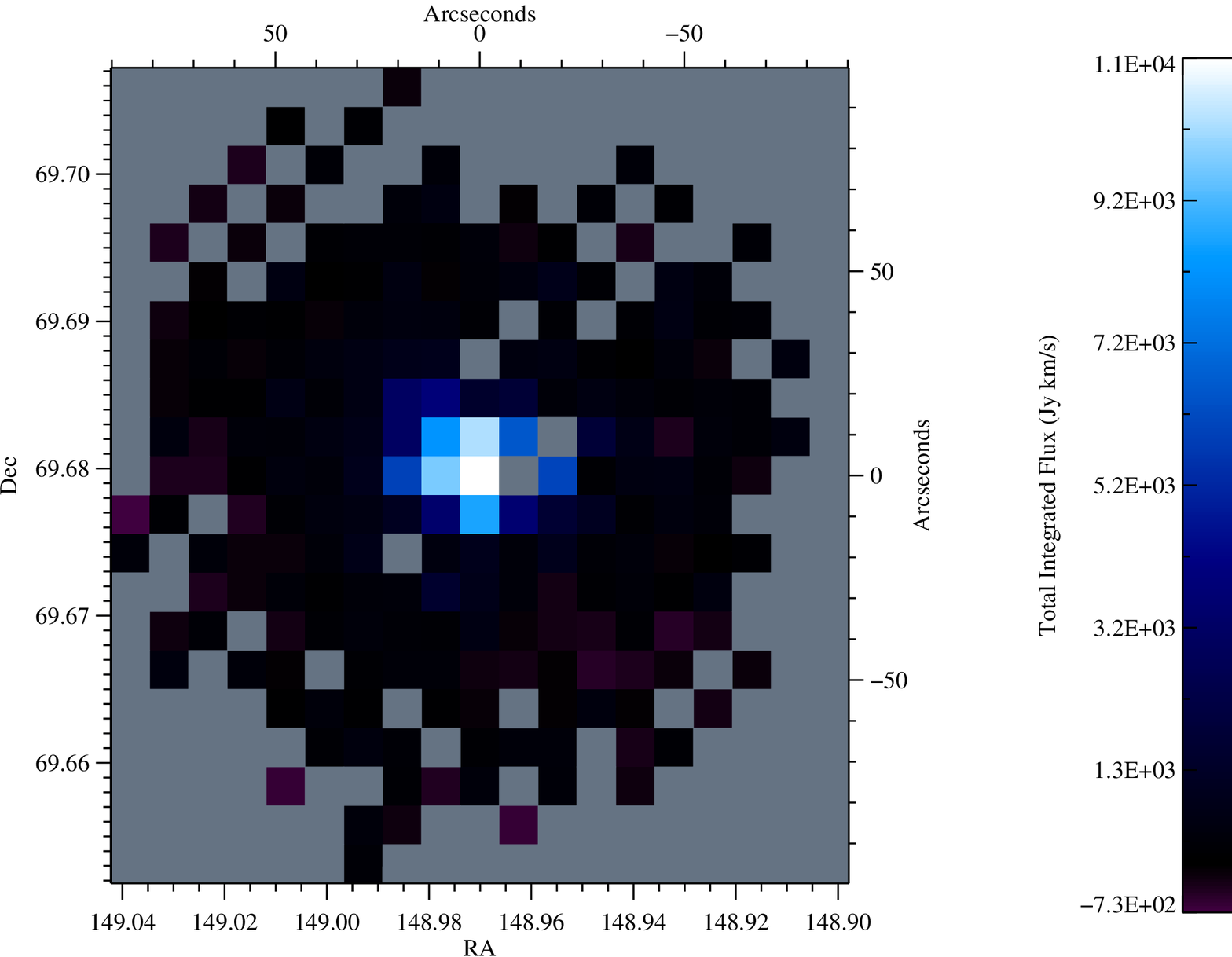}}
\subfigure{\includegraphics[width=0.8\textwidth]{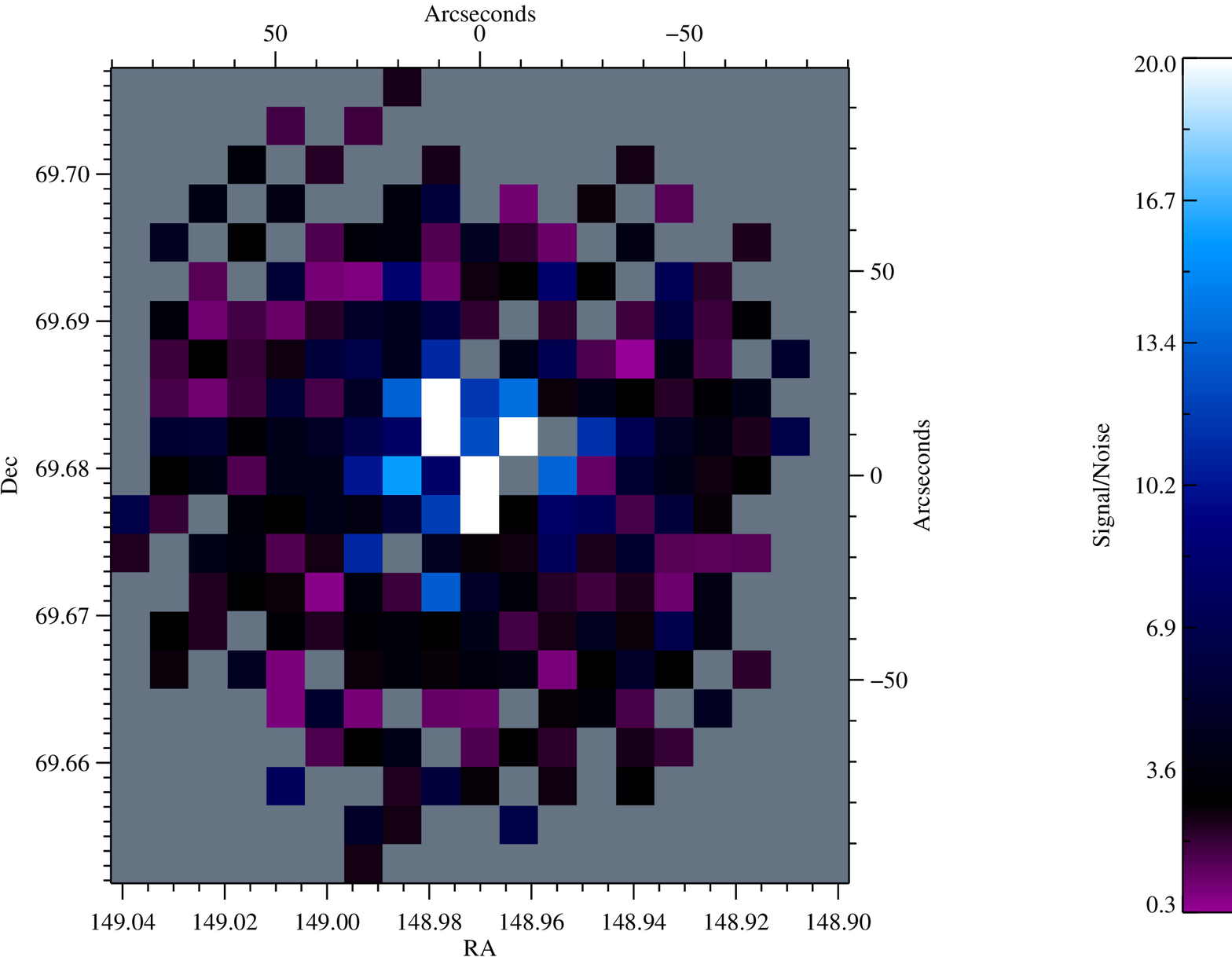}}
\caption{Integrated Flux (top) and Signal/Noise (bottom) maps for CO \jten.}\label{fig:intflux9}
\end{figure*}

\clearpage

\begin{figure*}[th]
\centering
\subfigure{\includegraphics[width=0.8\textwidth]{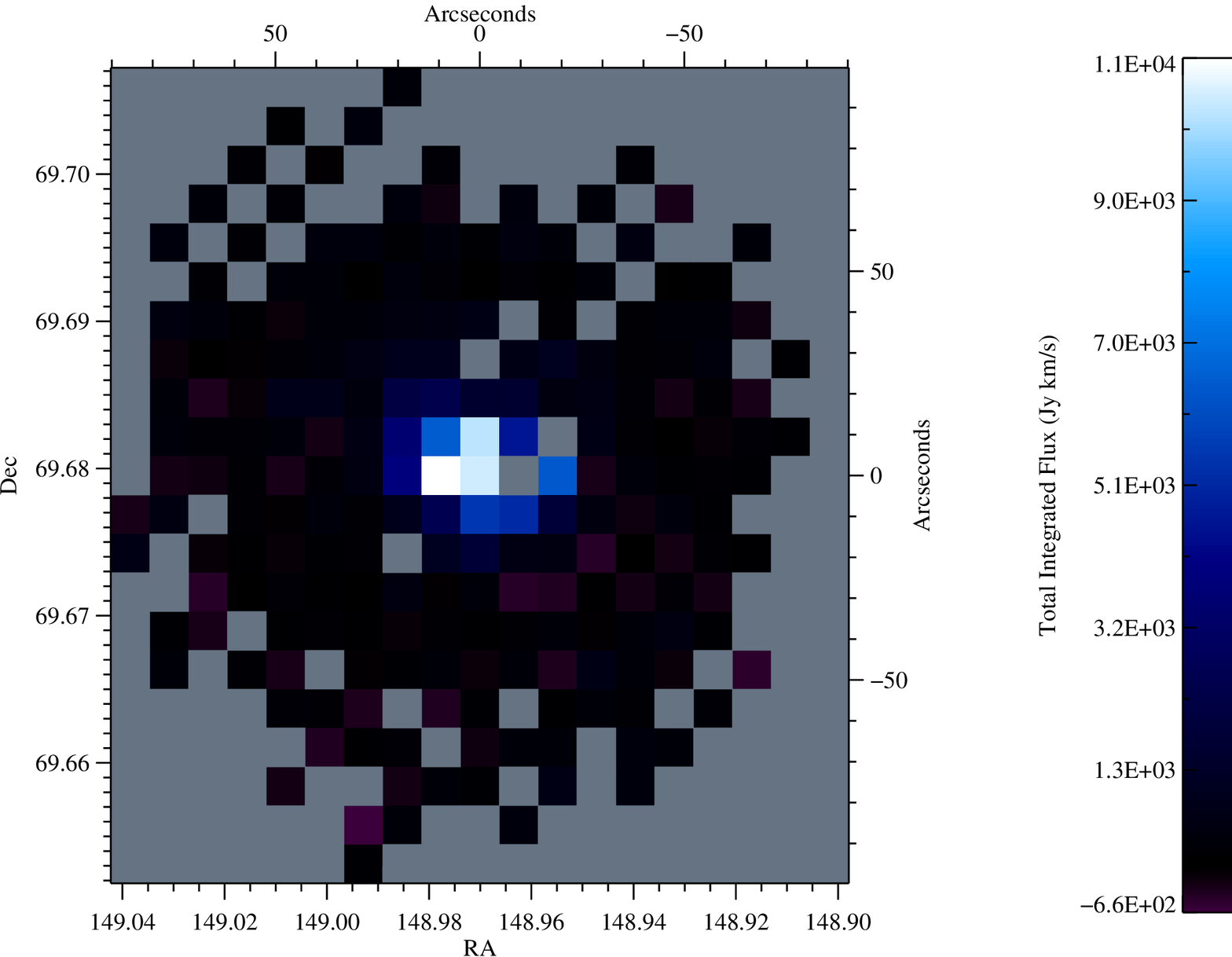}}
\subfigure{\includegraphics[width=0.8\textwidth]{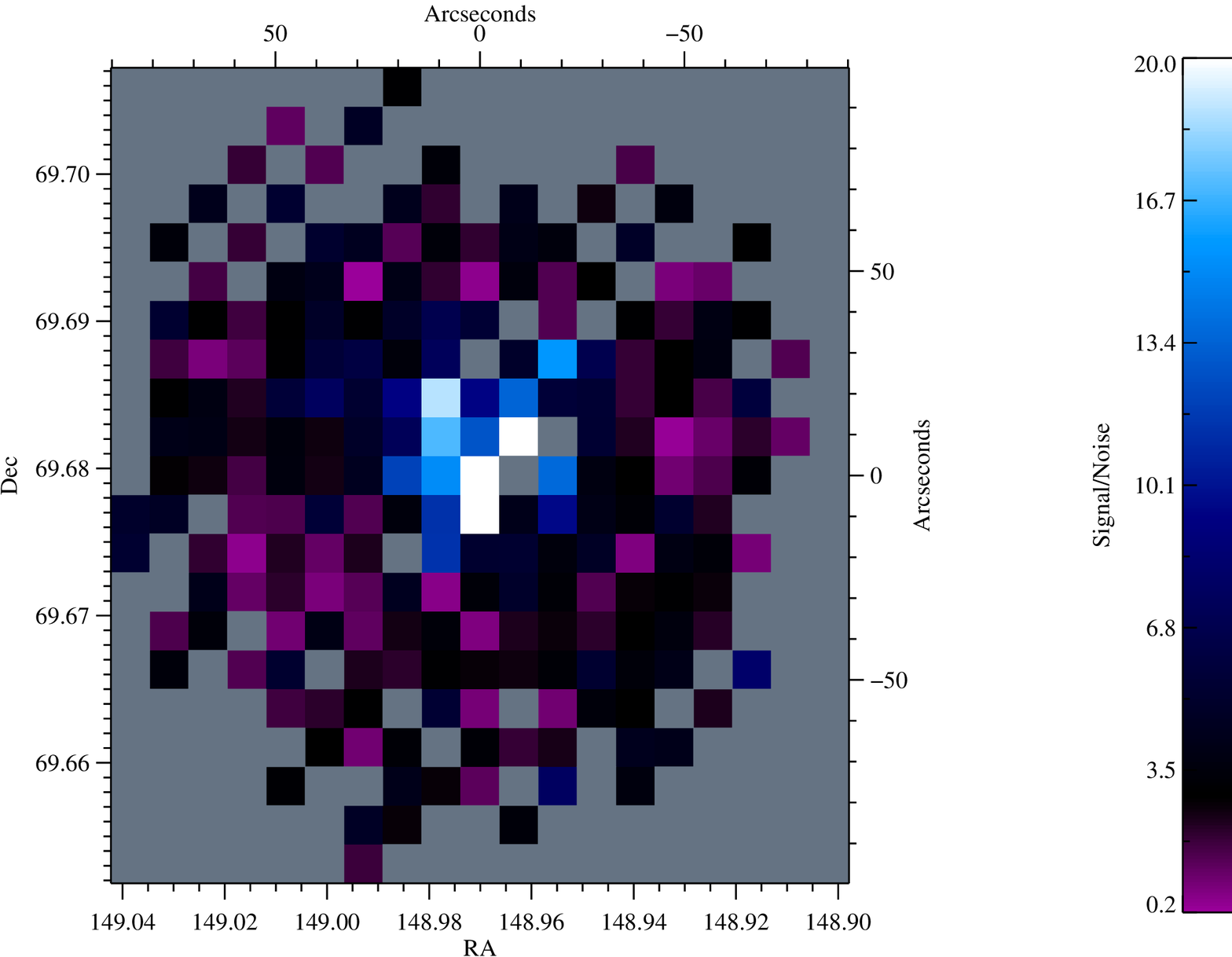}}
\caption{Integrated Flux (top) and Signal/Noise (bottom) maps for CO \jeleven.}\label{fig:intflux10}
\end{figure*}

\clearpage

\begin{figure*}[th]
\centering
\subfigure{\includegraphics[width=0.8\textwidth]{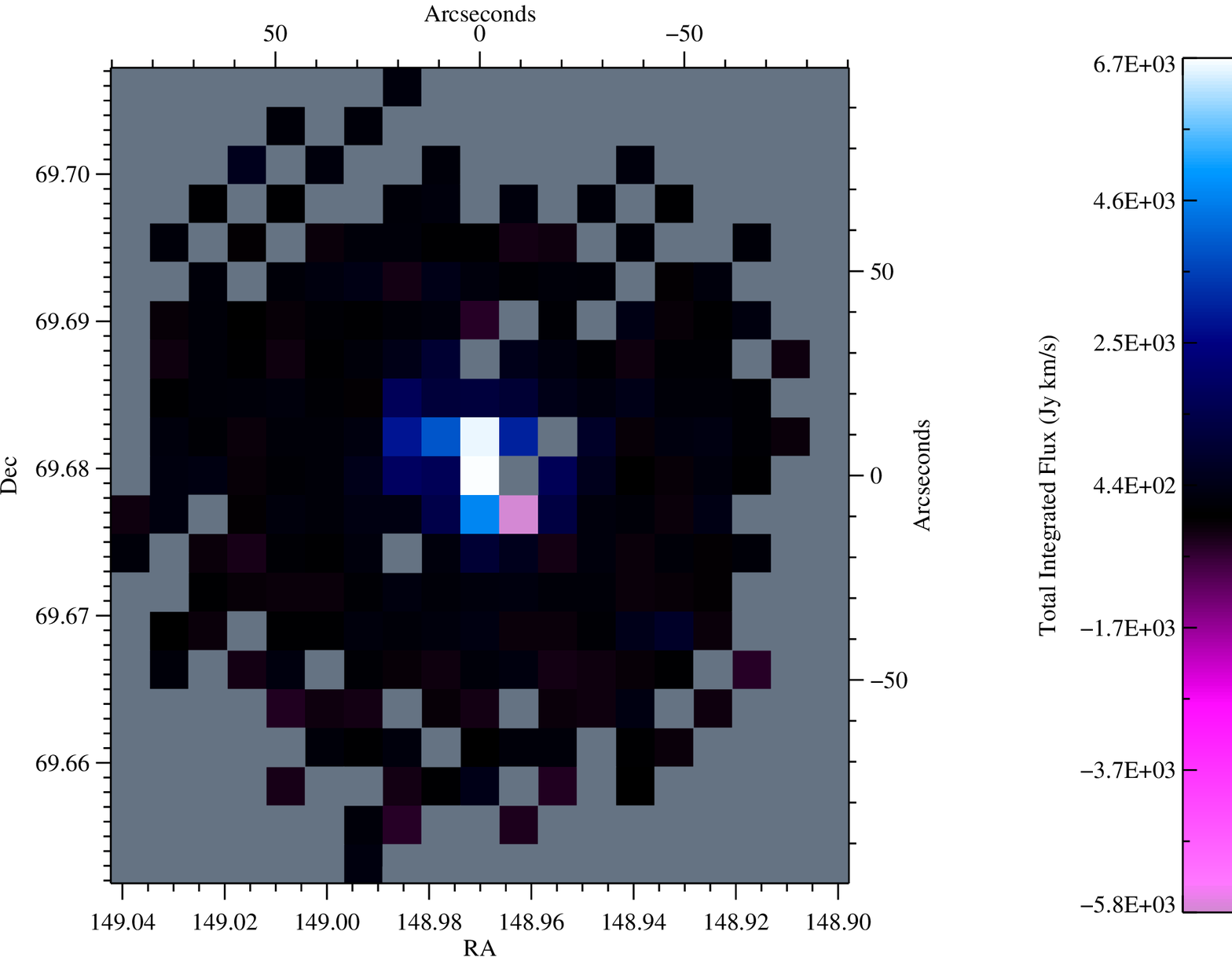}}
\subfigure{\includegraphics[width=0.8\textwidth]{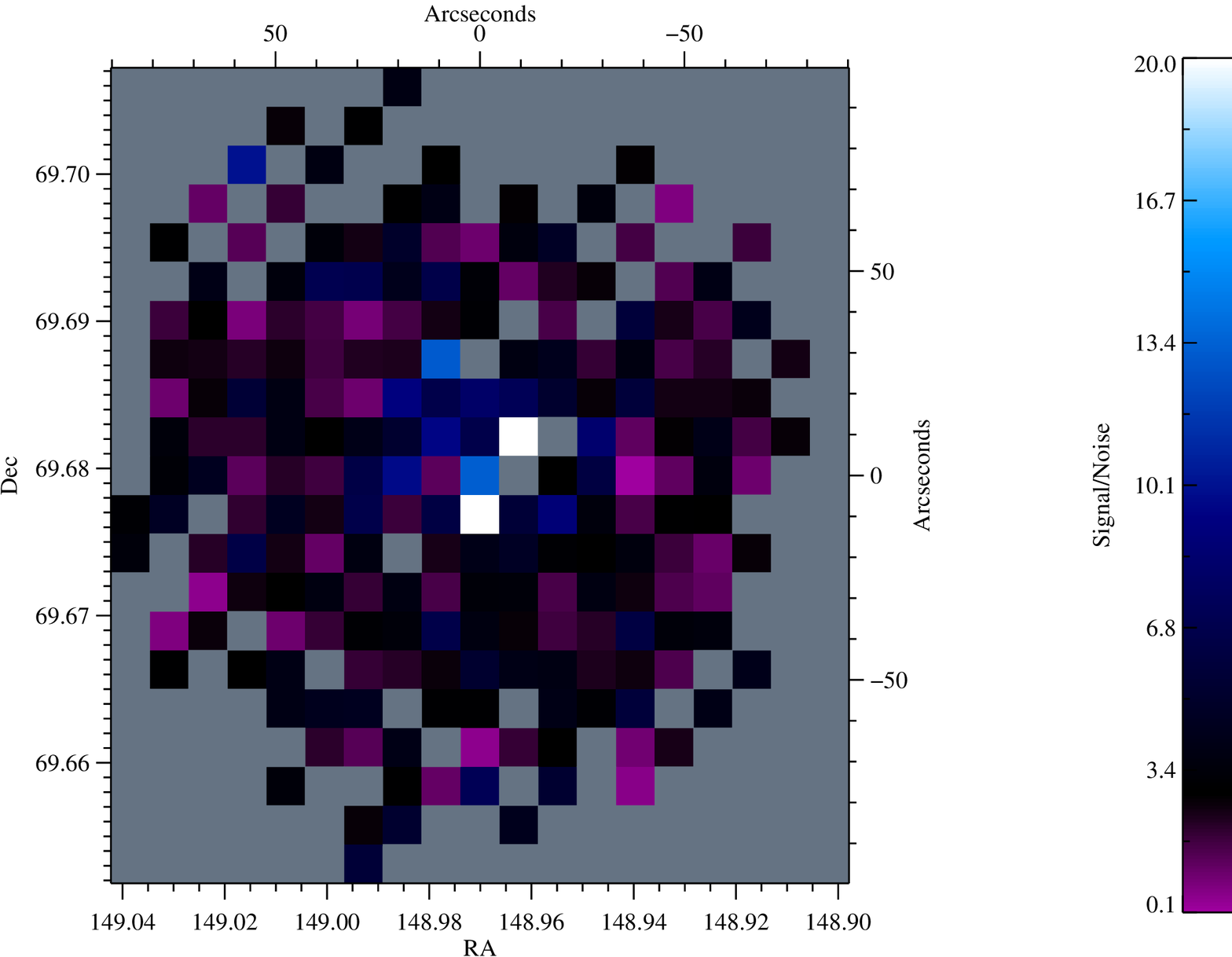}}
\caption{Integrated Flux (top) and Signal/Noise (bottom) maps for CO \jtwelve.}\label{fig:intflux11}
\end{figure*}

\clearpage

\begin{figure*}[th]
\centering
\subfigure{\includegraphics[width=0.8\textwidth]{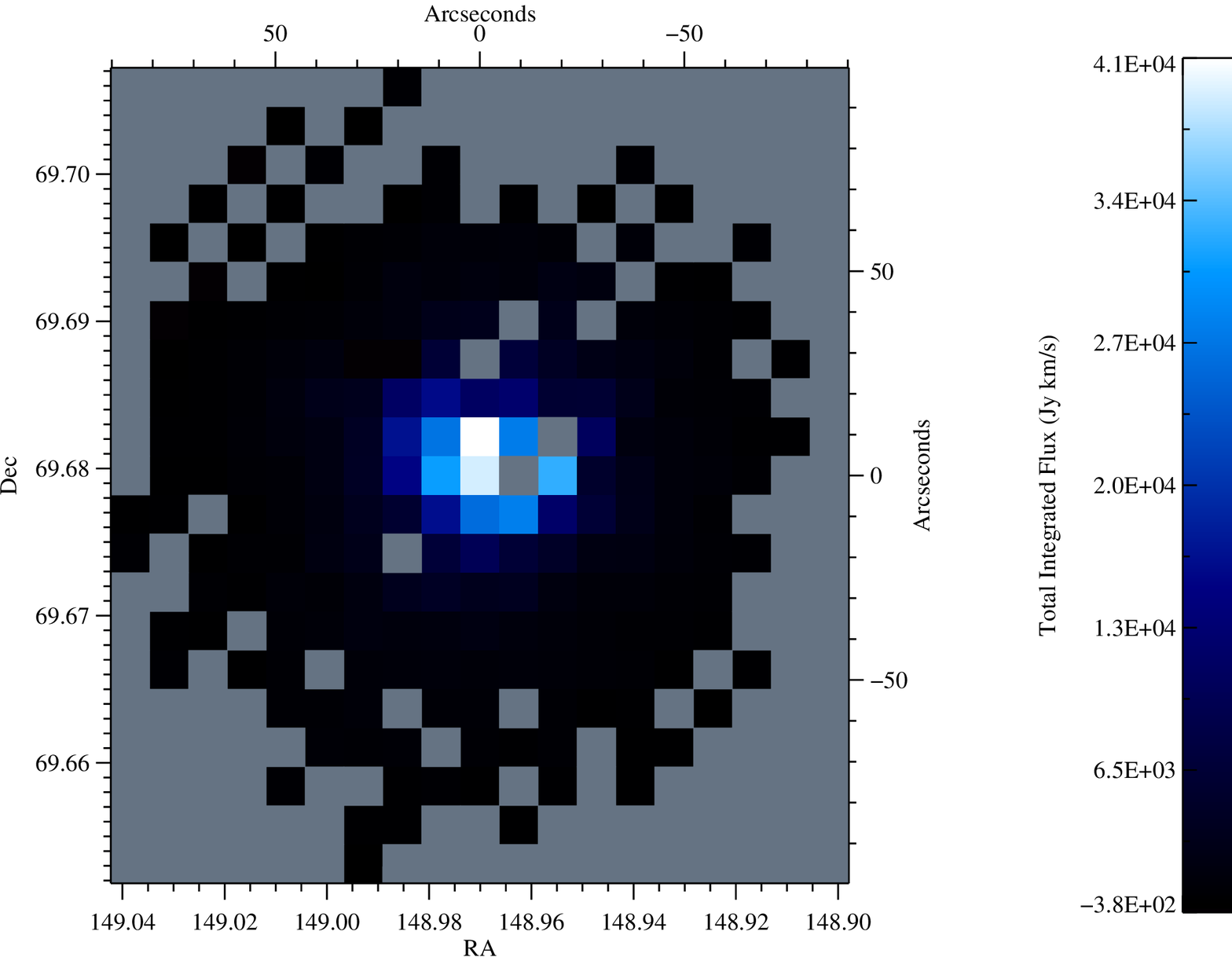}}
\subfigure{\includegraphics[width=0.8\textwidth]{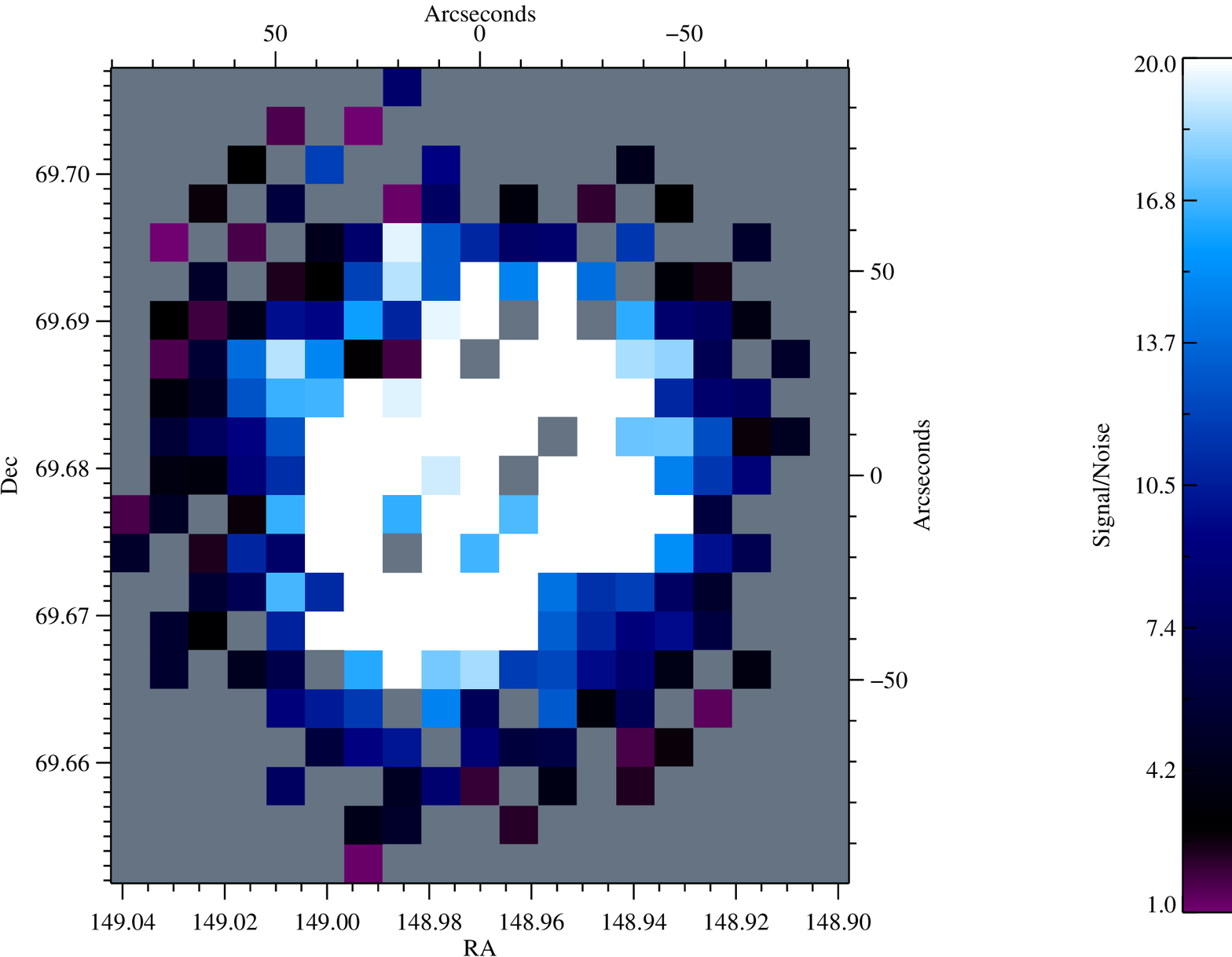}}
\caption{Integrated Flux (top) and Signal/Noise (bottom) maps for NII.}\label{fig:intflux12}
\end{figure*}

\clearpage

\begin{figure*}[th]
\centering
\subfigure{\includegraphics[width=0.8\textwidth]{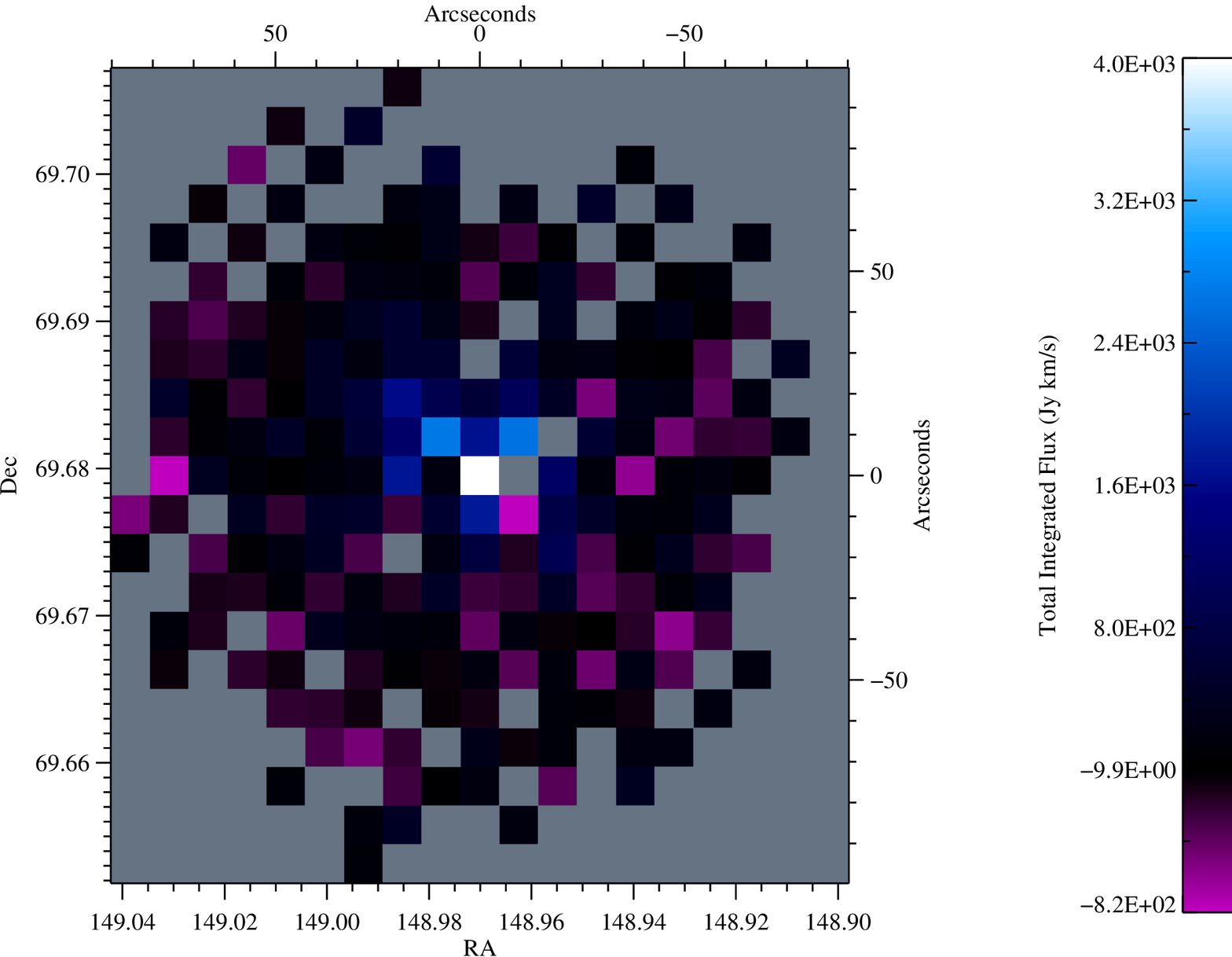}}
\subfigure{\includegraphics[width=0.8\textwidth]{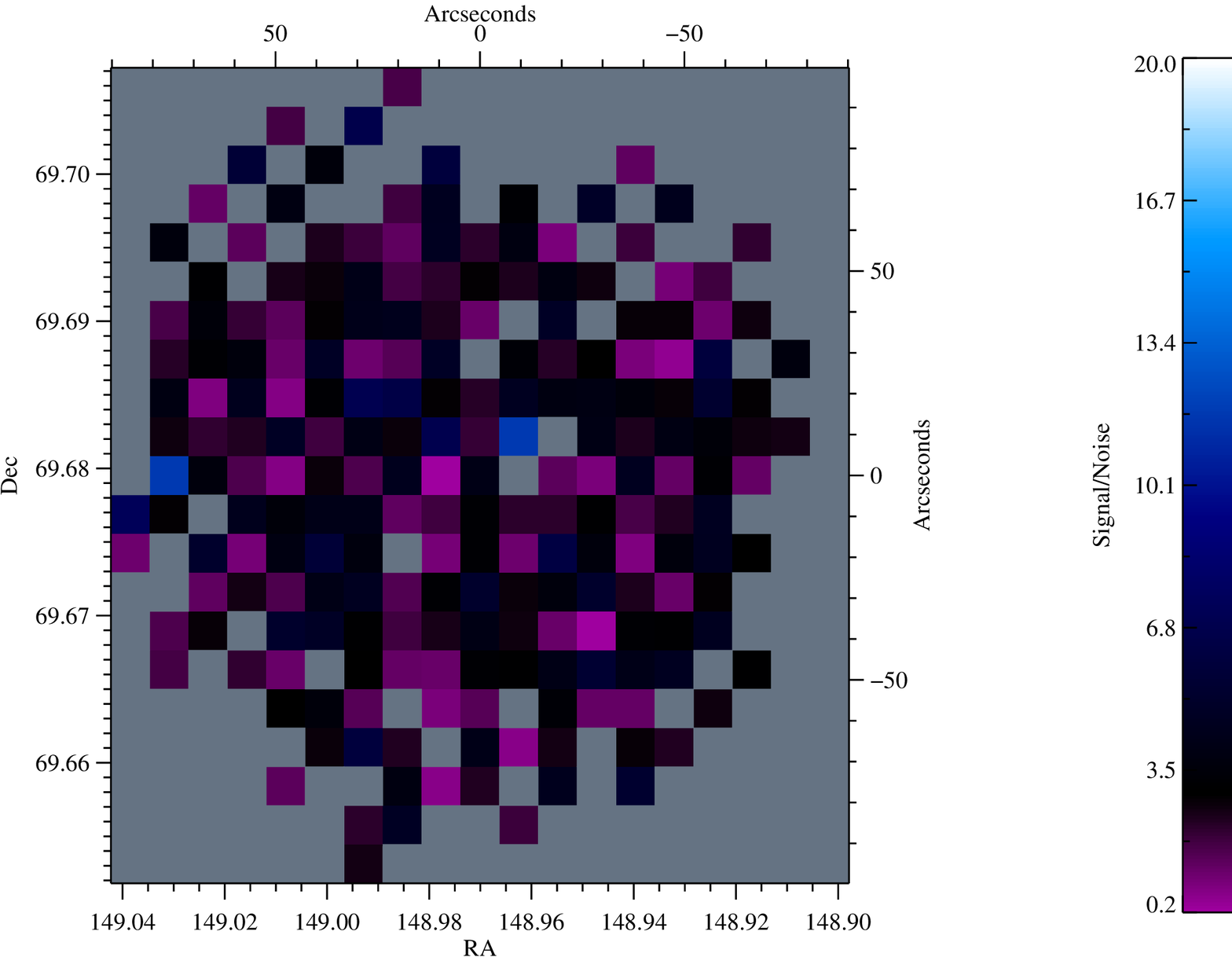}}
\caption{Integrated Flux (top) and Signal/Noise (bottom) maps for CO \jthirteen.}\label{fig:intflux13}
\end{figure*}

\clearpage

\begin{figure*}[th]
\centering
\subfigure{\includegraphics[width=0.8\textwidth]{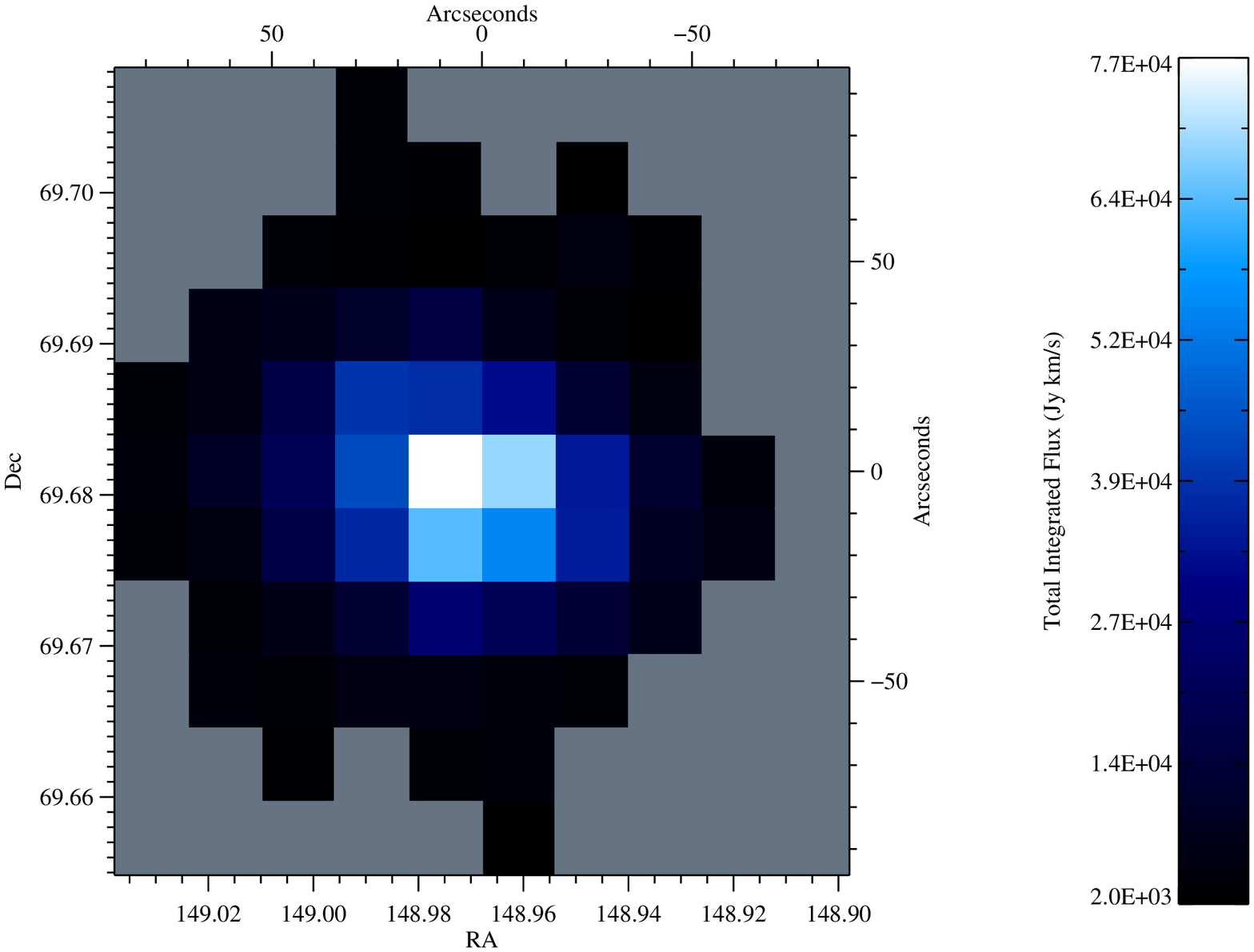}}
\subfigure{\includegraphics[width=0.8\textwidth]{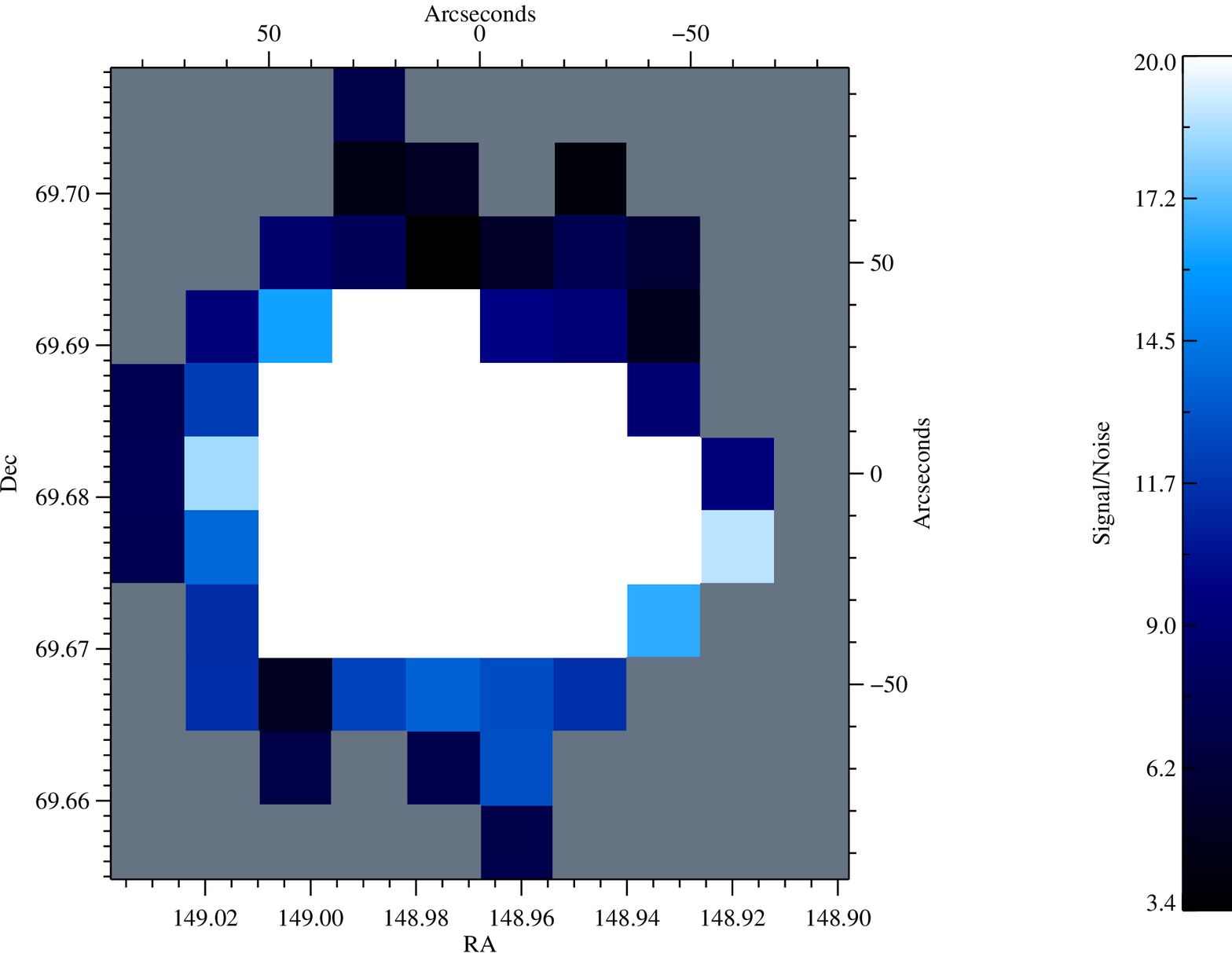}}
\caption{Integrated Flux (top) and Signal/Noise (bottom) maps for CO \jfour.}\label{fig:intflux_big1}
\end{figure*}

\clearpage

\begin{figure*}[th]
\centering
\subfigure{\includegraphics[width=0.8\textwidth]{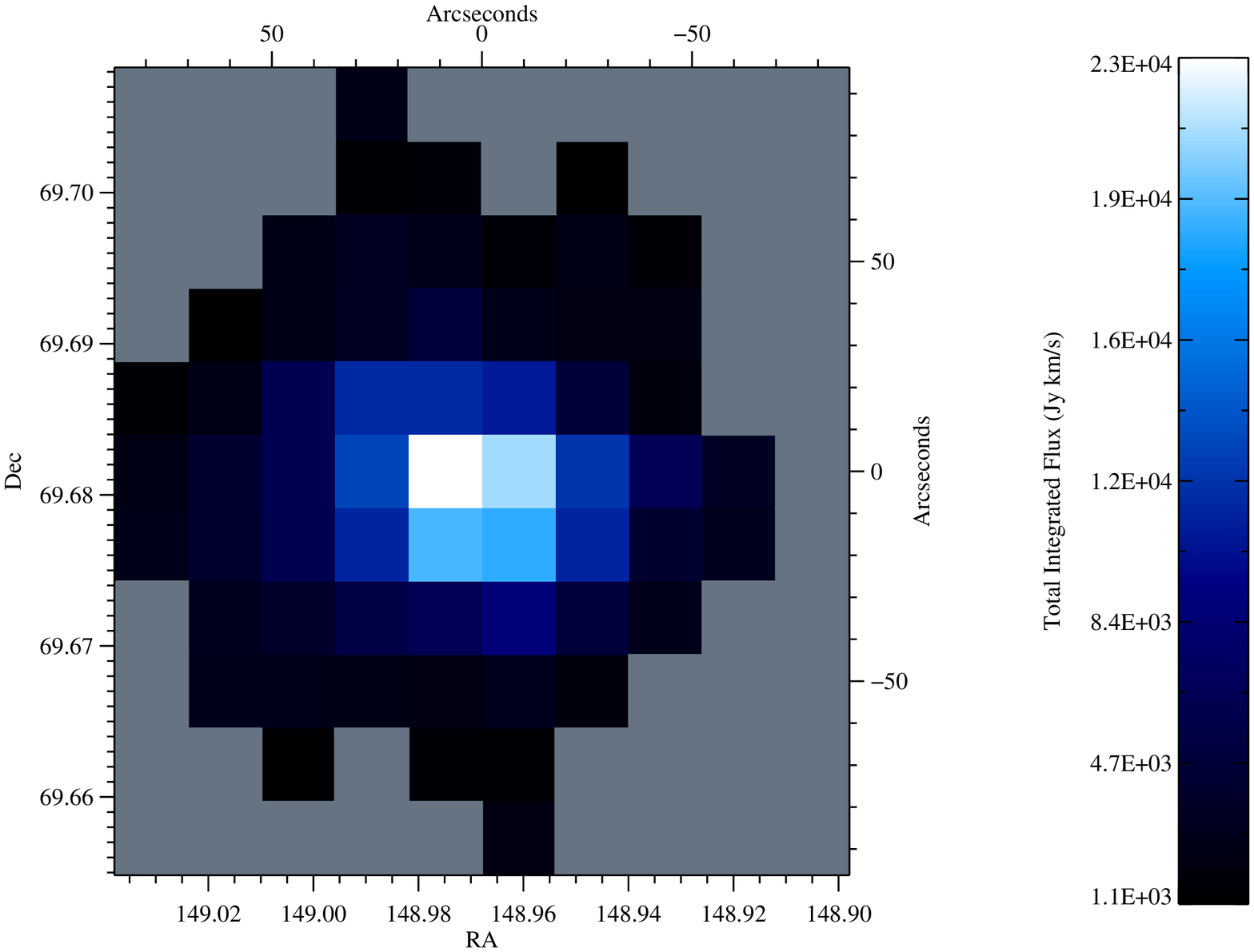}}
\subfigure{\includegraphics[width=0.8\textwidth]{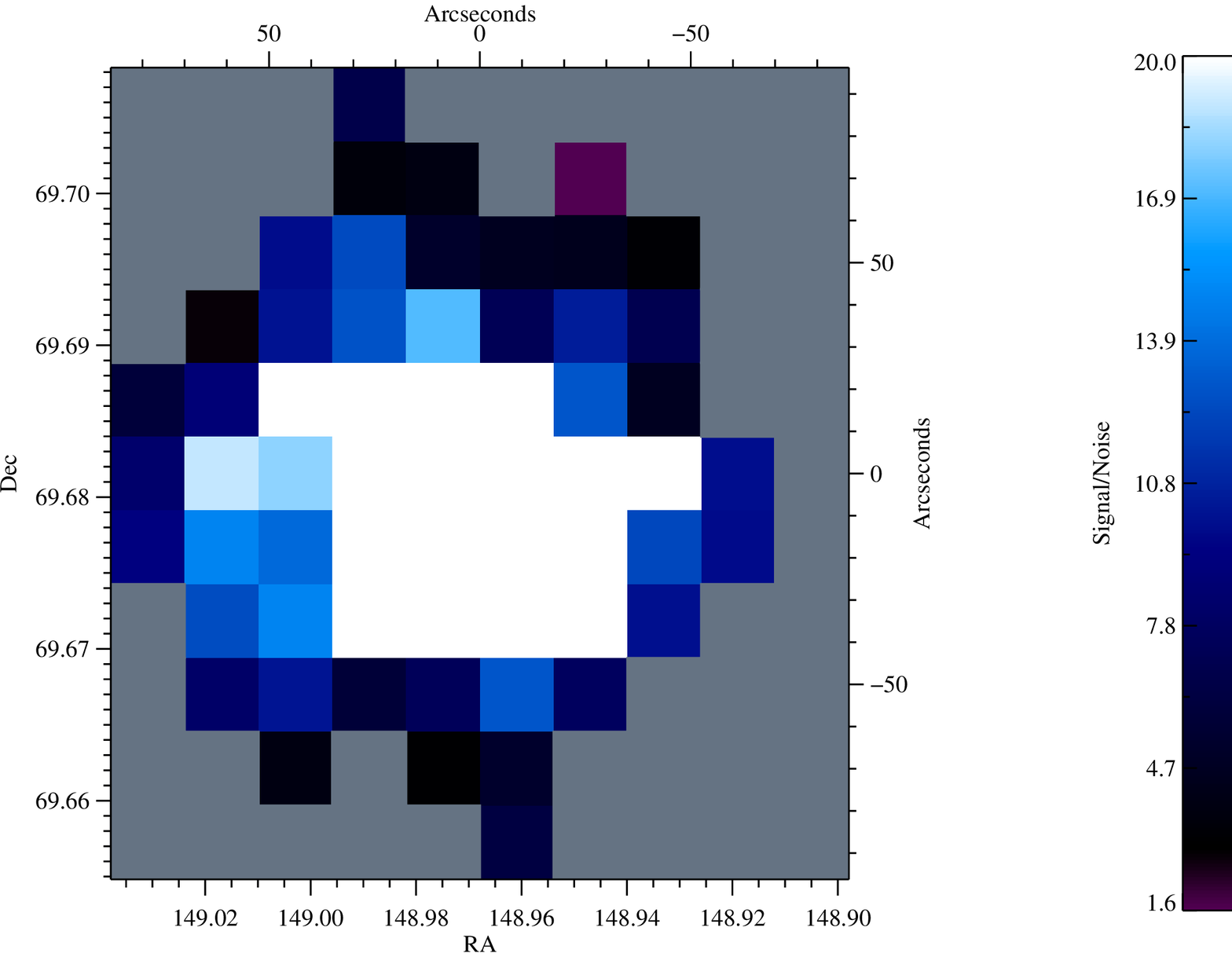}}
\caption{Integrated Flux (top) and Signal/Noise (bottom) maps for CI \jone.}\label{fig:intflux_big2}
\end{figure*}

\clearpage

\begin{figure*}[th]
\centering
\subfigure{\includegraphics[width=0.8\textwidth]{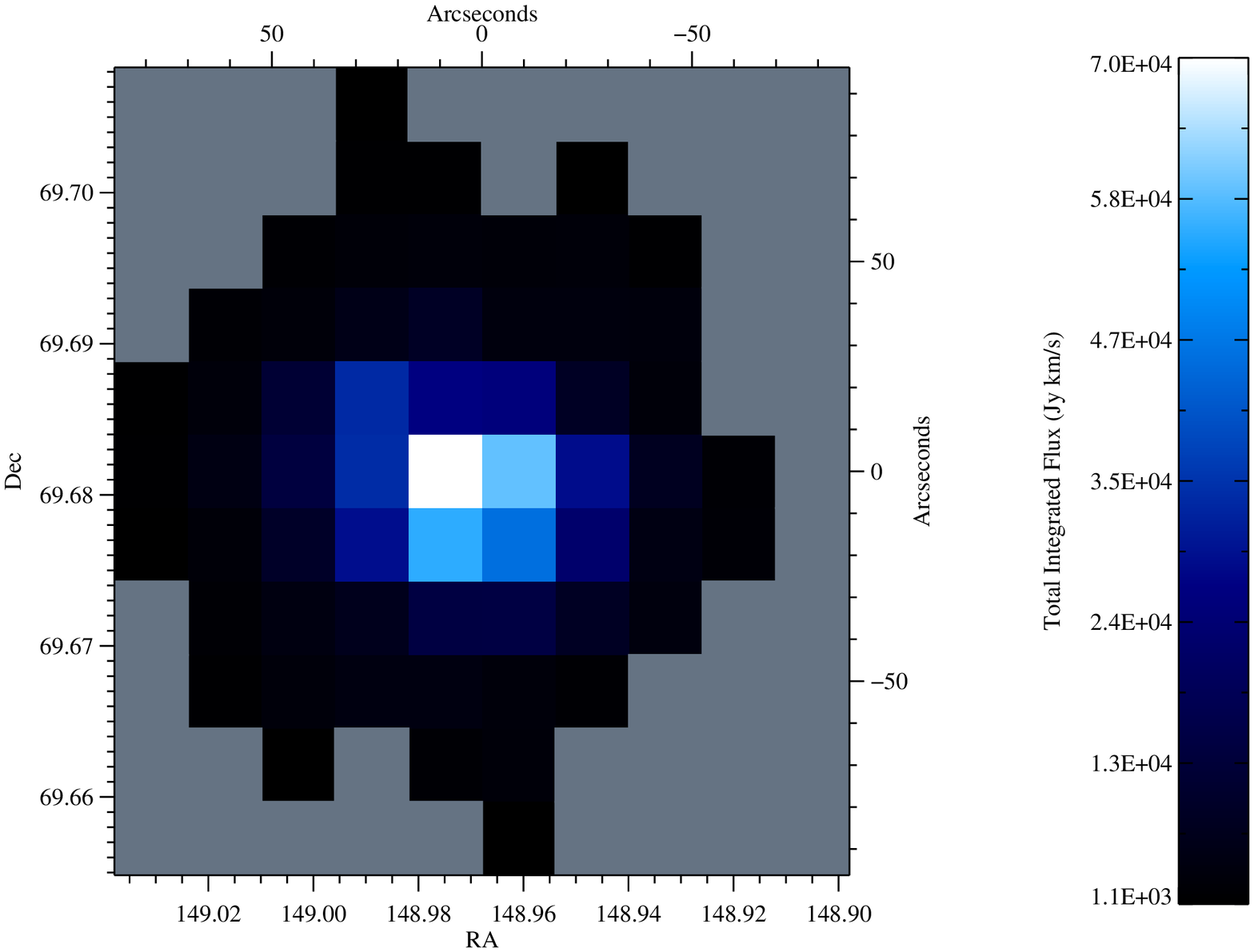}}
\subfigure{\includegraphics[width=0.8\textwidth]{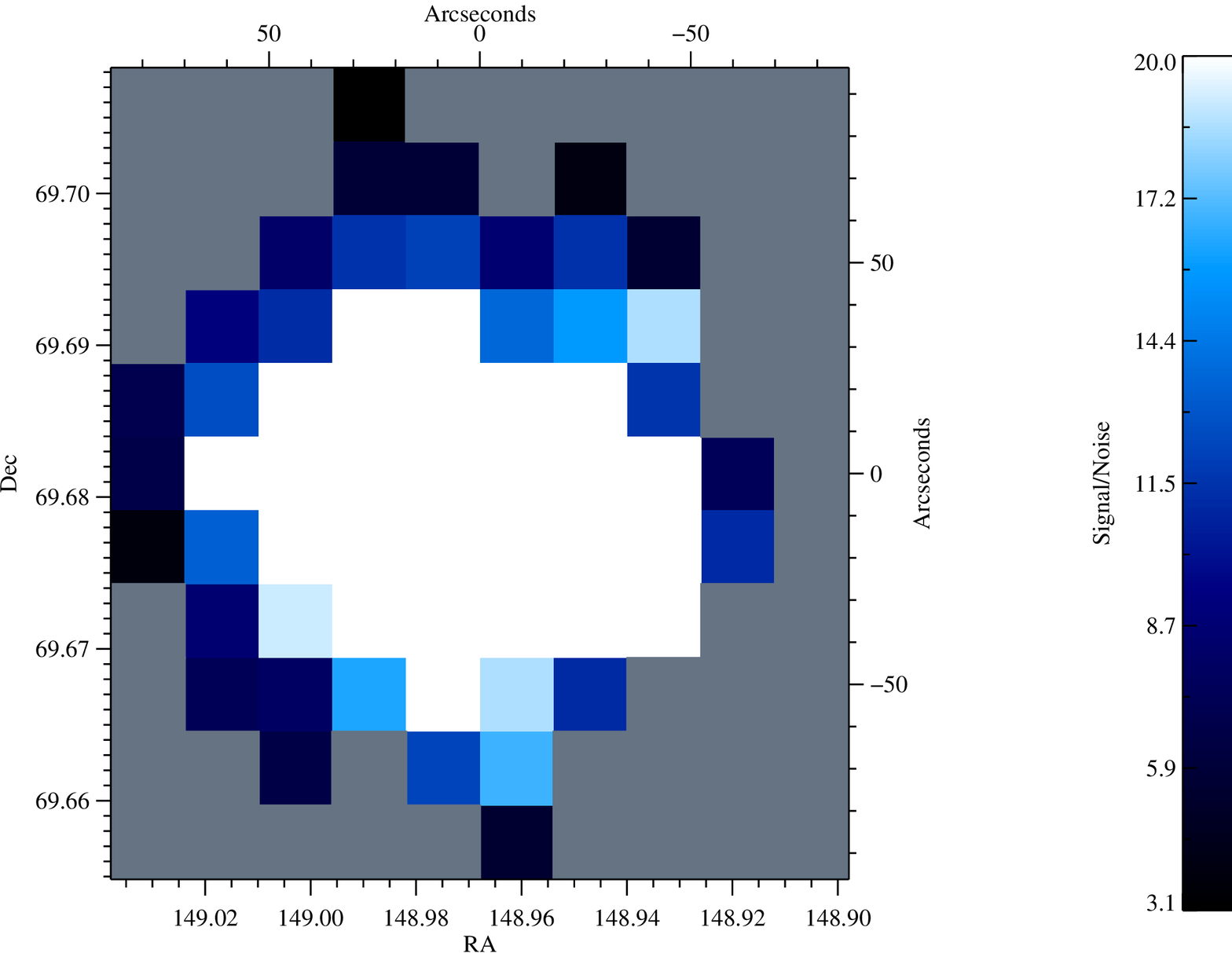}}
\caption{Integrated Flux (top) and Signal/Noise (bottom) maps for CO \jfive.}\label{fig:intflux_big3}
\end{figure*}

\clearpage

\begin{figure*}[th]
\centering
\subfigure{\includegraphics[width=0.8\textwidth]{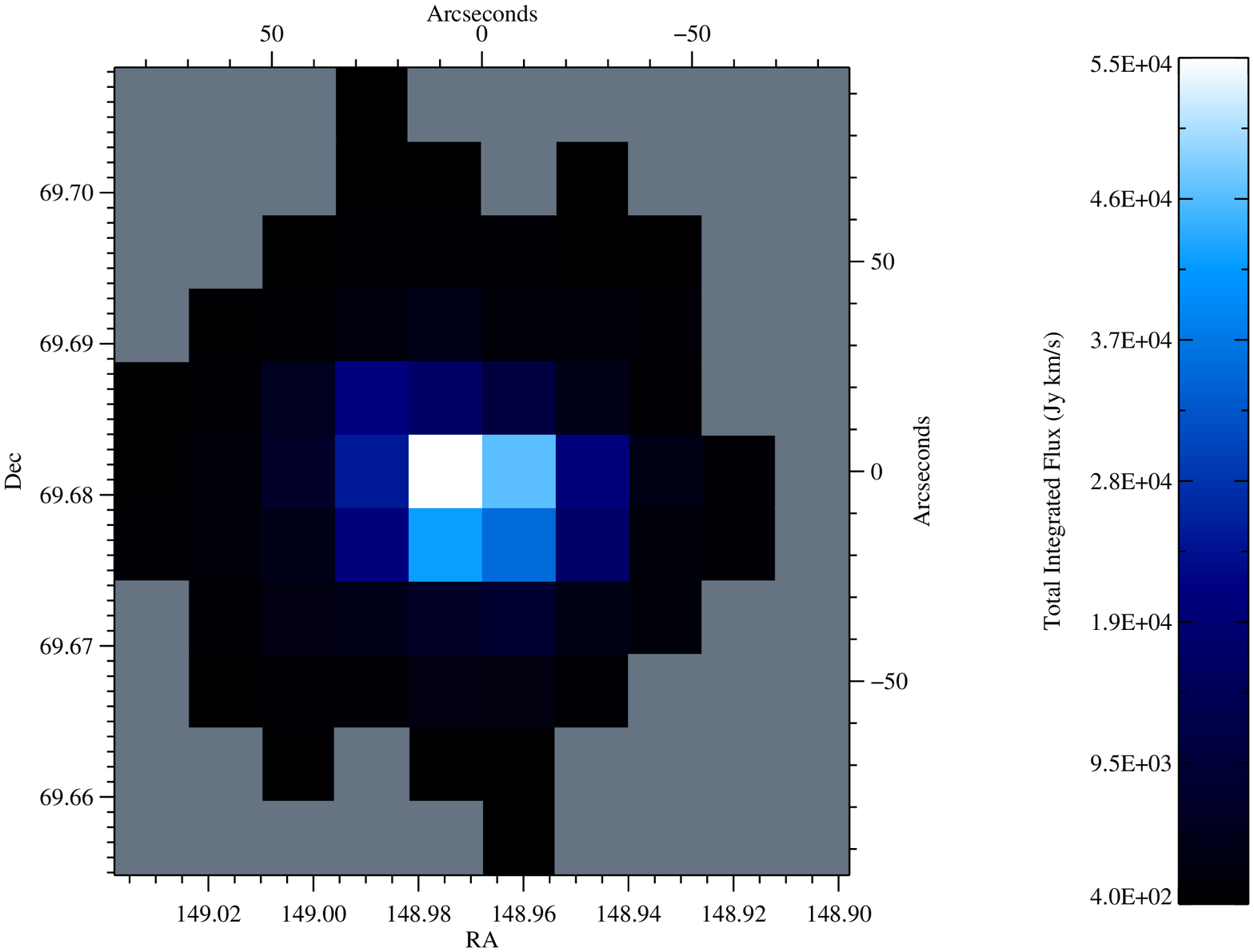}}
\subfigure{\includegraphics[width=0.8\textwidth]{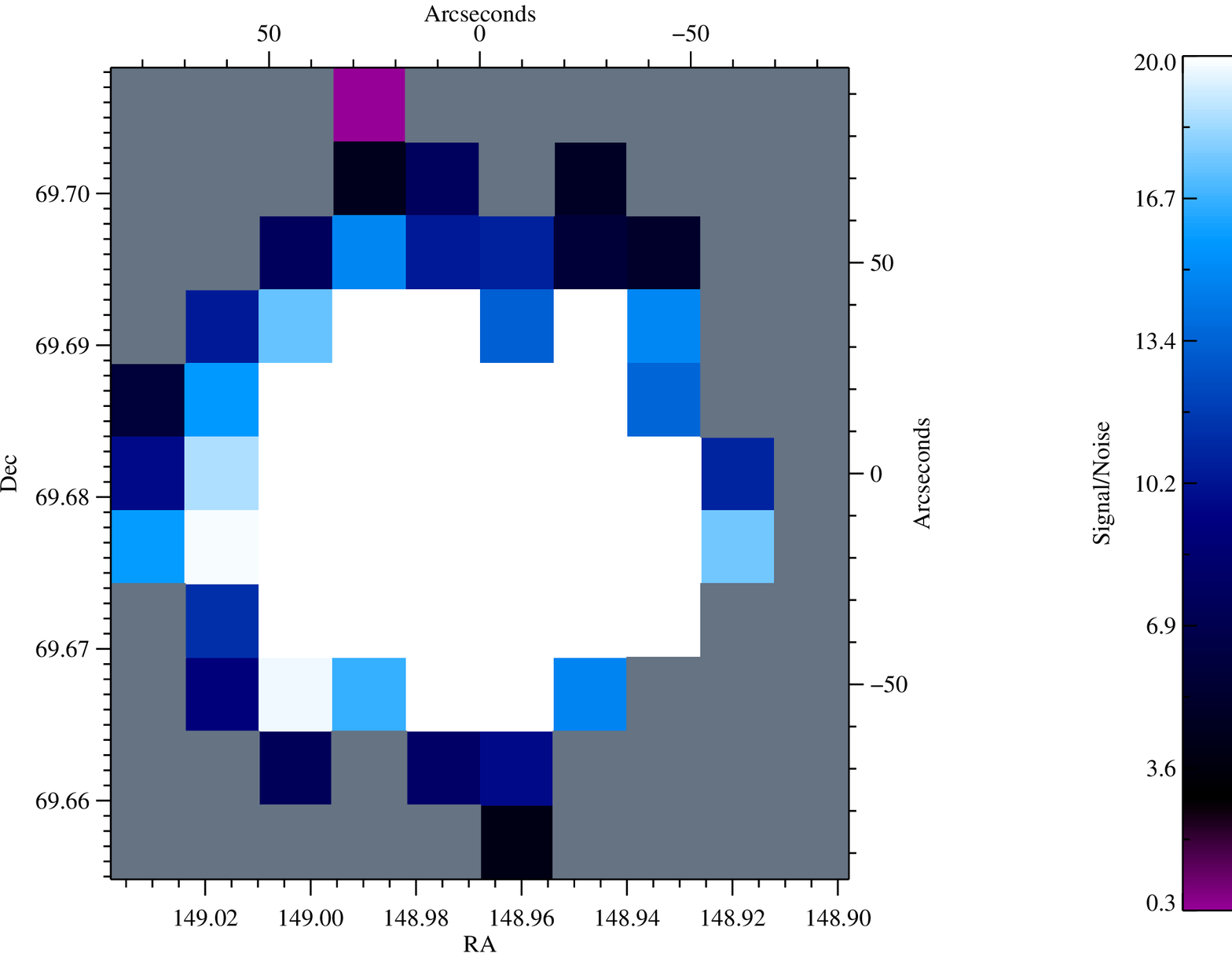}}
\caption{Integrated Flux (top) and Signal/Noise (bottom) maps for CO \jsix.}\label{fig:intflux_big4}
\end{figure*}

\clearpage

\begin{figure*}[th]
\centering
\subfigure{\includegraphics[width=0.8\textwidth]{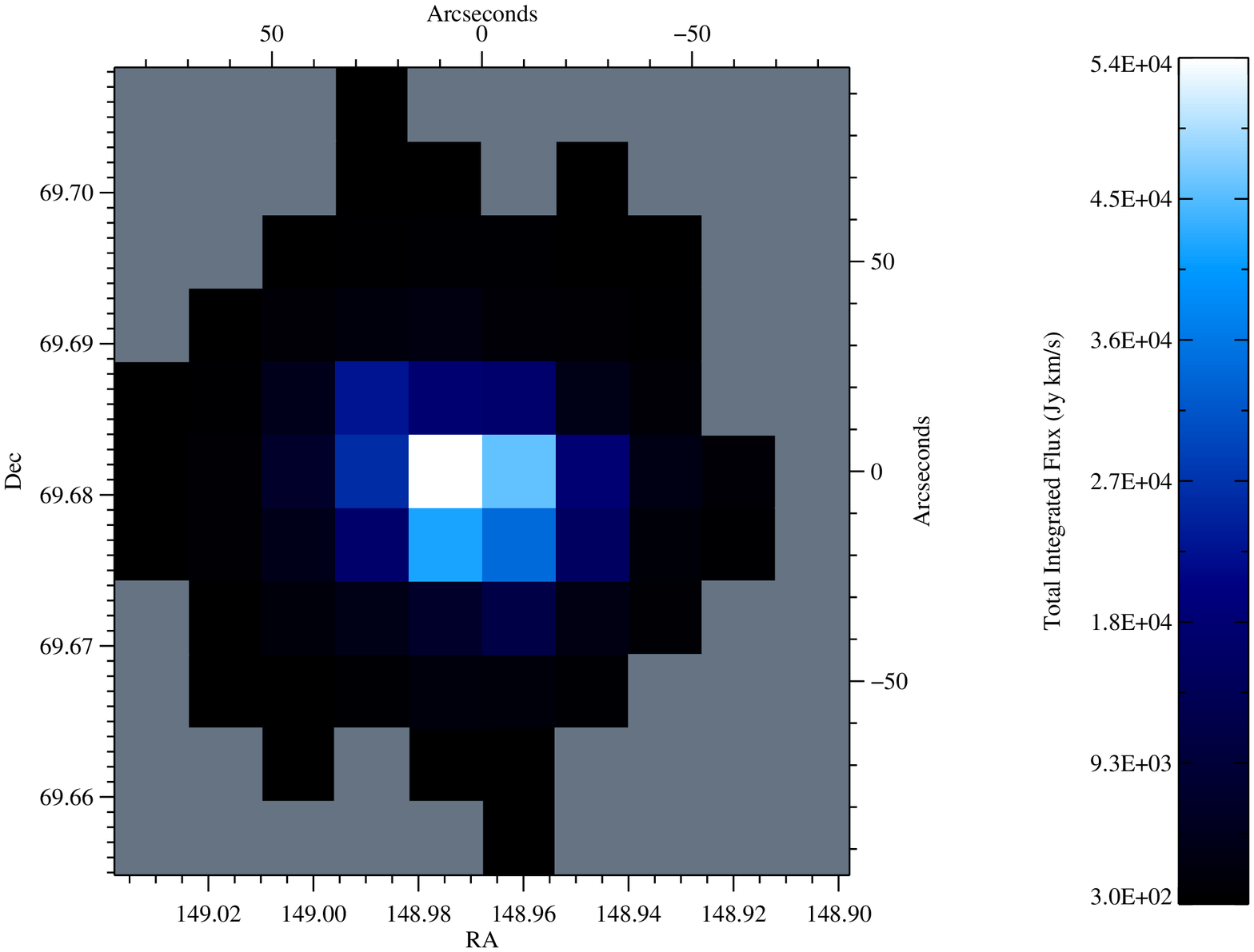}}
\subfigure{\includegraphics[width=0.8\textwidth]{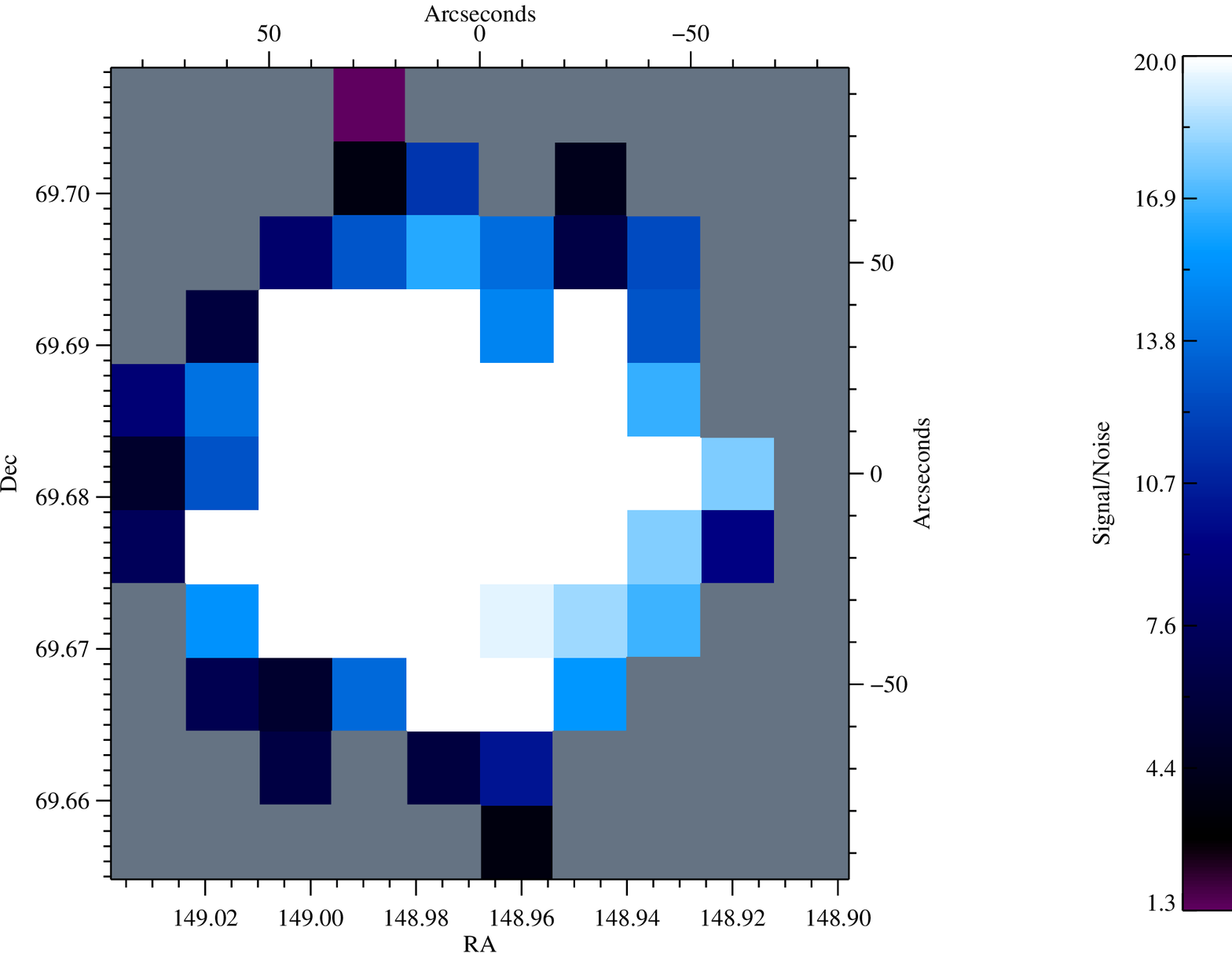}}
\caption{Integrated Flux (top) and Signal/Noise (bottom) maps for CO \jseven.}\label{fig:intflux_big5}
\end{figure*}

\clearpage

\begin{figure*}[th]
\centering
\subfigure{\includegraphics[width=0.8\textwidth]{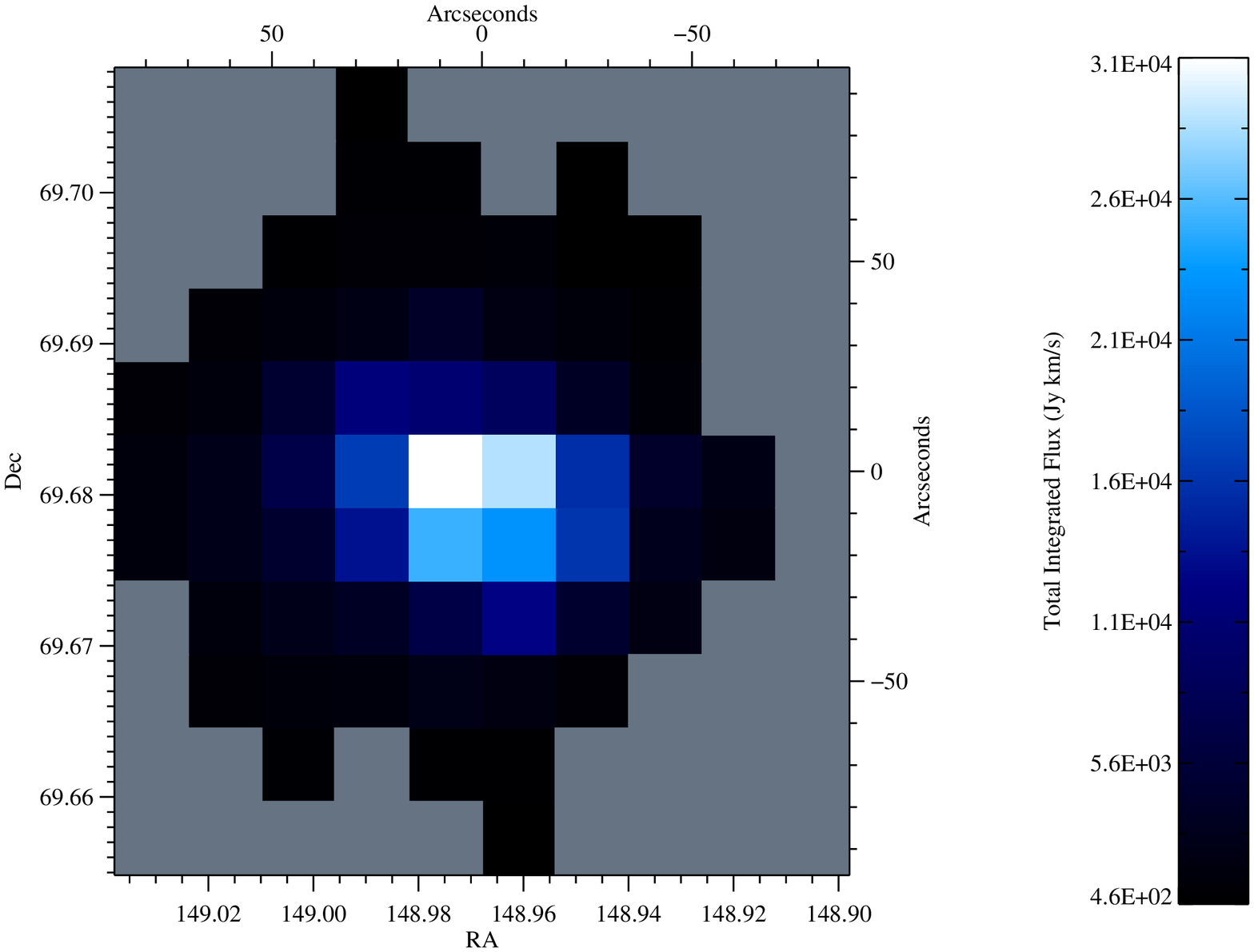}}
\subfigure{\includegraphics[width=0.8\textwidth]{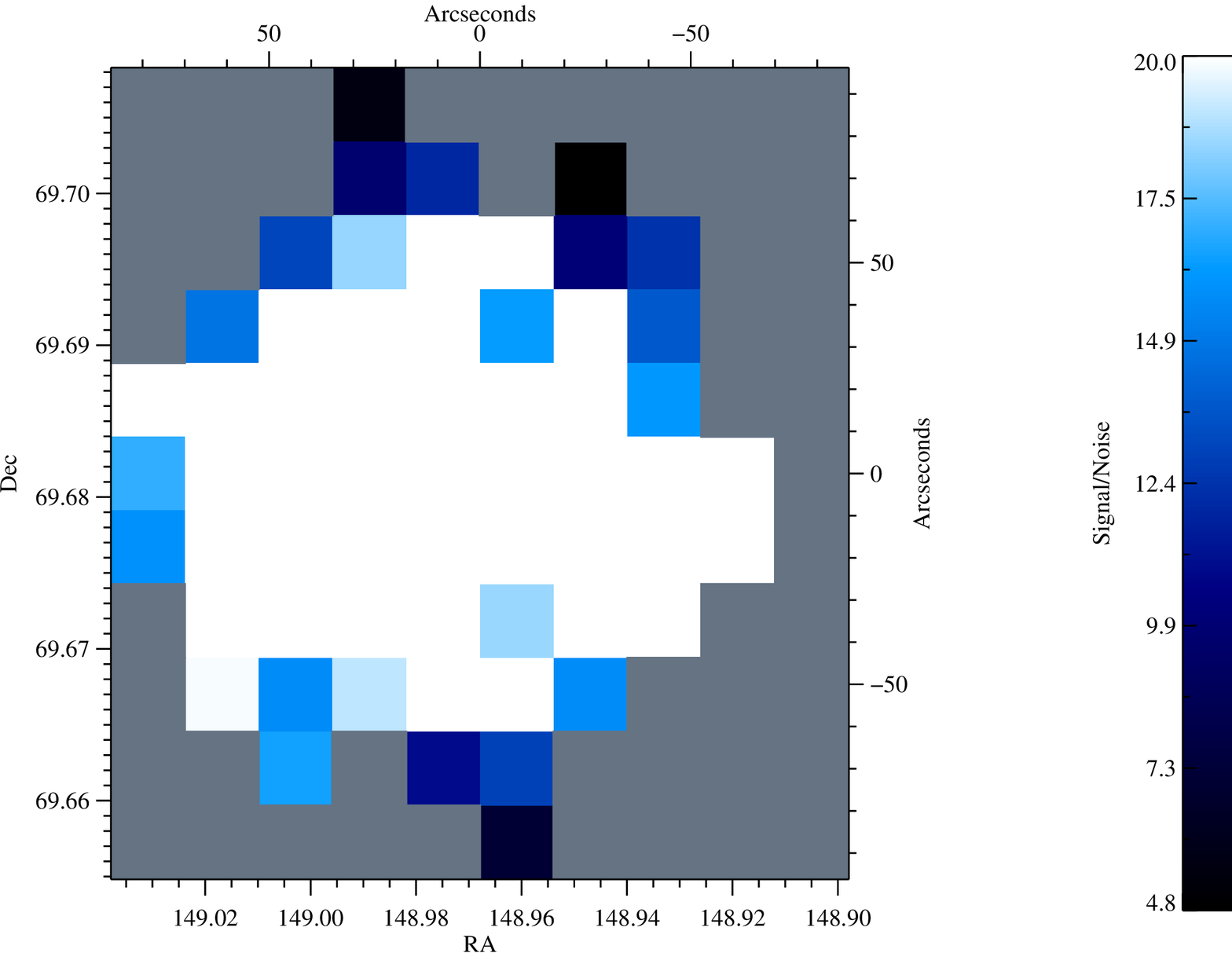}}
\caption{Integrated Flux (top) and Signal/Noise (bottom) maps for CI \jtwo.}\label{fig:intflux_big6}
\end{figure*}

\clearpage

\begin{figure*}[th]
\centering
\subfigure{\includegraphics[width=0.8\textwidth]{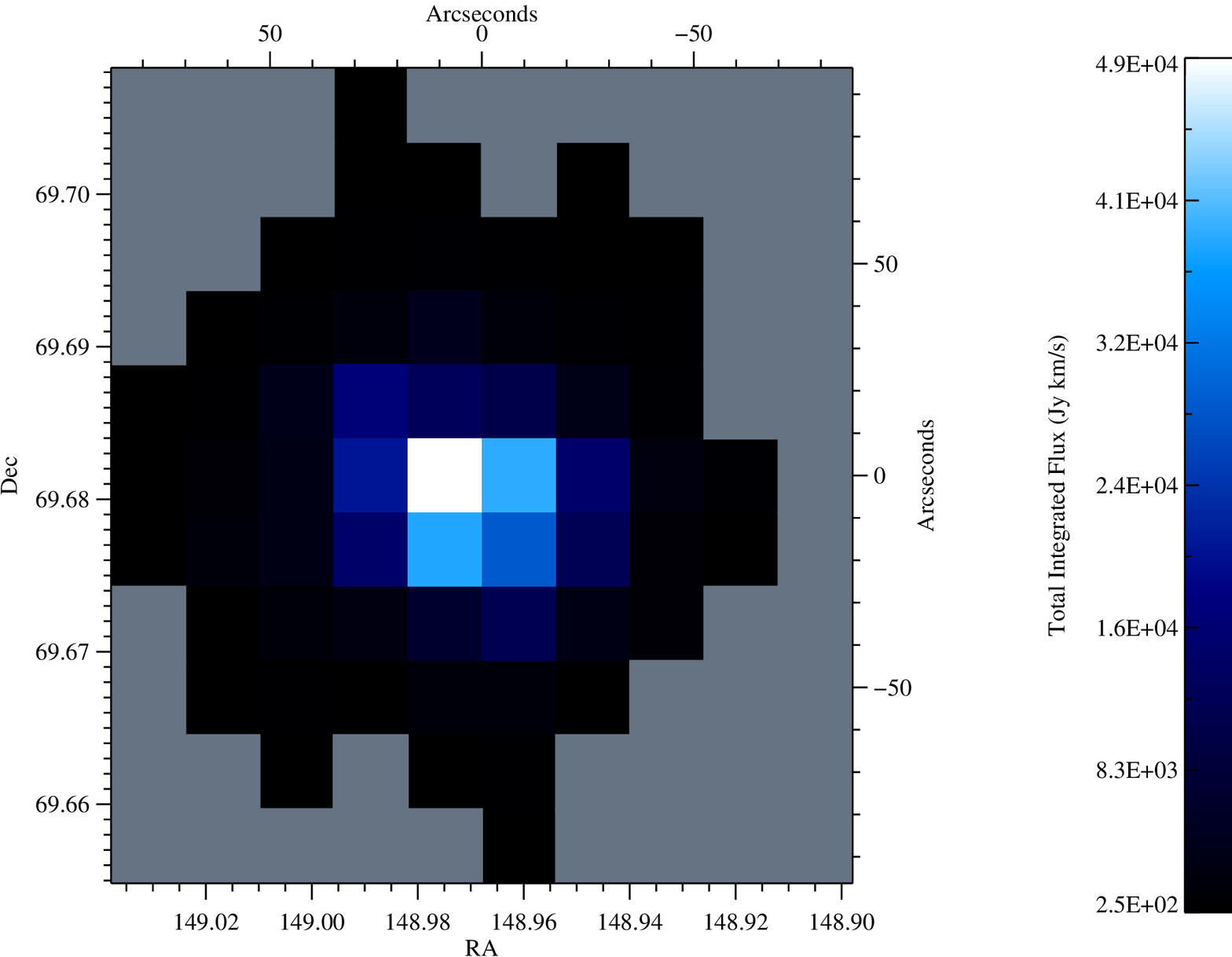}}
\subfigure{\includegraphics[width=0.8\textwidth]{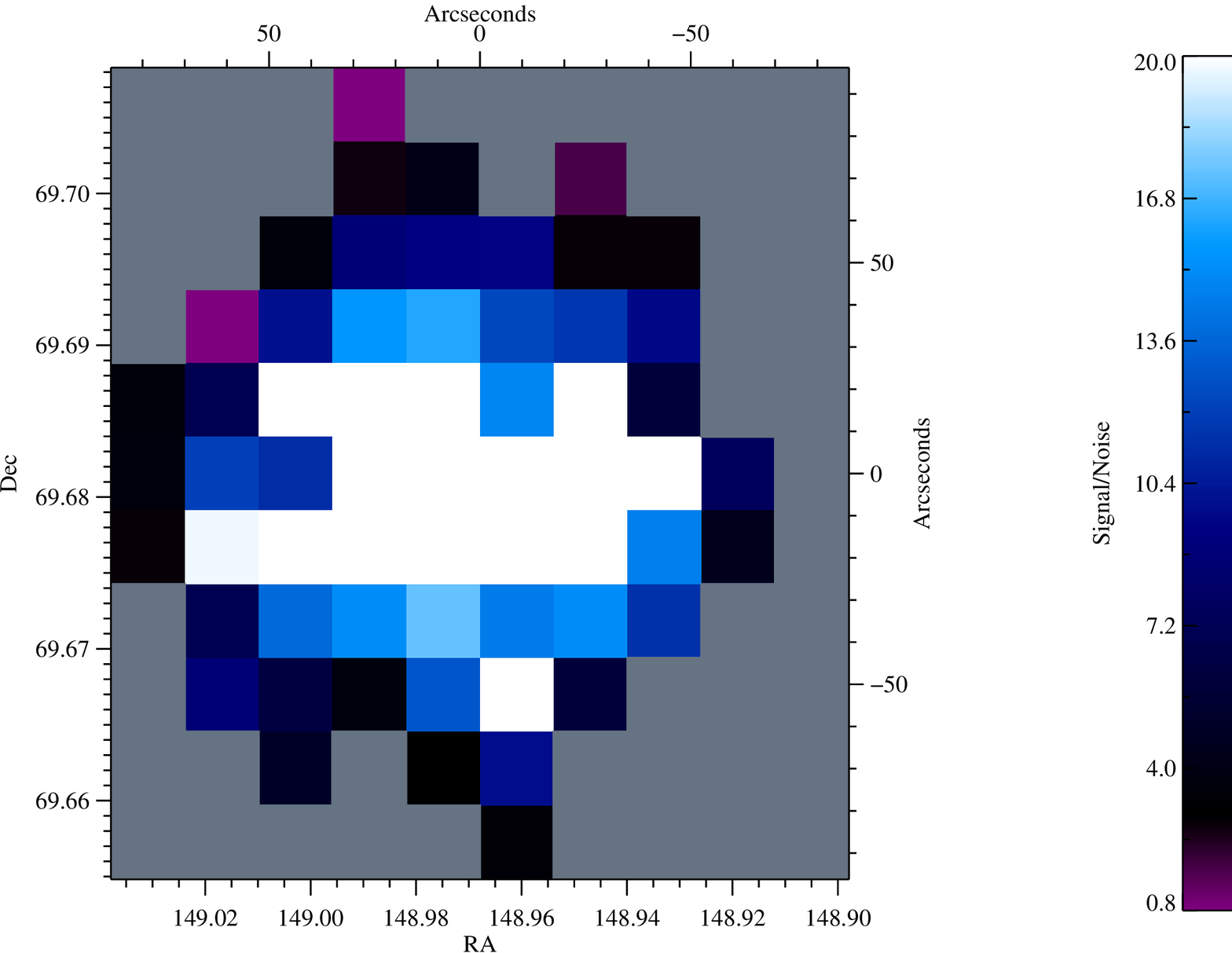}}
\caption{Integrated Flux (top) and Signal/Noise (bottom) maps for CO \jeight.}\label{fig:intflux_big7}
\end{figure*}

%% file: M82.bbl
\begin{thebibliography}{60}
\expandafter\ifx\csname natexlab\endcsname\relax\def\natexlab#1{#1}\fi

\bibitem[{Aalto {et~al.}(1995)Aalto, Booth, Black, \& Johansson}]{Aalto:1995}
Aalto, S., Booth, R.~S., Black, J.~H., \& Johansson, L. E.~B. 1995, Astronomy
  and Astrophysics, 300, 369

\bibitem[{{Aladro} {et~al.}(2011){Aladro}, {Mart{\'{\i}}n},
  {Mart{\'{\i}}n-Pintado}, {Mauersberger}, {Henkel}, {Oca{\~n}a Flaquer}, \&
  {Amo-Baladr{\'o}n}}]{Aladro:2011}
{Aladro}, R., {Mart{\'{\i}}n}, S., {Mart{\'{\i}}n-Pintado}, J., {Mauersberger},
  R., {Henkel}, C., {Oca{\~n}a Flaquer}, B., \& {Amo-Baladr{\'o}n}, M.~A. 2011,
  \aap, 535, A84

\bibitem[{Beir{\~a}o {et~al.}(2008)Beir{\~a}o, Brandl, Appleton, Groves, Armus,
  Schreiber, Smith, Charmandaris, \& Houck}]{Beirao:2008}
Beir{\~a}o, P. {et~al.} 2008, The Astrophysical Journal, 676, 304

\bibitem[{{Bendo} {et~al.}(2011){Bendo}, {Boselli}, {Dariush}, {Pohlen},
  {Roussel}, {Sauvage}, {Smith}, {Wilson}, {Baes}, {Cooray}, {Clements},
  {Cortese}, {Foyle}, {Galametz}, {Gomez}, {Lebouteiller}, {Lu}, {Madden},
  {Mentuch}, {O'Halloran}, {Page}, {Remy}, {Schulz}, \&
  {Spinoglio}}]{Bendo:2011}
{Bendo}, G.~J. {et~al.} 2011, ArXiv e-prints

\bibitem[{{Bradford} {et~al.}(2009){Bradford}, {Aguirre}, {Aikin}, {Bock},
  {Earle}, {Glenn}, {Inami}, {Maloney}, {Matsuhara}, {Naylor}, {Nguyen}, \&
  {Zmuidzinas}}]{Bradford:2009}
{Bradford}, C.~M. {et~al.} 2009, \apj, 705, 112

\bibitem[{{Bradford} {et~al.}(2005){Bradford}, {Stacey}, {Nikola}, {Bolatto},
  {Jackson}, {Savage}, \& {Davidson}}]{Bradford:2005}
{Bradford}, C.~M., {Stacey}, G.~J., {Nikola}, T., {Bolatto}, A.~D., {Jackson},
  J.~M., {Savage}, M.~L., \& {Davidson}, J.~A. 2005, \apj, 623, 866

\bibitem[{Chattopadhyay {et~al.}(2003)Chattopadhyay, Glenn, Bock, Rownd,
  Caldwell, \& Griffin}]{Chattopadhyay:2003}
Chattopadhyay, G., Glenn, J., Bock, J.~J., Rownd, B.~K., Caldwell, M., \&
  Griffin, M.~J. 2003, IEEE Transactions on Microwave Theory and Techniques,
  51, 2139

\bibitem[{{Cheng} {et~al.}(1997){Cheng}, {Collins}, {Angione}, {Talbert},
  {Hintzen}, {Smith}, {Stecher}, \& {The UIT Team}}]{Cheng:1997}
{Cheng}, K.~P., {Collins}, N., {Angione}, R., {Talbert}, F., {Hintzen}, P.,
  {Smith}, E.~P., {Stecher}, T., \& {The UIT Team}, eds. 1997, {UV/Visible Sky
  Gallery on CDROM}

\bibitem[{{Conley} {et~al.}(2011){Conley}, {Cooray}, {Vieira}, {Gonz{\'a}lez
  Solares}, {Kim}, {Aguirre}, {Amblard}, {Auld}, {Baker}, {Beelen}, {Blain},
  {Blundell}, {Bock}, {Bradford}, {Bridge}, {Brisbin}, {Burgarella},
  {Carpenter}, {Chanial}, {Chapin}, {Christopher}, {Clements}, {Cox},
  {Djorgovski}, {Dowell}, {Eales}, {Earle}, {Ellsworth-Bowers}, {Farrah},
  {Franceschini}, {Frayer}, {Fu}, {Gavazzi}, {Glenn}, {Griffin}, {Gurwell},
  {Halpern}, {Ibar}, {Ivison}, {Jarvis}, {Kamenetzky}, {Krips}, {Levenson},
  {Lupu}, {Mahabal}, {Maloney}, {Maraston}, {Marchetti}, {Marsden},
  {Matsuhara}, {Mortier}, {Murphy}, {Naylor}, {Neri}, {Nguyen}, {Oliver},
  {Omont}, {Page}, {Papageorgiou}, {Pearson}, {P{\'e}rez-Fournon}, {Pohlen},
  {Rangwala}, {Rawlings}, {Raymond}, {Riechers}, {Rodighiero}, {Roseboom},
  {Rowan-Robinson}, {Schulz}, {Scott}, {Scott}, {Serra}, {Seymour}, {Shupe},
  {Smith}, {Symeonidis}, {Tugwell}, {Vaccari}, {Valiante}, {Valtchanov},
  {Verma}, {Viero}, {Vigroux}, {Wang}, {Wiebe}, {Wright}, {Xu}, {Zeimann},
  {Zemcov}, \& {Zmuidzinas}}]{Conley:2011}
{Conley}, A. {et~al.} 2011, \apjl, 732, L35

\bibitem[{{Dalcanton} {et~al.}(2009){Dalcanton}, {Williams}, {Seth}, {Dolphin},
  {Holtzman}, {Rosema}, {Skillman}, {Cole}, {Girardi}, {Gogarten},
  {Karachentsev}, {Olsen}, {Weisz}, {Christensen}, {Freeman}, {Gilbert},
  {Gallart}, {Harris}, {Hodge}, {de Jong}, {Karachentseva}, {Mateo}, {Stetson},
  {Tavarez}, {Zaritsky}, {Governato}, \& {Quinn}}]{Dalcanton:2009}
{Dalcanton}, J.~J. {et~al.} 2009, \apjs, 183, 67

\bibitem[{{de Graauw} {et~al.}(2010){de Graauw}, {Helmich}, {Phillips},
  {Stutzki}, {Caux}, {Whyborn}, {Dieleman}, {Roelfsema}, {Aarts}, {Assendorp},
  {Bachiller}, {Baechtold}, {Barcia}, {Beintema}, {Belitsky}, {Benz}, {Bieber},
  {Boogert}, {Borys}, {Bumble}, {Ca{\"i}s}, {Caris}, {Cerulli-Irelli},
  {Chattopadhyay}, {Cherednichenko}, {Ciechanowicz}, {Coeur-Joly}, {Comito},
  {Cros}, {de Jonge}, {de Lange}, {Delforges}, {Delorme}, {den Boggende},
  {Desbat}, {Diez-Gonz{\'a}lez}, {di Giorgio}, {Dubbeldam}, {Edwards},
  {Eggens}, {Erickson}, {Evers}, {Fich}, {Finn}, {Franke}, {Gaier}, {Gal},
  {Gao}, {Gallego}, {Gauffre}, {Gill}, {Glenz}, {Golstein}, {Goulooze},
  {Gunsing}, {G{\"u}sten}, {Hartogh}, {Hatch}, {Higgins}, {Honingh}, {Huisman},
  {Jackson}, {Jacobs}, {Jacobs}, {Jarchow}, {Javadi}, {Jellema}, {Justen},
  {Karpov}, {Kasemann}, {Kawamura}, {Keizer}, {Kester}, {Klapwijk}, {Klein},
  {Kollberg}, {Kooi}, {Kooiman}, {Kopf}, {Krause}, {Krieg}, {Kramer},
  {Kruizenga}, {Kuhn}, {Laauwen}, {Lai}, {Larsson}, {Leduc}, {Leinz}, {Lin},
  {Liseau}, {Liu}, {Loose}, {L{\'o}pez-Fernandez}, {Lord}, {Luinge}, {Marston},
  {Mart{\'{\i}}n-Pintado}, {Maestrini}, {Maiwald}, {McCoey}, {Mehdi}, {Megej},
  {Melchior}, {Meinsma}, {Merkel}, {Michalska}, {Monstein}, {Moratschke},
  {Morris}, {Muller}, {Murphy}, {Naber}, {Natale}, {Nowosielski}, {Nuzzolo},
  {Olberg}, {Olbrich}, {Orfei}, {Orleanski}, {Ossenkopf}, {Peacock}, {Pearson},
  {Peron}, {Phillip-May}, {Piazzo}, {Planesas}, {Rataj}, {Ravera}, {Risacher},
  {Salez}, {Samoska}, {Saraceno}, {Schieder}, {Schlecht}, {Schl{\"o}der},
  {Schm{\"u}lling}, {Schultz}, {Schuster}, {Siebertz}, {Smit}, {Szczerba},
  {Shipman}, {Steinmetz}, {Stern}, {Stokroos}, {Teipen}, {Teyssier}, {Tils},
  {Trappe}, {van Baaren}, {van Leeuwen}, {van de Stadt}, {Visser}, {Wildeman},
  {Wafelbakker}, {Ward}, {Wesselius}, {Wild}, {Wulff}, {Wunsch}, {Tielens},
  {Zaal}, {Zirath}, {Zmuidzinas}, \& {Zwart}}]{deGraauw:2010}
{de Graauw}, T. {et~al.} 2010, \aap, 518, L6

\bibitem[{{de Vaucouleurs} {et~al.}(1991){de Vaucouleurs}, {de Vaucouleurs},
  {Corwin}, {Buta}, {Paturel}, \& {Fouque}}]{deVaucouleurs:1999}
{de Vaucouleurs}, G., {de Vaucouleurs}, A., {Corwin}, Jr., H.~G., {Buta},
  R.~J., {Paturel}, G., \& {Fouque}, P. 1991, {Third Reference Catalogue of
  Bright Galaxies}, ed. {Roman, N.~G., de Vaucouleurs, G., de Vaucouleurs, A.,
  Corwin, H.~G., Jr., Buta, R.~J., Paturel, G., \& Fouqu{\'e}, P.}

\bibitem[{{Downes} \& {Solomon}(1998)}]{Downes:1998}
{Downes}, D. \& {Solomon}, P.~M. 1998, \apj, 507, 615

\bibitem[{{Draine}(1989)}]{Draine:1989}
{Draine}, B.~T. 1989, in ESA Special Publication, Vol. 290, Infrared
  Spectroscopy in Astronomy, ed. {E.~B{\"o}hm-Vitense}, 93--98

\bibitem[{Flower \& {Pineau des For{\^e}ts}(2010)}]{Flower:2010}
Flower, D.~R. \& {Pineau des For{\^e}ts}, G. 2010, Monthly Notices of the Royal
  Astronomical Society, 406, 1745

\bibitem[{{Fulton} {et~al.}(2010){Fulton}, {Baluteau}, {Bendo}, {Benielli},
  {Gastaud}, {Griffin}, {Guest}, {Imhof}, {Lim}, {Lu}, {Naylor}, {Panuzzo},
  {Polehampton}, {Schwartz}, {Surace}, {Swinyard}, \& {Xu}}]{Fulton:2010}
{Fulton}, T.~R. {et~al.} 2010, in Society of Photo-Optical Instrumentation
  Engineers (SPIE) Conference Series, Vol. 7731, Society of Photo-Optical
  Instrumentation Engineers (SPIE) Conference Series

\bibitem[{Goldsmith \& Langer(1978)}]{Goldsmith:1978}
Goldsmith, P.~F. \& Langer, W.~D. 1978, Astrophysical Journal, 222, 881

\bibitem[{{Gordon} {et~al.}(2008){Gordon}, {Engelbracht}, {Rieke}, {Misselt},
  {Smith}, \& {Kennicutt}}]{Gordon:2008}
{Gordon}, K.~D., {Engelbracht}, C.~W., {Rieke}, G.~H., {Misselt}, K.~A.,
  {Smith}, J.-D.~T., \& {Kennicutt}, Jr., R.~C. 2008, \apj, 682, 336

\bibitem[{Griffin {et~al.}(2010)Griffin, Abergel, Abreu, Ade, Andr{\'e},
  Augueres, Babbedge, Bae, Baillie, Baluteau, Barlow, Bendo, Benielli, Bock,
  Bonhomme, Brisbin, Brockley-Blatt, Caldwell, Cara, Castro-Rodriguez, Cerulli,
  Chanial, Chen, Clark, Clements, Clerc, Coker, Communal, Conversi, Cox, Crumb,
  Cunningham, Daly, Davis, de~Antoni, Delderfield, Devin, di~Giorgio,
  Didschuns, Dohlen, Donati, Dowell, Dowell, Duband, Dumaye, Emery, Ferlet,
  Ferrand, Fontignie, Fox, Franceschini, Frerking, Fulton, Garcia, Gastaud,
  Gear, Glenn, Goizel, Griffin, Grundy, Guest, Guillemet, Hargrave, Harwit,
  Hastings, Hatziminaoglou, Herman, Hinde, Hristov, Huang, Imhof, Isaak,
  Israelsson, Ivison, Jennings, Kiernan, King, Lange, Latter, Laurent, Laurent,
  Leeks, Lellouch, Levenson, Li, Li, Lilienthal, Lim, Liu, Lu, Madden,
  Mainetti, Marliani, McKay, Mercier, Molinari, Morris, Moseley, Mulder, Mur,
  Naylor, Nguyen, O'Halloran, Oliver, Olofsson, Olofsson, Orfei, Page, Pain,
  Panuzzo, Papageorgiou, Parks, Parr-Burman, Pearce, Pearson,
  P{\'e}rez-Fournon, Pinsard, Pisano, Podosek, Pohlen, Polehampton, Pouliquen,
  Rigopoulou, Rizzo, Roseboom, Roussel, Rowan-Robinson, Rownd, Saraceno,
  Sauvage, Savage, Savini, Sawyer, Scharmberg, Schmitt, Schneider, Schulz,
  Schwartz, Shafer, Shupe, Sibthorpe, Sidher, Smith, Smith, Smith, Spencer,
  Stobie, Sudiwala, Sukhatme, Surace, Stevens, Swinyard, Trichas, Tourette,
  Triou, Tseng, Tucker, Turner, Vaccari, Valtchanov, Vigroux, Virique,
  Voellmer, Walker, Ward, Waskett, Weilert, Wesson, White, Whitehouse, Wilson,
  Winter, Woodcraft, Wright, Xu, Zavagno, Zemcov, Zhang, \&
  Zonca}]{Griffin:2010}
Griffin, M.~J. {et~al.} 2010, Astronomy and Astrophysics, 518, L3

\bibitem[{Howe {et~al.}(2000)Howe, Ashby, Bergin, Chin, Erickson, Goldsmith,
  Harwit, Hollenbach, Kaufman, Kleiner, Koch, Neufeld, Patten, Plume, Schieder,
  Snell, Stauffer, Tolls, Wang, Winnewisser, Zhang, \& Melnick}]{Howe:2000}
Howe, J.~E. {et~al.} 2000, The Astrophysical Journal, 539, L137

\bibitem[{Imara \& Blitz(2011)}]{Imara:2011}
Imara, N. \& Blitz, L. 2011, The Astrophysical Journal, 732, 78

\bibitem[{Kamenetzky {et~al.}(2011)Kamenetzky, Glenn, Maloney, Aguirre, Bock,
  Bradford, Earle, Inami, Matsuhara, Murphy, Naylor, Nguyen, \&
  Zmuidzinas}]{Kamenetzky:2011}
Kamenetzky, J. {et~al.} 2011, ApJ, 731, 83

\bibitem[{{Le Bourlot} {et~al.}(1999){Le Bourlot}, {Pineau des For{\^e}ts}, \&
  {Flower}}]{LeBourlot:1999}
{Le Bourlot}, J., {Pineau des For{\^e}ts}, G., \& {Flower}, D.~R. 1999, \mnras,
  305, 802

\bibitem[{Li {et~al.}(2004)Li, Goldsmith, \& Melnick}]{Li:2004}
Li, D., Goldsmith, P.~F., \& Melnick, G. 2004, Milky Way Surveys: The Structure
  and Evolution of our Galaxy, 317, 82

\bibitem[{Loenen {et~al.}(2010)Loenen, van~der Werf, G{\"u}sten, Meijerink,
  Israel, Requena-Torres, Garc{\'\i}a-Burillo, Harris, Klein, Kramer, Lord,
  Mart{\'\i}n-Pintado, R{\"o}llig, Stutzki, Szczerba, Wei{\ss}, Philipp-May,
  Yorke, Caux, Delforge, Helmich, Lorenzani, Morris, Philips, Risacher, \&
  Tielens}]{Loenen:2010}
Loenen, A.~F. {et~al.} 2010, Astronomy and Astrophysics, 521, L2

\bibitem[{Mao {et~al.}(2000)Mao, Henkel, Schulz, Zielinsky, Mauersberger,
  St{\"o}rzer, Wilson, \& Gensheimer}]{Mao:2000}
Mao, R.~Q., Henkel, C., Schulz, A., Zielinsky, M., Mauersberger, R.,
  St{\"o}rzer, H., Wilson, T.~L., \& Gensheimer, P. 2000, Astronomy and
  Astrophysics, 358, 433

\bibitem[{Mayya {et~al.}(2005)Mayya, Carrasco, \& Luna}]{Mayya:2005}
Mayya, Y.~D., Carrasco, L., \& Luna, A. 2005, The Astrophysical Journal, 628,
  L33

\bibitem[{Meijerink {et~al.}(2006)Meijerink, Spaans, \&
  Israel}]{Meijerink:2006}
Meijerink, R., Spaans, M., \& Israel, F.~P. 2006, The Astrophysical Journal,
  650, L103

\bibitem[{Monje {et~al.}(2011)Monje, Emprechtinger, Phillips, Lis, Goldsmith,
  Bergin, Bell, Neufeld, \& Sonnentrucker}]{Monje:2011}
Monje, R.~R. {et~al.} 2011, The Astrophysical Journal Letters, 734, L23

\bibitem[{Naylor {et~al.}(2010)Naylor, Bradford, Aguirre, Bock, Earle, Glenn,
  Inami, Kamenetzky, Maloney, Matsuhara, Nguyen, \& Zmuidzinas}]{Naylor:2010}
Naylor, B.~J. {et~al.} 2010, The Astrophysical Journal, 722, 668

\bibitem[{Nikola {et~al.}(2011)Nikola, Stacey, Brisbin, Ferkinhoff,
  Hailey-Dunsheath, Parshley, \& Tucker}]{Nikola:2011}
Nikola, T., Stacey, G.~J., Brisbin, D., Ferkinhoff, C., Hailey-Dunsheath, S.,
  Parshley, S., \& Tucker, C. 2011, eprint arXiv, 1108, 3213

\bibitem[{Panuzzo {et~al.}(2010)Panuzzo, Rangwala, Rykala, Isaak, Glenn,
  Wilson, Auld, Baes, Barlow, Bendo, Bock, Boselli, Bradford, Buat,
  Castro-Rodr{\'\i}guez, Chanial, Charlot, Ciesla, Clements, Cooray, Cormier,
  Cortese, Davies, Dwek, Eales, Elbaz, Fulton, Galametz, Galliano, Gear, Gomez,
  Griffin, Hony, Levenson, Lu, Madden, O'Halloran, Okumura, Oliver, Page,
  Papageorgiou, Parkin, P{\'e}rez-Fournon, Pohlen, Polehampton, Rigby, Roussel,
  Sacchi, Sauvage, Schulz, Schirm, Smith, Spinoglio, Stevens, Srinivasan,
  Symeonidis, Swinyard, Trichas, Vaccari, Vigroux, Wozniak, Wright, \&
  Zeilinger}]{Panuzzo:2010}
Panuzzo, P. {et~al.} 2010, Astronomy and Astrophysics, 518, L37

\bibitem[{Papadopoulos \& Greve(2004)}]{Papadopoulos:2004b}
Papadopoulos, P.~P. \& Greve, T.~R. 2004, The Astrophysical Journal, 615, L29

\bibitem[{Papadopoulos {et~al.}(2004)Papadopoulos, Thi, \&
  Viti}]{Papadopoulos:2004a}
Papadopoulos, P.~P., Thi, W.-F., \& Viti, S. 2004, Monthly Notices of the Royal
  Astronomical Society, 351, 147

\bibitem[{Pilbratt {et~al.}(2010)Pilbratt, Riedinger, Passvogel, Crone, Doyle,
  Gageur, Heras, Jewell, Metcalfe, Ott, \& Schmidt}]{Pilbratt:2010}
Pilbratt, G.~L. {et~al.} 2010, Astronomy and Astrophysics, 518, L1

\bibitem[{{Pon} {et~al.}(2012){Pon}, {Johnstone}, \& {Kaufman}}]{Pon:2012}
{Pon}, A., {Johnstone}, D., \& {Kaufman}, M.~J. 2012, ArXiv e-prints

\bibitem[{{Rangwala} {et~al.}(2011){Rangwala}, {Maloney}, {Glenn}, {Wilson},
  {Rykala}, {Isaak}, {Baes}, {Bendo}, {Boselli}, {Bradford}, {Clements},
  {Cooray}, {Fulton}, {Imhof}, {Kamenetzky}, {Madden}, {Mentuch}, {Sacchi},
  {Sauvage}, {Schirm}, {Smith}, {Spinoglio}, \& {Wolfire}}]{Rangwala:2011}
{Rangwala}, N. {et~al.} 2011, \apj, 743, 94

\bibitem[{Rigopoulou {et~al.}(2002)Rigopoulou, Kunze, Lutz, Genzel, \&
  Moorwood}]{Rigopoulou:2002}
Rigopoulou, D., Kunze, D., Lutz, D., Genzel, R., \& Moorwood, A. F.~M. 2002,
  Astronomy and Astrophysics, 389, 374

\bibitem[{Roussel {et~al.}(2010)Roussel, Wilson, Vigroux, Isaak, Sauvage,
  Madden, Auld, Baes, Barlow, Bendo, Bock, Boselli, Bradford, Buat,
  Castro-Rodriguez, Chanial, Charlot, Ciesla, Clements, Cooray, Cormier,
  Cortese, Davies, Dwek, Eales, Elbaz, Galametz, Galliano, Gear, Glenn, Gomez,
  Griffin, Hony, Levenson, Lu, O'Halloran, Okumura, Oliver, Page, Panuzzo,
  Papageorgiou, Parkin, Perez-Fournon, Pohlen, Rangwala, Rigby, Rykala, Sacchi,
  Schulz, Schirm, Smith, Spinoglio, Stevens, Srinivasan, Symeonidis, Trichas,
  Vaccari, Wozniak, Wright, \& Zeilinger}]{Roussel:2010}
Roussel, H. {et~al.} 2010, Astronomy and Astrophysics, 518, L66

\bibitem[{Sanders {et~al.}(2003)Sanders, Mazzarella, Kim, Surace, \&
  Soifer}]{Sanders:2003}
Sanders, D.~B., Mazzarella, J.~M., Kim, D.-C., Surace, J.~A., \& Soifer, B.~T.
  2003, The Astronomical Journal, 126, 1607

\bibitem[{Schilke {et~al.}(1993)Schilke, Carlstrom, Keene, \&
  Phillips}]{Schilke:1993}
Schilke, P., Carlstrom, J.~E., Keene, J., \& Phillips, T.~G. 1993,
  Astrophysical Journal Letters v.417, 417, L67

\bibitem[{{Scott} {et~al.}(2011){Scott}, {Lupu}, {Aguirre}, {Auld}, {Aussel},
  {Baker}, {Beelen}, {Bock}, {Bradford}, {Brisbin}, {Burgarella}, {Carpenter},
  {Chanial}, {Chapman}, {Clements}, {Conley}, {Cooray}, {Cox}, {Dowell},
  {Eales}, {Farrah}, {Franceschini}, {Frayer}, {Gavazzi}, {Glenn}, {Griffin},
  {Harris}, {Ibar}, {Ivison}, {Kamenetzky}, {Kim}, {Krips}, {Maloney},
  {Matsuhara}, {Mortier}, {Murphy}, {Naylor}, {Neri}, {Nguyen}, {Oliver},
  {Omont}, {Page}, {Papageorgiou}, {Pearson}, {P{\'e}rez-Fournon}, {Pohlen},
  {Rawlings}, {Raymond}, {Riechers}, {Rodighiero}, {Roseboom},
  {Rowan-Robinson}, {Scott}, {Seymour}, {Smith}, {Symeonidis}, {Tugwell},
  {Vaccari}, {Vieira}, {Vigroux}, {Wang}, {Wright}, \&
  {Zmuidzinas}}]{Scott:2011}
{Scott}, K.~S. {et~al.} 2011, \apj, 733, 29

\bibitem[{{Strickland} \& {Heckman}(2007)}]{Strickland:2007}
{Strickland}, D.~K. \& {Heckman}, T.~M. 2007, \apj, 658, 258

\bibitem[{Stutzki {et~al.}(1997)Stutzki, Graf, Haas, Honingh, Hottgenroth,
  Jacobs, Schieder, Simon, Staguhn, Winnewisser, Martin, Peters, \&
  McMullin}]{Stutzki:1997}
Stutzki, J. {et~al.} 1997, Astrophysical Journal Letters v.477, 477, L33

\bibitem[{Swinyard {et~al.}(2010)Swinyard, Ade, Baluteau, Aussel, Barlow,
  Bendo, Benielli, Bock, Brisbin, Conley, Conversi, Dowell, Dowell, Ferlet,
  Fulton, Glenn, Glauser, Griffin, Griffin, Guest, Imhof, Isaak, Jones, King,
  Leeks, Levenson, Lim, Lu, Makiwa, Naylor, Nguyen, Oliver, Panuzzo,
  Papageorgiou, Pearson, Pohlen, Polehampton, Pouliquen, Rigopoulou, Ronayette,
  Roussel, Rykala, Savini, Schulz, Schwartz, Shupe, Sibthorpe, Sidher, Smith,
  Spencer, Trichas, Triou, Valtchanov, Wesson, Woodcraft, Xu, Zemcov, \&
  Zhang}]{Swinyard:2010}
Swinyard, B.~M. {et~al.} 2010, Astronomy and Astrophysics, 518, L4

\bibitem[{Taylor {et~al.}(2001)Taylor, Walter, \& Yun}]{Taylor:2001}
Taylor, C.~L., Walter, F., \& Yun, M.~S. 2001, The Astrophysical Journal, 562,
  L43

\bibitem[{van~der Tak {et~al.}(2007)van~der Tak, Black, Sch{\"o}ier, Jansen, \&
  van Dishoeck}]{vanderTak:2007}
van~der Tak, F. F.~S., Black, J.~H., Sch{\"o}ier, F.~L., Jansen, D.~J., \& van
  Dishoeck, E.~F. 2007, Astronomy and Astrophysics, 468, 627

\bibitem[{van~der Werf {et~al.}(2010)van~der Werf, Isaak, Meijerink, Spaans,
  Rykala, Fulton, Loenen, Walter, Wei{\ss}, Armus, Fischer, Israel, Harris,
  Veilleux, Henkel, Savini, Lord, Smith, Gonz{\'a}lez-Alfonso, Naylor, Aalto,
  Charmandaris, Dasyra, Evans, Gao, Greve, G{\"u}sten, Kramer,
  Mart{\'\i}n-Pintado, Mazzarella, Papadopoulos, Sanders, Spinoglio, Stacey,
  Vlahakis, Wiedner, \& Xilouris}]{vanderWerf:2010}
van~der Werf, P.~P. {et~al.} 2010, Astronomy and Astrophysics, 518, L42

\bibitem[{van Dishoeck \& Black(1986)}]{vanDishoeck:1986}
van Dishoeck, E.~F. \& Black, J.~H. 1986, Astrophysical Journal Supplement
  Series (ISSN 0067-0049), 62, 109

\bibitem[{van Dishoeck {et~al.}(2011)van Dishoeck, Kristensen, Benz, Bergin,
  Caselli, Cernicharo, Herpin, Hogerheijde, Johnstone, Liseau, Nisini, Shipman,
  Tafalla, van~der Tak, Wyrowski, Aikawa, Bachiller, Baudry, Benedettini,
  Bjerkeli, Blake, Bontemps, Braine, Brinch, Bruderer, Chavarr{\'\i}a, Codella,
  Daniel, de~Graauw, Deul, di~Giorgio, Dominik, Doty, Dubernet, Encrenaz,
  Feuchtgruber, Fich, Frieswijk, Fuente, Giannini, Goicoechea, Helmich,
  Herczeg, Jacq, J{\o}rgensen, Karska, Kaufman, Keto, Larsson, Lefloch, Lis,
  Marseille, McCoey, Melnick, Neufeld, Olberg, Pagani, Pani{\'c}, Parise,
  Pearson, Plume, Risacher, Salter, Santiago-Garc{\'\i}a, Saraceno,
  St{\"a}uber, van Kempen, Visser, Viti, Walmsley, Wampfler, \&
  Yıldız}]{vanDishoeck:2011}
van Dishoeck, E.~F. {et~al.} 2011, Publications of the Astronomical Society of
  the Pacific, 123, 138

\bibitem[{{VERITAS Collaboration} {et~al.}(2009){VERITAS Collaboration},
  Acciari, Aliu, ArleConn, Aune, Bautista, Beilicke, Benbow, Boltuch, Bradbury,
  Buckley, Bugaev, Byrum, Cannon, Celik, Cesarini, Chow, Ciupik, Cogan, Colin,
  Cui, Dickherber, Duke, Fegan, Finley, Finnegan, Fortin, Fortson, Furniss,
  Galante, Gall, Gibbs, Gillanders, Godambe, Grube, Guenette, Gyuk, Hanna,
  Holder, Horan, Hui, Humensky, Imran, Kaaret, Karlsson, Kertzman, Kieda,
  Kildea, Konopelko, Krawczynski, Krennrich, Lang, Lebohec, Maier, McArthur,
  McCann, McCutcheon, Millis, Moriarty, Mukherjee, Nagai, Ong, Otte, Pandel,
  Perkins, Pizlo, Pohl, Quinn, Ragan, Reyes, Reynolds, Roache, Rose,
  Schroedter, Sembroski, Smith, Steele, Swordy, Theiling, Thibadeau, Varlotta,
  Vassiliev, Vincent, Wagner, Wakely, Ward, Weekes, Weinstein, Weisgarber,
  Williams, Wissel, Wood, \& Zitzer}]{Veritas:2009}
{VERITAS Collaboration} {et~al.} 2009, Nature, 462, 770

\bibitem[{Walter {et~al.}(2002)Walter, Weiss, \& Scoville}]{Walter:2002}
Walter, F., Weiss, A., \& Scoville, N. 2002, The Astrophysical Journal, 580,
  L21

\bibitem[{Ward {et~al.}(2003)Ward, Zmuidzinas, Harris, \& Isaak}]{Ward:2003}
Ward, J.~S., Zmuidzinas, J., Harris, A.~I., \& Isaak, K.~G. 2003, The
  Astrophysical Journal, 587, 171

\bibitem[{Wei{\ss} {et~al.}(2001)Wei{\ss}, Neininger, H{\"u}ttemeister, \&
  Klein}]{Weiss:2001}
Wei{\ss}, A., Neininger, N., H{\"u}ttemeister, S., \& Klein, U. 2001, Astronomy
  and Astrophysics, 365, 571

\bibitem[{Wei{\ss} {et~al.}(2010)Wei{\ss}, Requena-Torres, G{\"u}sten,
  Garc{\'\i}a-Burillo, Harris, Israel, Klein, Kramer, Lord, Martin-Pintado,
  R{\"o}llig, Stutzki, Szczerba, van~der Werf, Philipp-May, Yorke, Akyilmaz,
  Gal, Higgins, Marston, Roberts, Schl{\"o}der, Schultz, Teyssier, Whyborn, \&
  Wunsch}]{Weiss:2010}
Wei{\ss}, A. {et~al.} 2010, Astronomy and Astrophysics, 521, L1

\bibitem[{Wei{\ss} {et~al.}(2005)Wei{\ss}, Walter, \& Scoville}]{Weiss:2005}
Wei{\ss}, A., Walter, F., \& Scoville, N.~Z. 2005, Astronomy and Astrophysics,
  438, 533

\bibitem[{White {et~al.}(1994)White, Ellison, Claude, Dent, \&
  Matheson}]{White:1994}
White, G.~J., Ellison, B., Claude, S., Dent, W. R.~F., \& Matheson, D.~N. 1994,
  Astronomy and Astrophysics (ISSN 0004-6361), 284, L23

\bibitem[{Wild {et~al.}(1992)Wild, Harris, Eckart, Genzel, Graf, Jackson,
  Russell, \& Stutzki}]{Wild:1992}
Wild, W., Harris, A.~I., Eckart, A., Genzel, R., Graf, U.~U., Jackson, J.~M.,
  Russell, A. P.~G., \& Stutzki, J. 1992, Astronomy and Astrophysics (ISSN
  0004-6361), 265, 447

\bibitem[{Wolfire {et~al.}(2010)Wolfire, Hollenbach, \& McKee}]{Wolfire:2010}
Wolfire, M.~G., Hollenbach, D., \& McKee, C.~F. 2010, The Astrophysical
  Journal, 716, 1191

\bibitem[{Yun {et~al.}(1993)Yun, Ho, \& Lo}]{Yun:1993}
Yun, M.~S., Ho, P. T.~P., \& Lo, K.~Y. 1993, Astrophysical Journal, 411, L17

\end{thebibliography}
